\documentclass[11pt,a4paper]{article}
\usepackage{jheppub}
\usepackage{natbib}
\usepackage{graphicx}
\usepackage{cases}
\usepackage[dvipsnames]{xcolor}

\usepackage{mathtools}
\usepackage{hyperref}
\usepackage{float}
\usepackage{comment}
\usepackage{multirow}
\usepackage[overload]{empheq}
\usepackage[normalem]{ulem}
\usepackage{orcidlink}
\usepackage{rorlink}
\usepackage{bm}
\usepackage[T1]{fontenc}

\definecolor{Blu}{rgb}{0.,0.,1.}

\author[a]{Marco Cirelli$^{\ \orcidlink{0000-0002-4264-6323}}$}
\author[b]{\ \ Mattia Di Mauro$^{\ \orcidlink{ 0000-0003-2759-5625}}$}
\author[a]{\ \ Arpan Kar$^{\, \orcidlink{0000-0002-2993-3336}}$}
\affiliation[a]{Laboratoire de Physique Théorique et Hautes Énergies (\href{https://www.lpthe.jussieu.fr}{LPTHE})$^{\ \rorlink{https://ror.org/02mph9k76}}$,\\ CNRS $\&$ Sorbonne Université,
4 Place Jussieu, Paris, France}
\affiliation[b]{Istituto Nazionale di Fisica Nucleare, via P. Giuria, 1, 10125 Torino, Italy$^{\ \rorlink{https://ror.org/01vj6ck58}}$}
\emailAdd{marco.cirelli@gmail.com}
\emailAdd{dimauro.mattia@gmail.com}
\emailAdd{arpankarphys@gmail.com}

\title{Refined anti-proton and anti-deuteron fluxes\\ from weak-scale Dark Matter}

\abstract{We provide the cosmic-ray (CR) fluxes of antiprotons and antideuterons
produced by the Galactic annihilation or decay
of weak-scale dark matter (DM) particles of masses in the range from a few GeV to 100 TeV.
We estimate these fluxes based on the updated models for the propagation of charged
particles in the Galaxy and using the improved $\bar{p}$ and $\bar{d}$ spectra
provided by \texttt{CosmiXs}. For the updated propagation models we consider
the MIN/MED/MAX sets under the new {\sc Slim}/{\sc Big}/{\sc Quaint} schemes.
We treat the Galactic propagation in a semi-analytic way including different
possible effects such as spatial diffusion, energy-losses, convection
and diffusive reacceleration. For the DM distribution in the Galactic halo we consider
NFW, Einasto and Burkert profiles with the most updated parameters.
Moreover, we also incorporate the latest models for the inelastic cross-sections of $\bar{p}$ and $\bar{d}$ based on {\sc Alice} data. We validate our calculations with those available
in the literature or those obtained from other publicly available numerical packages.
We compare the CR fluxes obtained in this work with those provided
previously by \texttt{PPPC4DMID} which were based on
old propagation scenarios. We find that the
CR fluxes obtained here with the new propagation models are much more robust
(compared to the older ones) under the variation MIN--MAX.
We also discuss the impact of this in the improvement
of the discovery potential of a possible DM signal in the light of the
present and upcoming CR observations.
We provide all our results for the DM-induced interstellar CR fluxes in a tabulated format
(for the kinetic energy range 0.1 GeV -- 100 TeV) in the \texttt{GitHub} repository of the newly created \href{https://github.com/CosmiXsPPPC}{\texttt{CosmiXsPPPC}} project.
The results are ready to be used for studies related to DM indirect searches.}

\begin{document}

\maketitle

\section{Introduction}
\label{sec:introduction}

Searches for the particle nature of dark matter (DM) follow several complementary strategies \cite{Cirelli:2024ssz}. Indirect detection aims at identifying stable Standard Model particles produced by DM annihilations or decays in the Milky Way halo, or in other astrophysical environments. Among charged cosmic-ray (CR) messengers, antiprotons are presently the most constraining antimatter channel for many hadronic DM scenarios, while antideuterons and, more generally, light antinuclei are especially appealing discovery channels because their expected secondary backgrounds are strongly suppressed at low kinetic energy per nucleon \cite{Cuoco2017NovelConstraints,DeLaTorreLuque2024DRAGON2,Donato2000Antideuterons,Donato2008AntideuteronFluxes,Stefanuto:2026wxy}. At the same time, current {\sc Ams-02} antiproton data are broadly consistent with a predominantly secondary origin, so any DM contribution is expected to appear as a subdominant component on top of a precisely modeled astrophysical background \cite{Giesen2015AMS02Antiprotons,Boudaud2020SecondaryOrigin,DeLaTorreLuque2024DRAGON2,DiMauro:2023jgg}. For weak-scale DM, roughly from a few GeV to several tens of TeV, cosmic antimatter nevertheless remains a prime indirect-detection probe and already provides some of the strongest constraints on broad classes of models \cite{Cuoco2017NovelConstraints,DeLaTorreLuque2024DRAGON2}.

The computation of the observable antimatter flux involves two main ingredients: the source spectra at production and the Galactic transport. In the first step, the particle-physics properties of DM---such as its mass, annihilation or decay channels, and annihilation cross section or decay rate---together with its Galactic spatial distribution determine the source term. The subsequent propagation step describes how the produced CRs travel through the turbulent and magnetized interstellar medium, where diffusion, convection, reacceleration, spallation and energy losses reshape the spectrum.

On the source side, the Poor Particle Physicist's Cookbook for Dark Matter Indirect Detection (\texttt{PPPC4DMID}) \cite{Cirelli:2010xx} has provided, since 2010, the CR spectra at production. These were obtained through extensive Monte Carlo simulations with \texttt{Pythia} \cite{Sjostrand:2007gs,Sjostrand:2014zea}, supplemented by electroweak corrections added at first order. \texttt{PPPC4DMID} has been widely used by the community thanks to its relatively {\it `plug\&play'} nature, which allows one to readily download and use interpolating functions or numerical tables for CR spectra (among other observables), for a large set of DM annihilation channels and arbitrary values of the DM mass.

In recent years, the \texttt{CosmiXs} results \cite{Arina:2023eic} have provided the same type of products, but with several significant improvements. In particular, the calculation of the source spectra for annihilating and decaying DM has been revisited using the Vincia shower algorithm in \texttt{Pythia}, in order to include QED and QCD final-state radiation, as well as diagrams relevant for electroweak (EW) corrections involving massive bosons, which are not present in the default \texttt{Pythia} shower model. Spin information is retained throughout the entire EW shower, and off-shell contributions from massive gauge bosons are also taken into account. Furthermore, a dedicated tuning of the Vincia and \texttt{Pythia} parameters to LEP data on pion, photon, and hyperon production at the $Z$ resonance is performed, and the associated uncertainties are discussed.

\medskip

On the galactic transport side, back in 2003 Ref.~\cite{Donato2004NeutralinoAntiprotons} introduced three benchmark parameter sets, conventionally denoted MIN, MED and  MAX. These benchmarks were selected among transport models compatible with the boron/carbon data available at the time, with  MED close to the best-fit configuration and MIN/MAX chosen to bracket the DM-induced antiproton signal. They were later used extensively in DM phenomenology, and analogous benchmark variants were also adopted for positron studies \cite{Donato2004NeutralinoAntiprotons,Delahaye2008Positrons}. Subsequent work showed that effects often neglected in early antiproton calculations---most notably energy losses, tertiary antiproton production and diffusive reacceleration---can have a non-negligible impact and should be included for accurate predictions \cite{Boudaud2015Fussy,Giesen2015AMS02Antiprotons}.

With the advent of precise {\sc Ams-02} data, the original 2003 benchmarks became outdated. Ref.~\cite{Genolini2019Transport} introduced the new transport schemes {\sc Big}, {\sc Slim}, and {\sc Quaint}, which differ in their treatment of diffusion, convection, and reacceleration while all providing good fits to modern secondary-to-primary data. Building on this framework, Ref.~\cite{Genolini2021MinMedMax} derived updated MIN-MED-MAX benchmark sets for each scheme, consistently accounting for the breaks in the diffusion coefficient inferred from recent data. These new benchmarks considerably reduce the transport uncertainty on DM-induced charged CR fluxes and provide an up-to-date basis for phenomenological analyses \cite{Genolini2019Transport,Genolini2021MinMedMax,Stefanuto:2026wxy}.

\medskip

The goal of this work is to combine these two advances: the improved \texttt{CosmiXs} source spectra and the updated Galactic transport benchmarks. Concretely, we compute and publicly provide propagated antiproton and antideuteron fluxes within the {\sc Big}, {\sc Slim}, and {\sc Quaint} schemes and for their corresponding updated MIN-MED-MAX realizations, while preserving the broad set of annihilation and decay channels available in \texttt{CosmiXs}. This framework enables up-to-date phenomenological studies for essentially any weak-scale DM scenario. In this paper we focus on antiprotons and, with only minor modifications on the transport side once the source term is specified, on antideuterons. Positrons will be addressed in future work.

\medskip

The rest of this paper is organized as follows. In Section~\ref{sec:propagation} we briefly review the main ingredients of CR propagation in the Galaxy. In Section~\ref{sec:results} we present our results for the propagated antiproton and antideuteron fluxes. In Section~\ref{sec:applications} we focus on the impact for future searches of the improvements that we obtain and finally, in Section~\ref{sec:conclusions}, we summarize our conclusions.

\section{Propagation of CR antinuclei in the Galaxy}
\label{sec:propagation}

In this section we briefly review the standard formalism adopted for the propagation of antiprotons and antideuterons in the Milky Way. In the spirit of making this paper self-contained, we lay out all the different ingredients and tools explicitly. We nevertheless refer to the cited classical references for more details.

\subsection{Transport equation}
\label{sec:transport_eq}

The transport equation of a hadronic CR species $j$ (e.g., $\bar{p}, \overline{d}, \overline{\rm He}$)
can be written as follows, assuming steady-state propagation and 2D cylindrical symmetry
\cite{Calore:2022stf, Fornengo:2013xda, Maurin:2002ua}: \\
\vspace{-3mm}
\begin{eqnarray}
&-& \left[ D(K) \left( \frac{\partial^2}{\partial z^2}  + \frac{1}{r} \frac{\partial}{\partial r}
\left(r \frac{\partial}{\partial r}\right) \right) - V_c \frac{\partial}{\partial z} \right] f_j \,
+ \, 2 h \, \delta(z) \frac{\partial}{\partial K} \left[ b_{\rm loss}(K) \, f_j - \beta^2 \, D_{pp} (K) \, \frac{\partial f_j}{\partial K} \right] \nonumber\\ &&
= Q_j \, - \, 2 h \, \delta(z) \, \Gamma^j_{\rm inel}(K) \, f_j,
\label{eq:Transport_Eq}
\end{eqnarray}
\vspace{1mm}
where $f_j(r, z, K) \equiv dn_j/dK$ is the number density per unit kinetic energy of the propagated species,
with kinetic energy $K$ (corresponding to a velocity $\beta c$),
at a position identified by the galactic cylindrical coordinates $(r,z)$. The CRs
are assumed to be confined within a galactic cylindrical geometry of
half-thickness $L$ and radius $R_G$.

The different components of the transport equation \eqref{eq:Transport_Eq} are as follows.
\begin{itemize}
\item[$\circ$] $D(K)$ is the spatial diffusion coefficient, assumed to be isotropic and homogeneous
within the Galactic cylinder, and is expressed in general as \cite{Genolini2019Transport, Genolini:2017dfb}:
\begin{equation}
D(R) = \beta^\eta D_0 \left[ 1 + \left( \frac{R_l}{R} \right)^{\frac{-\delta_l+\delta}{s_l}} \right]^{s_l} \left( \frac{R}{\rm 1 \, GV}\right)^{\delta}
\left[ 1 + \left( \frac{R}{R_h} \right)^{\frac{\delta-\delta_h}{s_h}} \right]^{-s_h},
\end{equation}
as a function of the particle rigidity $R = p / |Ze|$, with $p$ and $Z$ the momentum and charge
of the particle, respectively. Here, the parameters $R_{l(h)}$ indicate the positions of the spectral breaks, $\delta_l$, $\delta$ and $\delta_h$ denote the diffusion spectral indices in the low-, intermediate-, and high-rigidity regimes, respectively, and $s_{l(h)}$ characterizes the rate at which the spectral change takes place around $R_{l(h)}$.

\item[$\circ$] $V_c$ is the convection velocity, assumed here to be constant and
perpendicular to the Galactic disk.

\item[$\circ$] The quantity $h$ $(\ll L)$ denotes the half-thickness of the Galactic disk,
where the interstellar medium (ISM) gas particles are confined.

\item[$\circ$] $b_{\rm loss}(K)$ represents the total energy-loss term,
\begin{equation}
b_{\rm loss} = \left(\frac{dK}{dt}\right)_{\rm ion, \, Coul} + \left(\frac{dK}{dt}\right)_{\rm adia} +
\left(\frac{dK}{dt}\right)_{\rm reacc} + \left(\frac{dK}{dt}\right)_{\rm pion}.
\end{equation}
The expressions for ionization, Coulomb and adiabatic losses are taken from the literature.
Explicitly, for ionization and Coulomb losses \cite{Mannheim:1994sv,Strong:1998pw,Maurin:2002ua}:
\begin{align}
\left(\frac{dK}{dt}\right)_{\rm ion} &= - \frac{2\pi r_e^2 m_e c^3 Z^2}{\beta} \sum_{i= {\rm H,\, He}} n_i B_i, \\
&{\rm with} \ \ \ B_i = \ln\left( \frac{2 m_e c^2 \beta^2 \gamma^2 Q_{\rm max}}{I_i^2} \right) - 2 \beta^2, \\
&{\rm and} \ \ Q_{\rm max} = \frac{2 m_e c^2 \beta^2 \gamma^2 M}{M+2\gamma m_e}.  \\
\left(\frac{dK}{dt}\right)_{\rm Coul} &= -4 \pi \, r_e^2 m_e c^3 Z^2 n_e \, \frac{\beta^2}{x_m^3 + \beta^3} \, \ln\Lambda,\\
&{\rm with} \ \ \ln\Lambda = \frac12\ln \left(\frac{m_e^2 c^2}{\pi r_e \hbar^2 n_e}\frac{M \gamma^2 \beta^4}{M+2\gamma m_e}\right)\simeq 40-50,\\
&{\rm and} \ \ x_m = \left(\frac{3\sqrt{\pi}}{4}\right)^{1/3} \sqrt{2 k T_e/m_e c^2}.
\end{align}
The expression for ionization losses holds for $\beta\ge 0.01$. Here $r_e$ is the classical electron radius, $m_e$ is the electron mass, $n_e$ is the density of electrons in the ISM, and $T_e \sim 10^4$ K is their temperature.
The quantities $I_{\rm H} \simeq 19$ eV and $I_{\rm He} \simeq 44$ eV represent the effective ionization potentials of hydrogen and helium, respectively, while $\gamma$ is the Lorentz factor of the CR particle and $M$ its mass.

For the adiabatic losses, induced by the same convective processes mentioned above, one has \cite{Maurin:2002ua,Boudaud2015Fussy}
\begin{equation}
    \left(\frac{dK}{dt}\right)_{\rm adia} = - \frac{V_c}{3h}\frac{p^2}{E},
\end{equation}
where $p=\sqrt{K^2 +2 M K}$ is the momentum of the CR and $E = K + M$ its total energy.

For reacceleration \cite{Maurin:2002ua}, one has
\begin{equation}
    \left(\frac{dK}{dt}\right)_{\rm reacc} = D_{pp} \frac{1+\beta^2}{E},
\end{equation}
where $D_{pp}$ is defined below. The positive overall sign makes it clear that this term corresponds to an energy {\em gain}, rather than a loss.

Finally, for the loss due to pion production in interactions with nuclei of the ISM, we follow the parametrization in \cite{Krakau:2015bea,Evoli:2016xgn}:  
\begin{equation}
    \left(\frac{dK}{dt}\right)_{\rm pion} = - A^{0.79} \,\, 3.85 \times 10^{-16} \left(\frac{n_{\rm HI}+2 n_{\rm H_2}}{\rm cm^{-3}}\right) \left(\frac{E}{\rm GeV}\right)^{1.28} \left( \frac{E}{\rm GeV} +200 \right)^{-0.2} {\rm GeV/s}.
\label{eq:pion_loss}
\end{equation}
Here $n_{\rm HI}$ and $n_{\rm H_2}$ are the densities of atomic and molecular hydrogen, respectively. 
The scaling of the loss term for heavier CR species with a 
mass number $A>1$ (like $\bar{d}$) is obtained following \cite{Krakau_2015,Evoli:2016xgn}.

\item[$\circ$] The term with $D_{pp}(K)$ represents diffusion in momentum space \cite{Genolini2021MinMedMax}, with
\begin{equation}
D_{pp} (K) = \frac{4}{3} \frac{1}{\delta (4-\delta^2) (4-\delta)} \frac{V_a^2 p^2}{D(R)},
\end{equation}
where $V_a$ is the Alfv\'en speed responsible for diffusive reacceleration,
which is assumed to be confined to the Galactic plane; see \cite{Genolini2021MinMedMax, Genolini2019Transport}.

\item[$\circ$] $\Gamma^j_{\rm inel}(K)$ on the right-hand side of the transport equation
represents a sink term for the propagating particle $j$. It accounts for its inelastic interactions with the ISM gas \cite{Cirelli:2010xx},
\begin{equation}
\Gamma^j_{\rm inel}(K) = (n_{\rm H} + 4^{2/3} n_{\rm He}) \, \sigma^{\rm inel}_{jp} \, v_j,
\label{eq:Gamma_inel}
\end{equation}
where $v_j$ is the particle velocity and $\sigma^{\rm inel}_{jp}$ is the cross section for the inelastic interaction of
$j$ with ISM protons. It can be written as
$\sigma^{\rm inel}_{jp} = \sigma^{\rm inel-Ann}_{jp} + \sigma^{\rm inel-NonAnn}_{jp}$.

\item[$\circ$] The term $Q_j$ on the right-hand side of the transport equation corresponds to the
source term of the species $j$ and can contain different contributions. The primary contribution
comes from the annihilation (with cross section $\langle \sigma v \rangle$) or decay (with rate $\Gamma$) of Galactic DM particles:
\begin{equation}
Q^{\rm DM-Ann}_j (r,z,K) = \frac{1}{2} \langle \sigma v \rangle \frac{dN^{\rm Ann}_j}{dK}
\left( \frac{\rho_{\rm DM}(r,z)}{m_{\rm DM}} \right)^2 \, ,
\end{equation}
\begin{equation}
Q^{\rm DM-Dec}_j (r,z,K) = \Gamma \frac{dN^{\rm Dec}_j}{dK}
\left( \frac{\rho_{\rm DM}(r,z)}{m_{\rm DM}} \right) \, .
\end{equation}
Here $dN_j/dK$ is the energy spectrum of particle $j$ produced in
DM annihilation/decay, and $\rho_{\rm DM}$ is the DM density distribution
in the Galaxy.

Apart from the primary DM contribution, $Q_j$ can also contain an additional source term
corresponding to `tertiary' antiparticles emerging from inelastic non-annihilating
interactions of primary antiparticles, produced by DM,
with ISM nuclei~\cite{Boudaud2015Fussy}.
This usually gives rise to an enhancement in the DM-induced flux and can be relevant
mainly at low energies ($K \lesssim$ a few GeV). In this work we ignore
such an extra contribution to the DM-induced flux in order to avoid computational complications and remain conservative.
\end{itemize}

\subsection{Solution of the transport equation}
\label{sec:solution}

We solve the transport equation \eqref{eq:Transport_Eq} to obtain the density of the
CR species $j$ at the location of the Sun, $f_j(r_\odot, 0, K)$.
For this purpose, we follow an approach initially implemented in \cite{Maurin_2001, Donato:2001ms};
see also \cite{Fornengo:2013xda} for the different steps.
Given that processes such as energy losses and diffusive reacceleration
take place only in the thin disk and not in the Galactic halo, the solution is obtained
in the following two main steps:
\begin{enumerate}
\item First, we obtain the solution $f^{(0)}(r_\odot, 0, K)$ for the CR species $j$
without including energy losses and diffusive reacceleration,
i.e., with $b_{\rm loss} = 0$ and $D_{pp} = 0$ in eq.~\eqref{eq:Transport_Eq}. This solution
can be expressed as
\begin{equation}
f^{(0)} (r_\odot, 0, K) = \sum^{\infty}_{n=1} f^{(0)}_n (0, K) \,
J_0 \left(\frac{\zeta_n r_\odot}{R_G} \right) \, ,
\label{eq:Bessel_expansion}
\end{equation}
with the coefficients of the Bessel expansion given by
\begin{equation}
f^{(0)}_n (0,K) = {\rm exp}\left[ -\frac{V_c L}{2 \, D(K)} \right] \,
\frac{y_n(L, K)}{A_n \, {\rm Sinh}\left(\frac{S_n L}{2}\right)} \,\, \times
Q^{\rm DM} (r_\odot, 0, K) \, ,
\end{equation}
where $J_0$ is the zeroth-order Bessel function of the first kind and $\zeta_n$ is its $n$-th zero.
The explicit analytical expressions for $y_n$, $A_n$, and $S_n$ are as follows~\cite{Cirelli:2010xx}:
\begin{eqnarray}
y_n (L, K) &=& \frac{4}{J^2_1(\zeta_n) R^2_G} \, \int^{R_G}_0 dr \, r \, J_0 \left(\frac{\zeta_n r}{R_G}\right)
\, \int^{L}_0 dz \,\, {\rm exp}\left[\frac{V_c (L-z)}{2 \, D(K)} \right] \nonumber\\
&& \times {\rm Sinh}\left(\frac{S_n (L-z)}{2}\right) \,
\left(\frac{\rho_{\rm DM}(r,z)}{\rho_\odot}\right)^2 \, , \\
A_n &=& 2h\Gamma_{\rm inel} \,+\, V_c \,+\, D(K) \, S_n \,
{\rm Coth}\left(\frac{S_n L}{2}\right) \, , \\
S_n &=& \left( \frac{V^2_c}{D^2(K)} + \frac{4 \zeta^2_n}{R^2_G} \right)^{1/2} \, .
\end{eqnarray}

\item The final solution including energy losses and reacceleration is obtained by solving:
\begin{equation}
f_n + \frac{2 h}{A_n} \frac{d}{dK} \left[b_{\rm loss}(K) \, f_n - \beta^2 D_{pp}(K) \frac{df_n}{dK} \right]
= f^{(0)}_n \, .
\end{equation}
This differential equation can be discretized over the energy values $K_i$ as
\begin{equation}
\sum_{i'} \mathcal{A}^{i i'}_n \, f^{i'}_n = f^{(0) i}_n \, .
\label{eq:MatrixEq}
\end{equation}
Here $\mathcal{A}$ denotes a tridiagonal matrix whose form can be
obtained from \cite{Donato:2001ms, Derome:2019jfs}.

We solve eq.~\eqref{eq:MatrixEq} by numerically inverting the matrix $\mathcal{A}$.
Solving this equation also requires specifying the boundary conditions at the lowest and highest energies.
Among the several possibilities tested in \cite{Derome:2019jfs}, we adopt the `2nd order' condition at low energy
and the `pure diffusive limit' condition at high energy, as given in Table C.1 of that reference.

Using the coefficients $f^{i}_n$ (for each value of $K_i$), the final number density
$f(r_\odot, z=0, K_i)$ can be obtained like in eq.~\eqref{eq:Bessel_expansion}, i.~e.~explicitly:
\begin{equation}
f(r_\odot, 0, K) = \sum^{\infty}_{n=1} f_n (0, K) \,
J_0 \left(\frac{\zeta_n r_\odot}{R_G} \right) \, ,
\label{eq:Bessel_expansion_final}
\end{equation}

\end{enumerate}

Finally, the CR flux (per unit energy, area, time, and solid angle) is obtained as
\begin{equation}
\frac{d\phi_j}{dK} (r_\odot, 0, K) = \frac{\beta}{4\pi} f_j (r_\odot, 0, K) \, .
\end{equation}

For completeness, we also include the average solar-modulation effect \cite{Gleeson:1967juf,Gleeson:1968zza}, which is
mainly relevant for low-energy CR particles. Taking this into account, one obtains
the top-of-the-atmosphere (ToA) fluxes (denoted by the symbol $_\oplus$) from the interstellar (IS) fluxes as
\begin{equation}
\frac{d\phi_\oplus}{dK_\oplus} = \frac{p^2_\oplus}{p^2} \frac{d\phi}{dK} \, ,
\hspace{8mm} K = K_\oplus + |Ze| \Phi_F \, ,
\end{equation}
where $\Phi_F$ is the Fisk potential.

In Appendix~\ref{sec:consistency_checks}, we present several consistency checks
for the $\bar{p}$ fluxes obtained in the present analysis, in particular by comparing them
with those from reference \cite{Genolini2021MinMedMax} where the publicly available package \texttt{USINE}~\cite{Maurin:2018rmm} was used.

\subsection{Models and parameters}
\label{sec:Models}

\begin{table}[!t]
\hspace*{-1.4cm}
\begin{tabular}{c|cccccccccccc}
        & $\delta$      & $\log_{10} D_0$   & $\eta$    & $R_\mathrm{l}$    & $\delta_\mathrm{l}$   & $s_\mathrm{l}$    & $R_\mathrm{h}$   & $\delta_\mathrm{h}$    & $s_\mathrm{h}$    & $V_a$ & $V_c$ & $L$ \\[-0.15cm]
 	      &               & {\scriptsize [kpc$^2$/Myr]} & & {\scriptsize [GV]} & &  & {\scriptsize [GV]} &  &  & {\scriptsize [km/s]} & {\scriptsize [km/s]} & {\scriptsize [kpc]}     \\
\hline\hline
\textbf{\textsc{Big}} &&&&&&&&&&&&\\
\hline
MAX     &  0.529        & -1.286             & 1         & 4.755              & -0.742              & 0.05              & 247               & 0.349                 & 0.04              & 6.002     & 1.819     & 6.637   \\
MED     &  0.498        & -1.446             & 1         & 4.490              & -1.102              & 0.05              & 247               & 0.318                 & 0.04              & 4.741     & 0.459     & 4.645   \\
MIN     &  0.465        & -1.616             & 1         & 4.208              & -1.455              & 0.05              & 247               & 0.285                 & 0.04              & 4.277     & 0.066     & 3.206   \\

\hline\hline
\textbf{\textsc{Slim}} &&&&&&&&&&&&\\
\hline
MAX     &     0.490     & -1.18             & 1         & 4.74              & -0.776                & 0.05              & 237               & 0.300                 & 0.04              & 0     & 0     & 8.40   \\
MED     &     0.499     & -1.44             & 1         & 4.48              & -1.11                 & 0.05              & 237               & 0.309                 & 0.04              & 0     & 0     & 4.67   \\
MIN     &     0.509     & -1.71             & 1         & 4.21              & -1.45                 & 0.05              & 237               & 0.319                 & 0.04              & 0     & 0     & 2.56   \\

\hline\hline
\textbf{\textsc{Quaint}} &&&&&&&&&&&&\\
\hline
MAX     &     0.504     & -1.092             & -1.001    & 0                & $-$                   & $-$              & 270               & 0.334                 & 0.04              & 83.929     & 0.469     & 6.840   \\
MED     &     0.451     & -1.367             & -2.156    & 0                & $-$                   & $-$              & 270               & 0.281                 & 0.04              & 52.066     & 0.239     & 4.080   \\
MIN     &     0.403     & -1.643             & -3.412    & 0                & $-$                   & $-$              & 270               & 0.233                 & 0.04              & 18.389     & 0.151     & 2.630   \\
\hline
\end{tabular}

\caption{
Antiproton propagation parameters for the MIN, MED and MAX benchmark sets, in the three schemes {\sc Big}, {\sc Slim} and {\sc Quaint}.}
\label{tab:propparams}
\end{table}

\begin{itemize}
\item[$\circ$] \underline{\bf Propagation models:}
We consider the MIN/MED/MAX propagation sets from \cite{Genolini2021MinMedMax},
obtained within the new propagation schemes {\sc Big/Slim/Quaint}~\cite{Genolini2019Transport}.
The values of the different propagation parameters described in sec.~\ref{sec:transport_eq}
are taken from \cite{Genolini2021MinMedMax} and reported in table \ref{tab:propparams} for self-containedness.

The value of the half-thickness $h$ is taken to be $h = 200$ pc, since we checked that
this value of $h$ gives a better agreement between the
antiparticle flux estimated here and that obtained from numerical simulations such as
\texttt{DRAGON}~\cite{Evoli:2016xgn, Evoli:2017vim} (see Appendix~\ref{sec:consistency_checks}).

To describe the interaction of the propagated antinuclei with the ISM gas
(encoded by $\sigma^{\rm inel}_{jp}$ in eq.~\eqref{eq:Gamma_inel}),
we adopt an analytic Glauber-eikonal framework,
following refs.~\cite{Glauber:1955qq, Glauber:1970jm, Miller:2007ri, Shukla:2003mb}.
This gives an estimate consistent with the {\sc Alice} measurement \cite{ALICE:2020mso}.
The estimated rates $\Gamma^j_{\rm inel}$ for the different species are shown in fig.~\ref{fig:Gamma_inel}.
In this figure we report the comparison with respect to the simpler model used also in \cite{Cirelli:2010xx}.
See Appendix \ref{sec:inelastic} for more details.

\begin{figure*}[!t]
\centering
\includegraphics[width=0.6\textwidth]{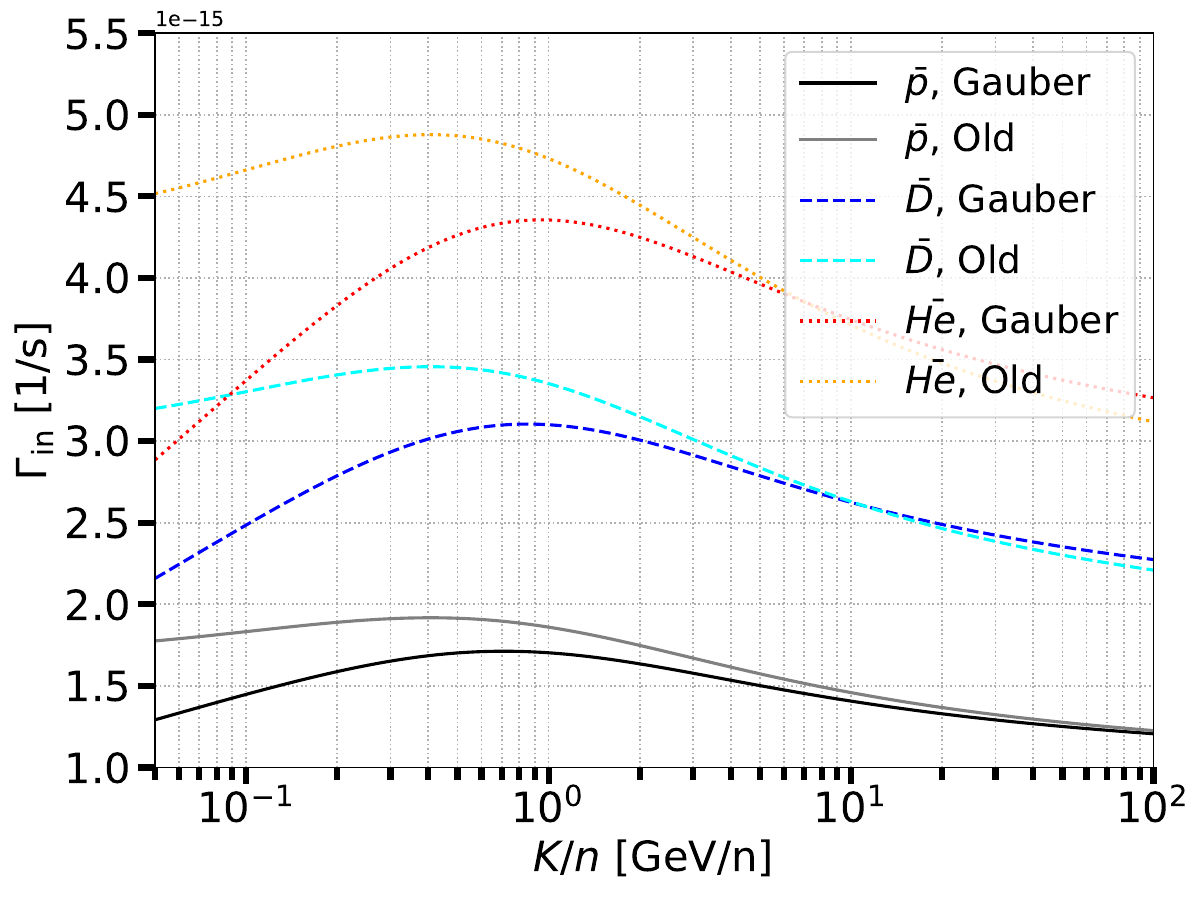}
\caption{$\Gamma_{\rm inel}$ for different species shown as a function of $K/n$ (kinetic energy divided by the mass number, i.e.~kinetic energy per nucleon).}
\label{fig:Gamma_inel}
\end{figure*}

Note that, in the propagation of different DM-induced species, the only differences arise from
their masses ($m_{j}$), charge number ($Z$), and inelastic interaction rate ($\Gamma^{j}_{\rm inel}$).

\item[$\circ$] \underline{\bf DM-induced spectra:} We consider the following two-body
DM annihilation and decay channels:
\begin{eqnarray*}
 &&   e^+ e^-, \mu^+ \mu^-, \tau^+ \tau^-, \nu_e \bar{\nu}_e, \nu_\mu \bar{\nu}_\mu,
\nu_\tau \bar{\nu}_\tau,\\
 && u \bar{u}, d \bar{d}, s \bar{s}, c \bar{c}, b \bar{b},
t \bar{t}, \\
 && \gamma \gamma, g g, W^+ W^-, Z Z, H H, Z \gamma, H Z,
\end{eqnarray*}

and take the corresponding tabulated energy spectra at production
from \texttt{CosmiXs} \cite{Arina:2023eic, DiMauro:2024kml}.
For the $\bar{d}$  species, we consider the \texttt{CosmiXs} spectra corresponding to the
`Wigner coalescence model with an Argonne wavefunction'.

\item[$\circ$] \underline{\bf Galactic DM distribution:}
For the DM distribution in the Galaxy, we consider the NFW, Einasto and Burkert profiles
from \cite{Cirelli:2024ssz}. All these profiles are normalized in order to have,  at the location of the Sun $\vec r_\odot =(r_\odot,z)$ = (8.33, 0) kpc,
a DM density $\rho_\odot = 0.4~\rm GeV\,cm^{-3}$. This value is currently accepted as the conventional best average, based on recent determinations; the typical associated error is $\pm 0.1 \ {\rm GeV}/{\rm cm}^3$ (see \cite{Cirelli:2024ssz} for a more detailed discussion).
\end{itemize}

We consider the same $m_{\rm DM}$ bins provided in \texttt{PPPC4DMID} and \texttt{CosmiXs} in the
range 5 GeV--100 TeV (10 GeV--200 TeV) for DM annihilation (decay),
and sample the final CR flux (for all species) in the kinetic-energy
range $0.1 \, {\rm GeV} \le K \le 10^5 \, {\rm GeV}$, with 20 bins per decade in energy.

\section{Results for the CR fluxes of antiprotons and anti-deuterons}
\label{sec:results}

In this section we present and discuss our main results, i.e.
the DM induced $\bar{p}$ and $\bar{d}$ CR fluxes
obtained in this work using the models discussed in Sec.~\ref{sec:Models},
and compare them with those provided in \texttt{PPPC4DMID}~\cite{PPPC4DMID}, which correspond to previous propagation scenarios.

\medskip

\begin{figure*}[!ht]
\vspace{1cm}
\hspace{-10mm}
\begin{tabular}{ccc}
\includegraphics[width=0.36\textwidth]{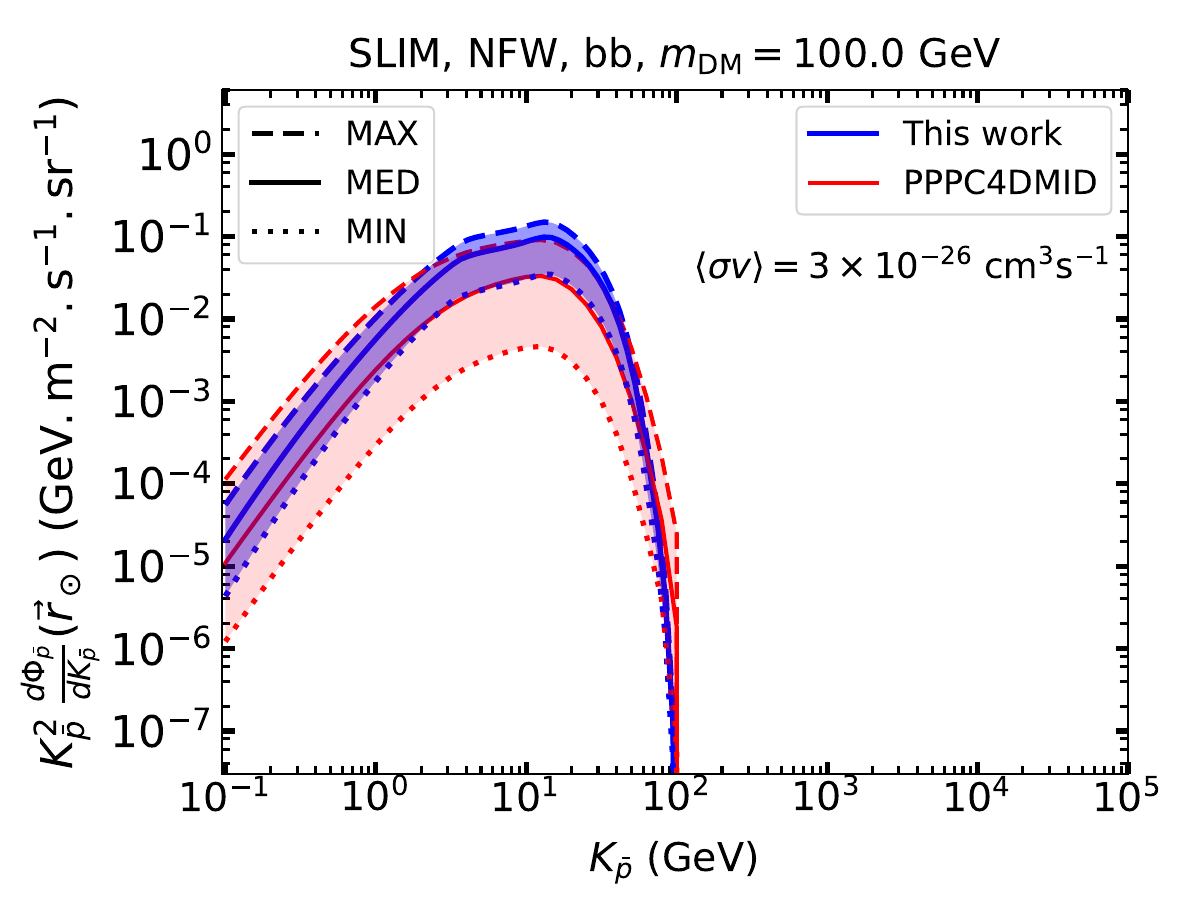} & \hspace{-6mm}
\includegraphics[width=0.36\textwidth]{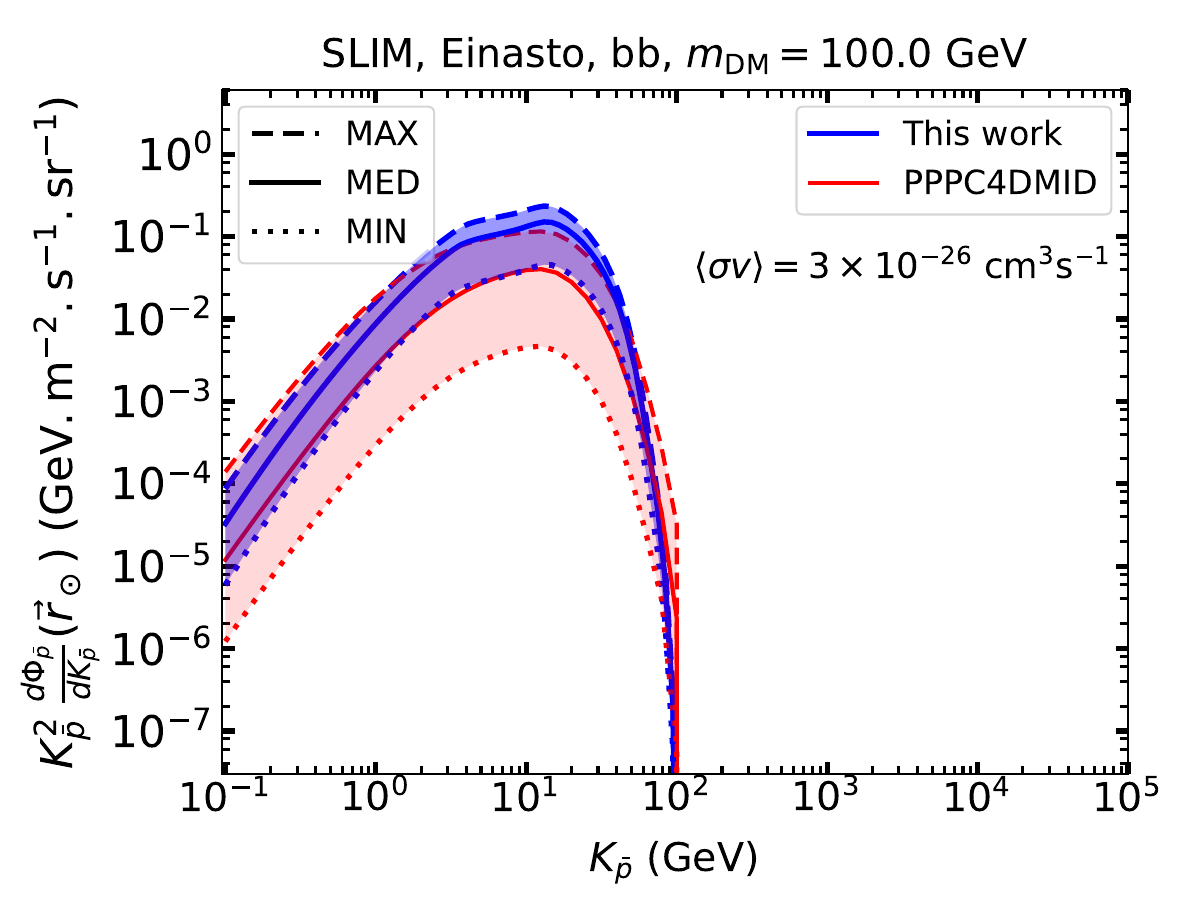} & \hspace{-6mm}
\includegraphics[width=0.36\textwidth]{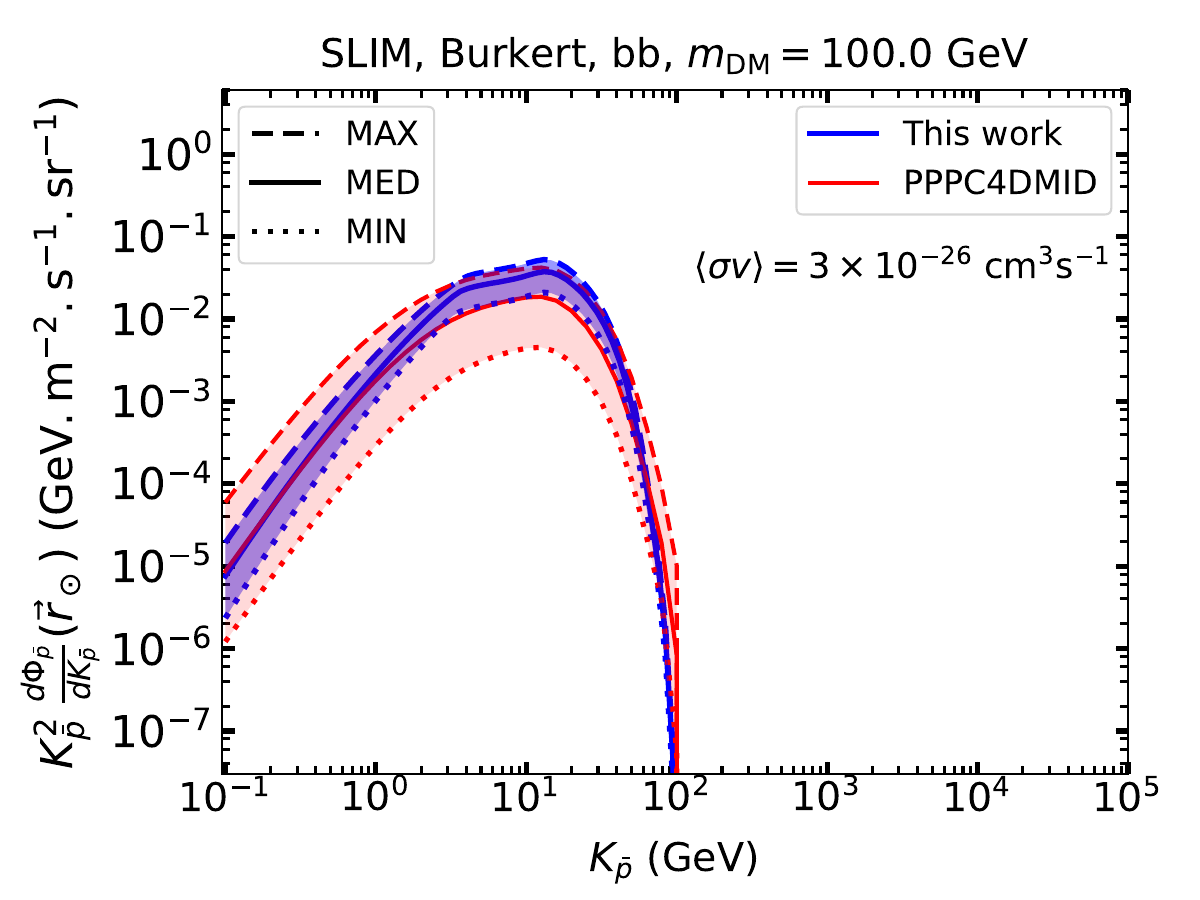}  \hspace{-6mm}
\end{tabular}

\hspace{-10mm}
\begin{tabular}{ccc}
\includegraphics[width=0.36\textwidth]{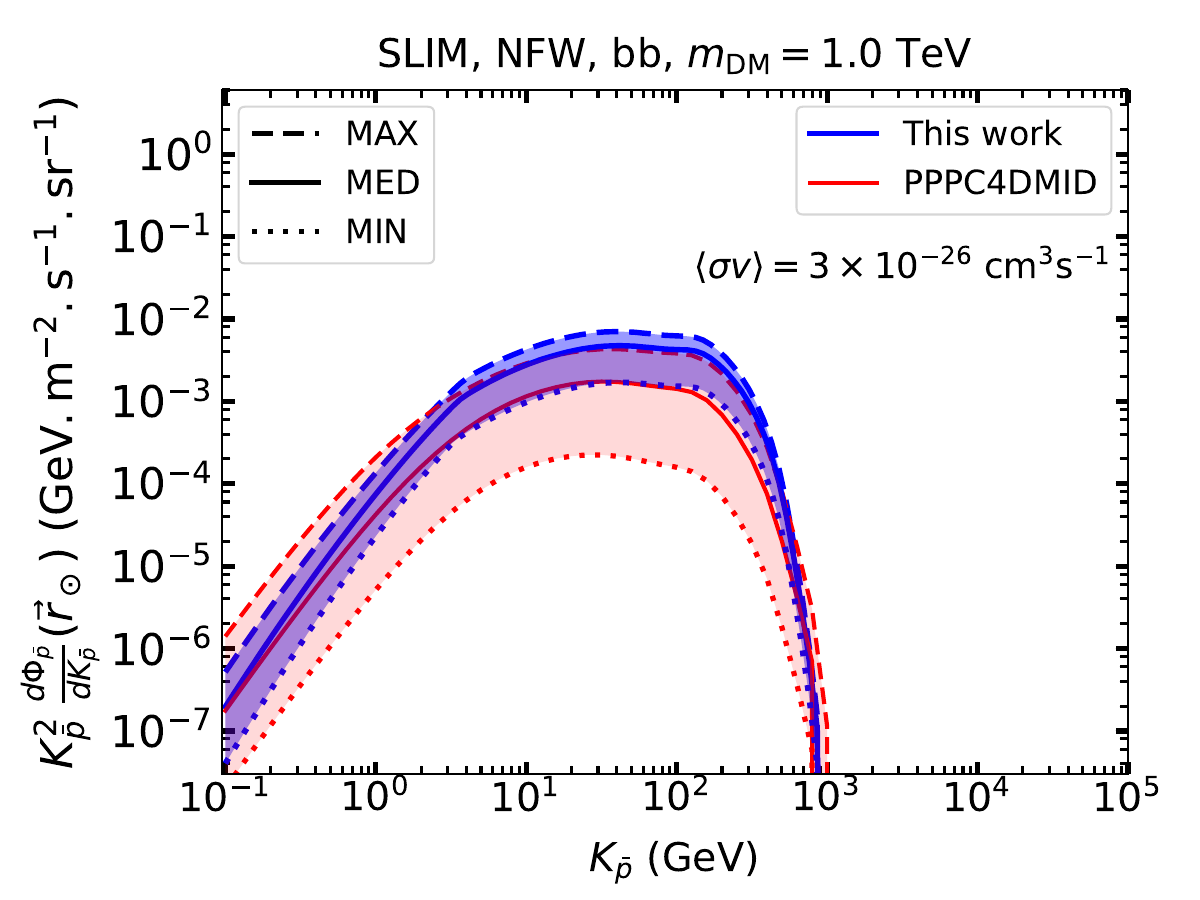} & \hspace{-6mm}
\includegraphics[width=0.36\textwidth]{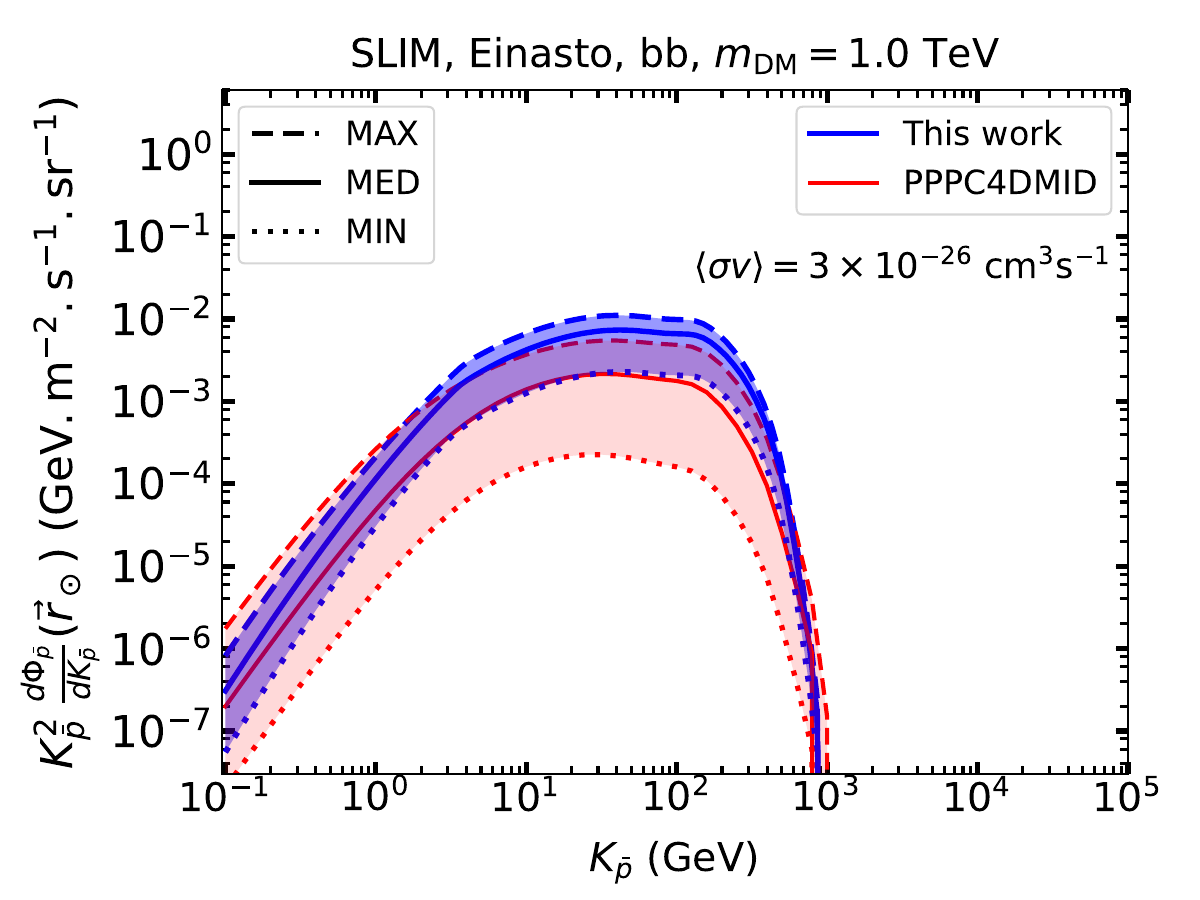} & \hspace{-6mm}
\includegraphics[width=0.36\textwidth]{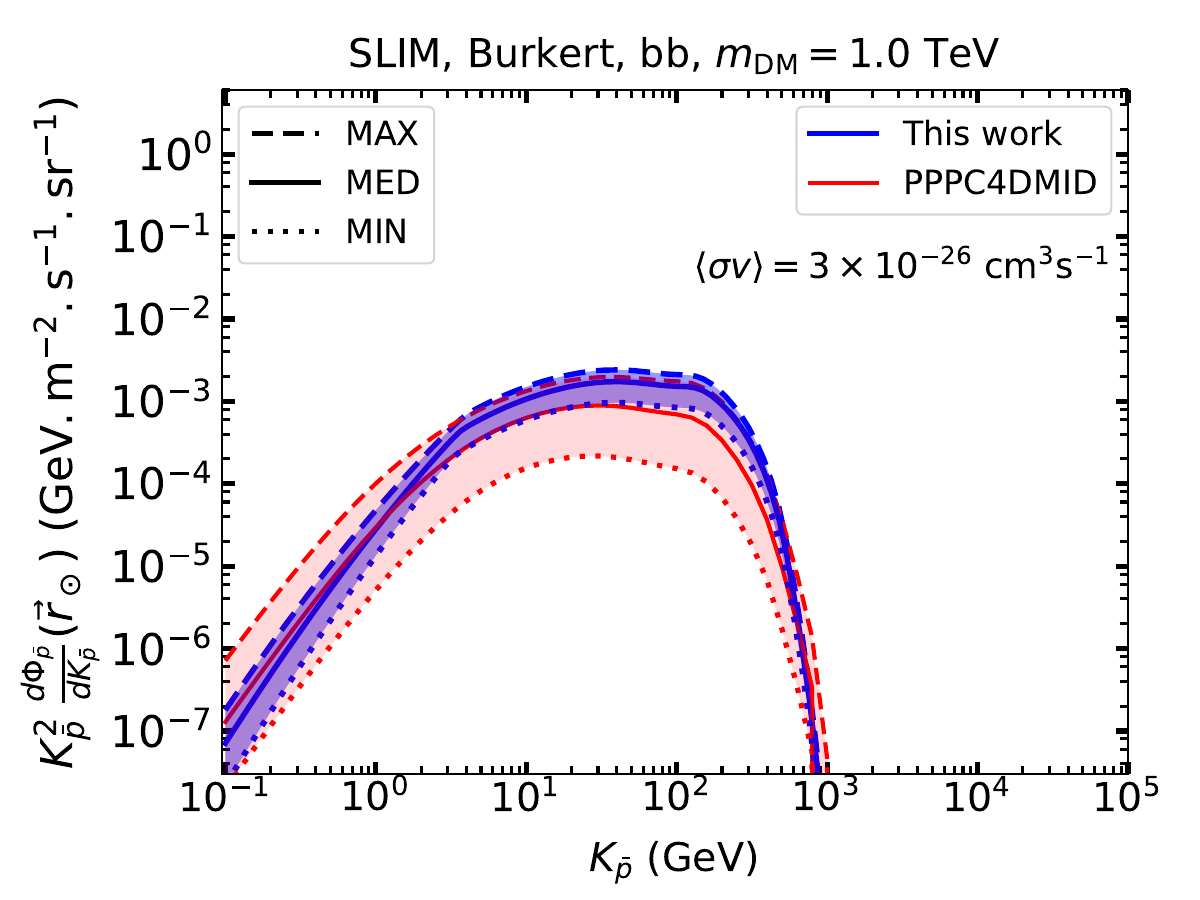}  \hspace{-6mm}
\end{tabular}

\hspace{-10mm}
\begin{tabular}{ccc}
\includegraphics[width=0.36\textwidth]{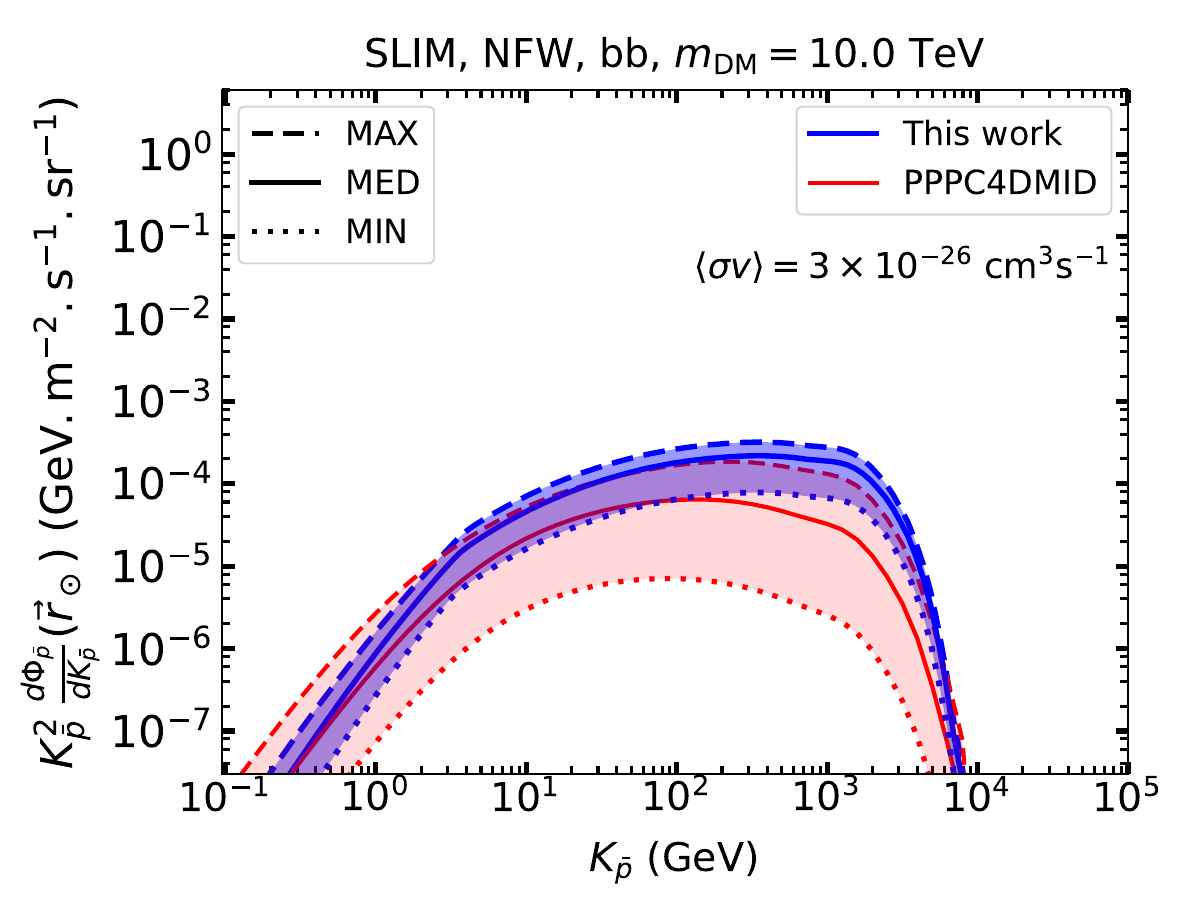} & \hspace{-6mm}
\includegraphics[width=0.36\textwidth]{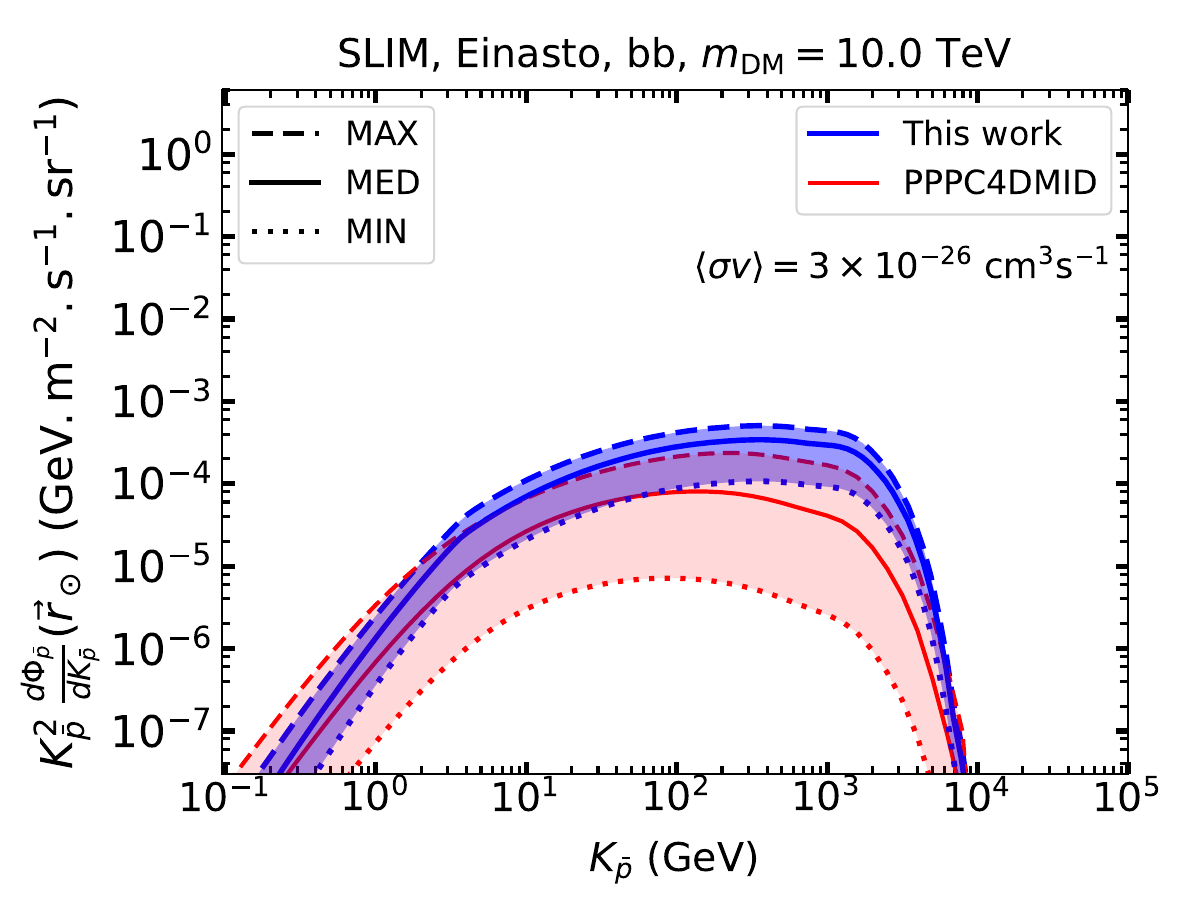} & \hspace{-6mm}
\includegraphics[width=0.36\textwidth]{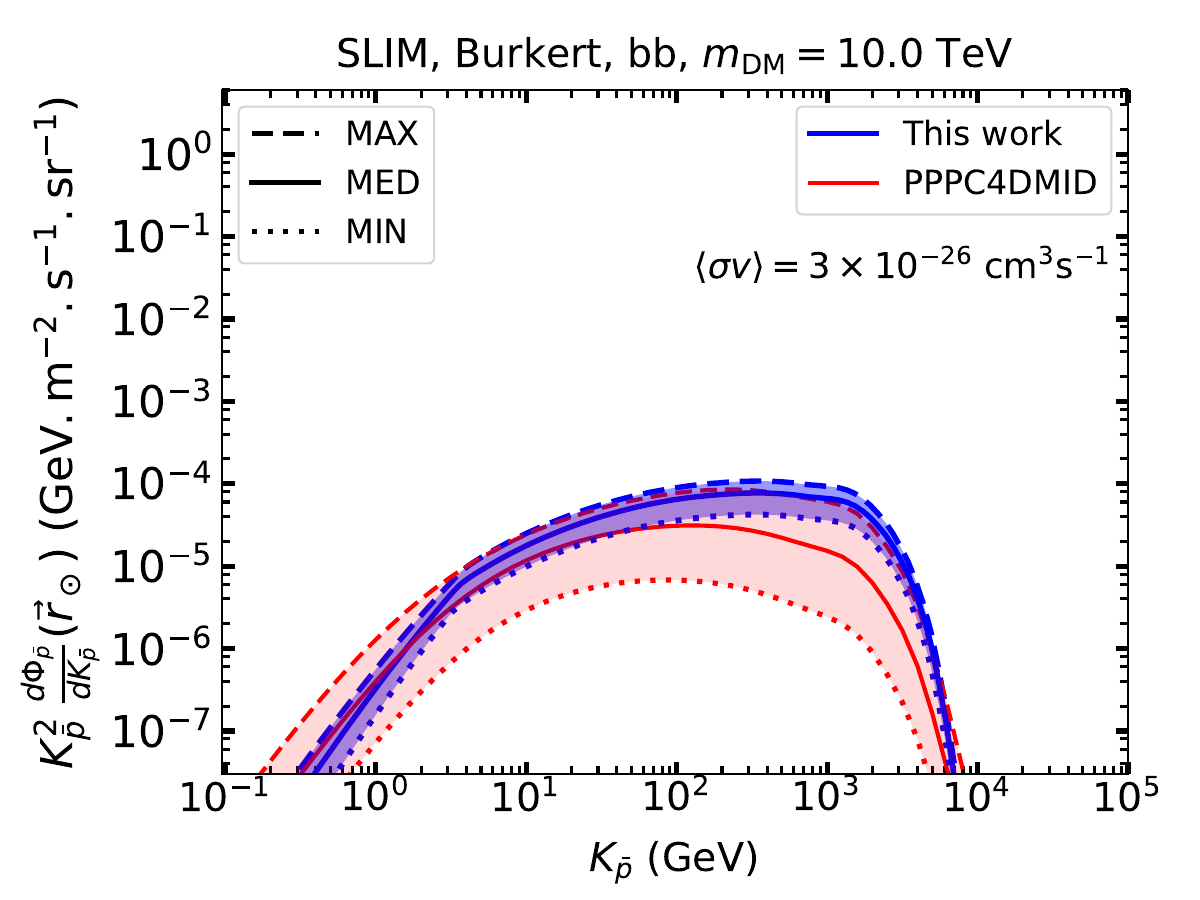}  \hspace{-6mm}
\end{tabular}

\hspace{-10mm}
\begin{tabular}{ccc}
\includegraphics[width=0.36\textwidth]{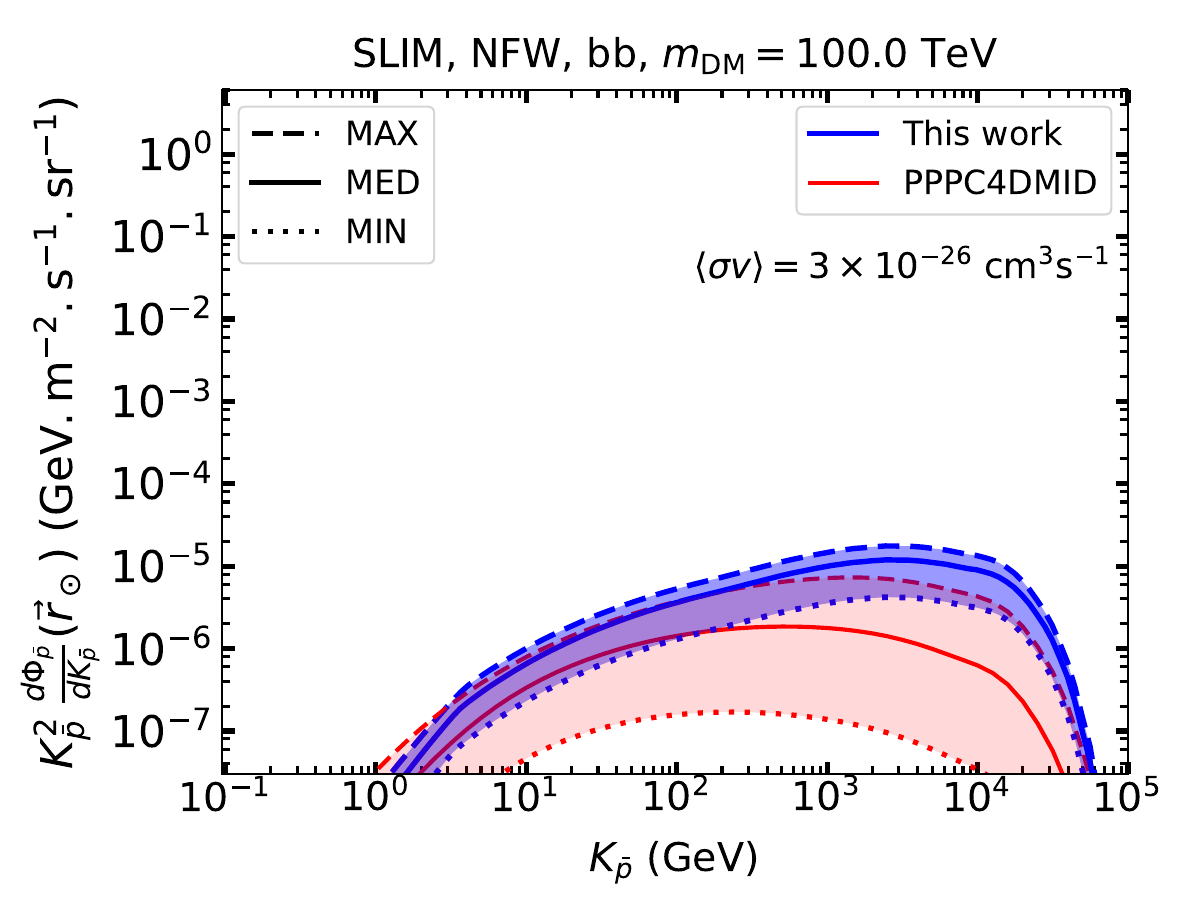} & \hspace{-6mm}
\includegraphics[width=0.36\textwidth]{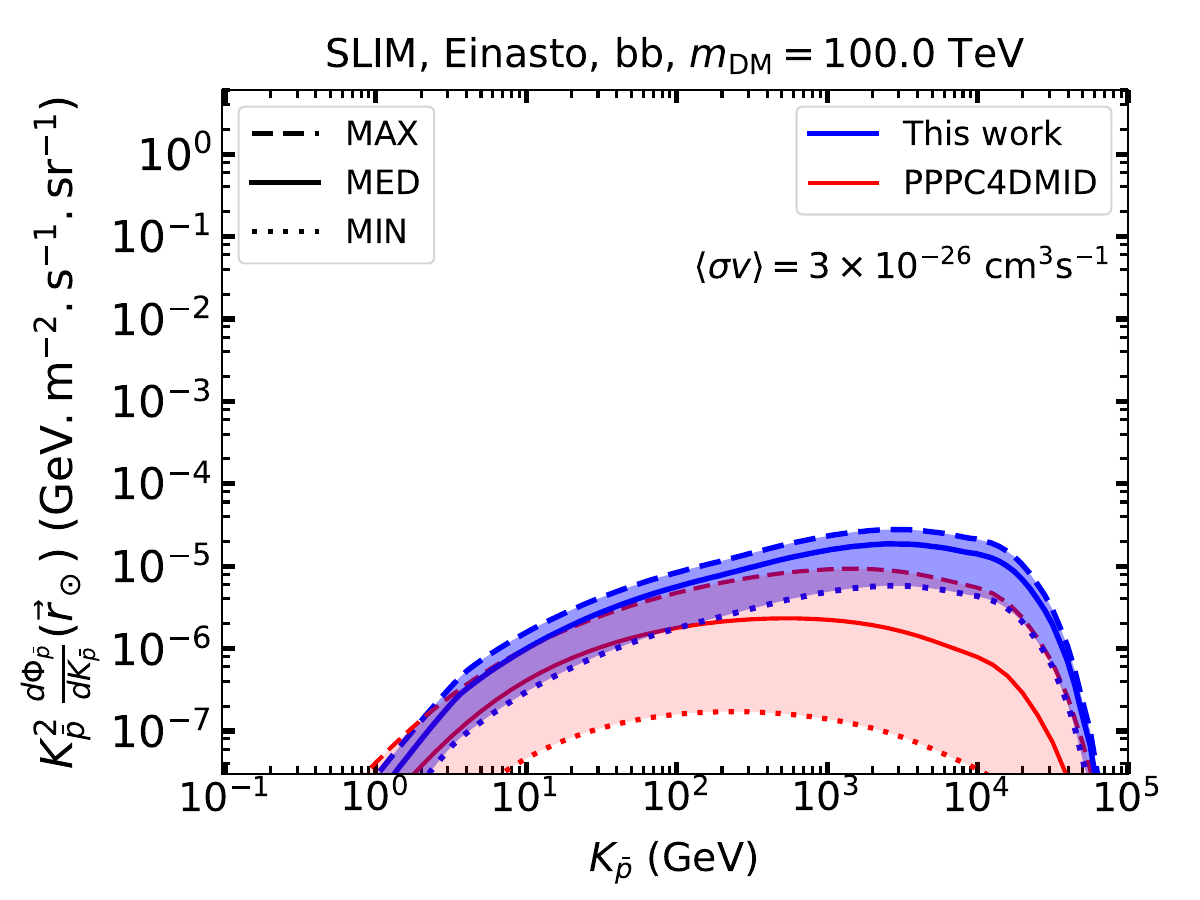} & \hspace{-6mm}
\includegraphics[width=0.36\textwidth]{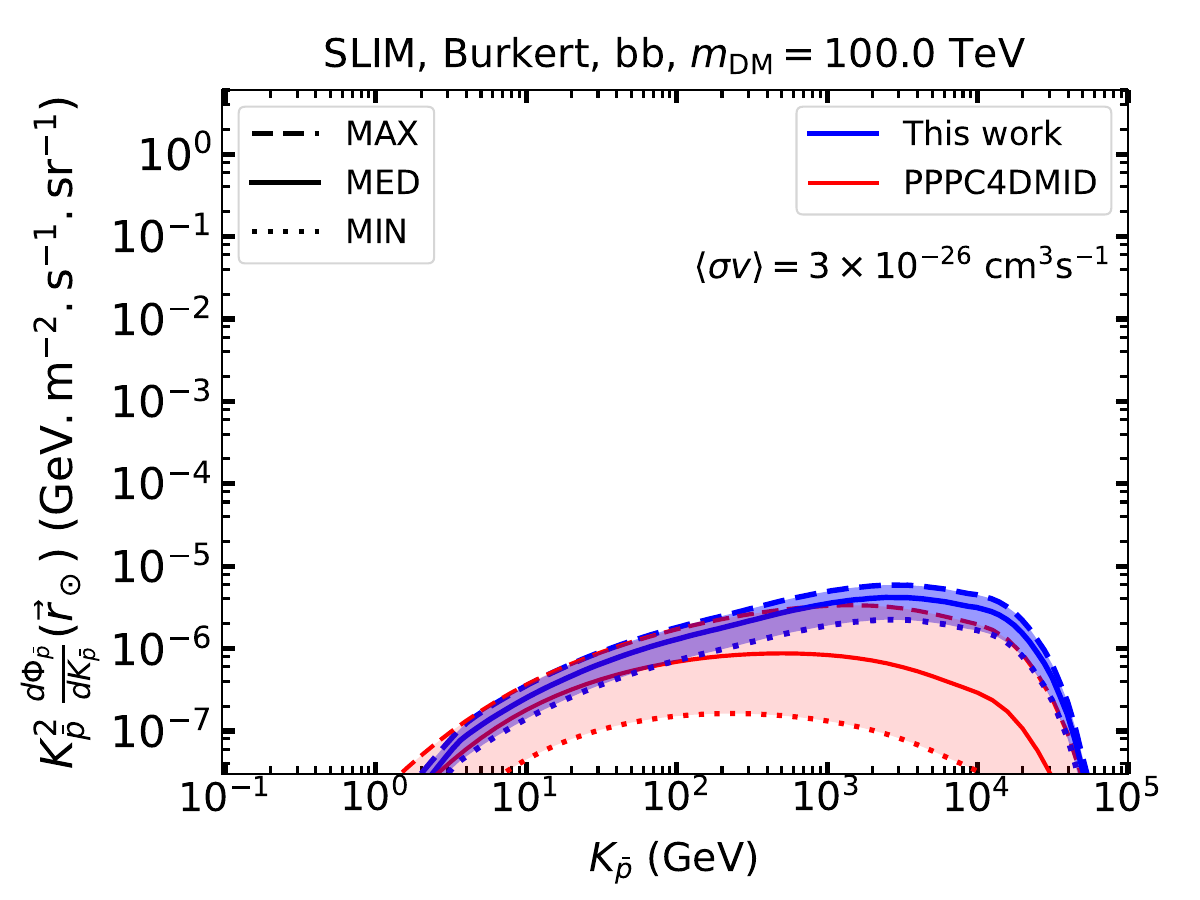}  \hspace{-6mm}
\end{tabular}

\caption{
{\it Antiproton fluxes} (IS) from ${\rm DM \, DM} \rightarrow \bm{b\bar{b}}$ annihilations, under the {\sc Slim} propagation scheme, compared to the previous \texttt{PPPC4DMID} results.
The three columns correspond to NFW, Einasto and Burkert DM profiles, respectively, while the four lines correspond to
DM masses of $10^2$, $10^3$, $10^4$ and $10^5$ GeV, respectively.}
\label{fig:pbar_SLIM_bb_Ann}
\vspace{1cm}
\end{figure*}

\begin{figure}[!ht]
\vspace{1cm}
\hspace{-10mm}
\begin{tabular}{ccc}
\includegraphics[width=0.36\textwidth]{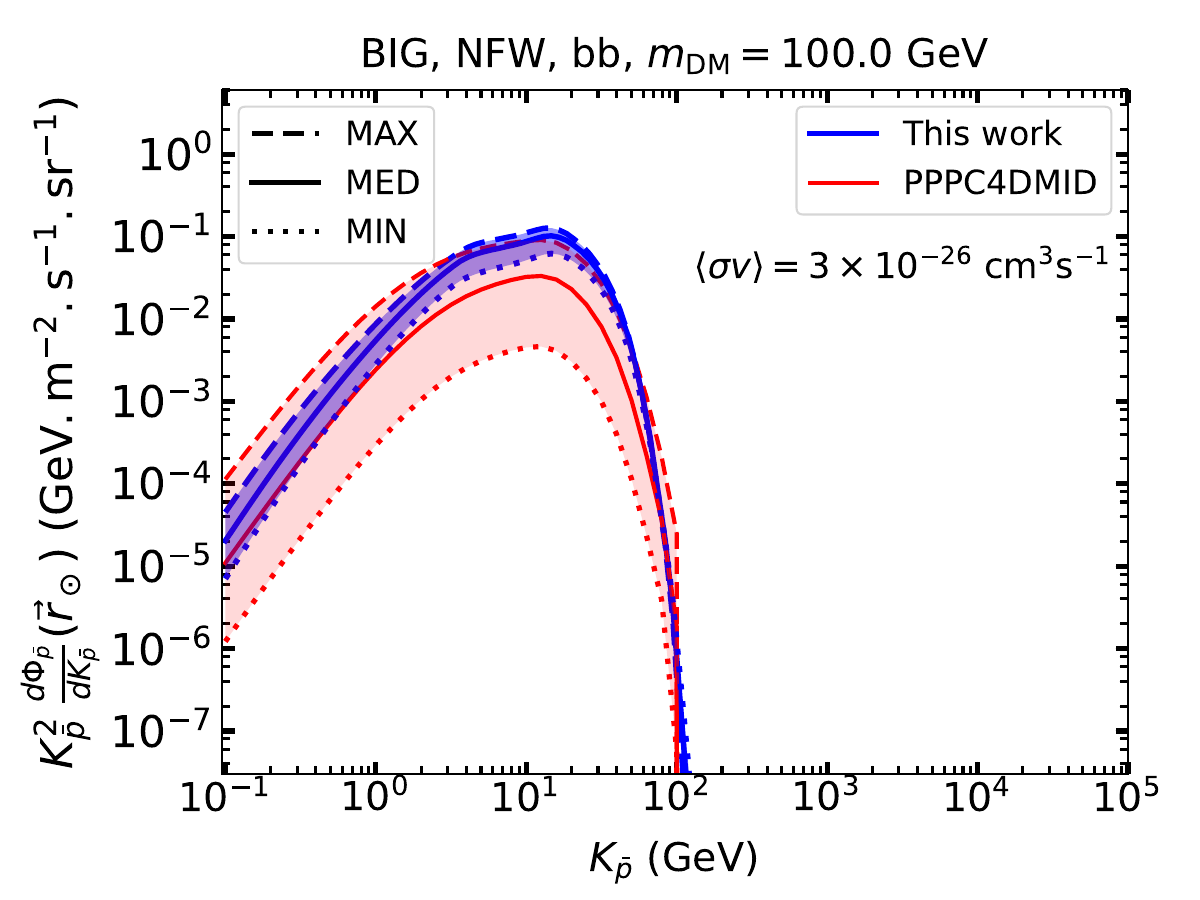} & \hspace{-6mm}
\includegraphics[width=0.36\textwidth]{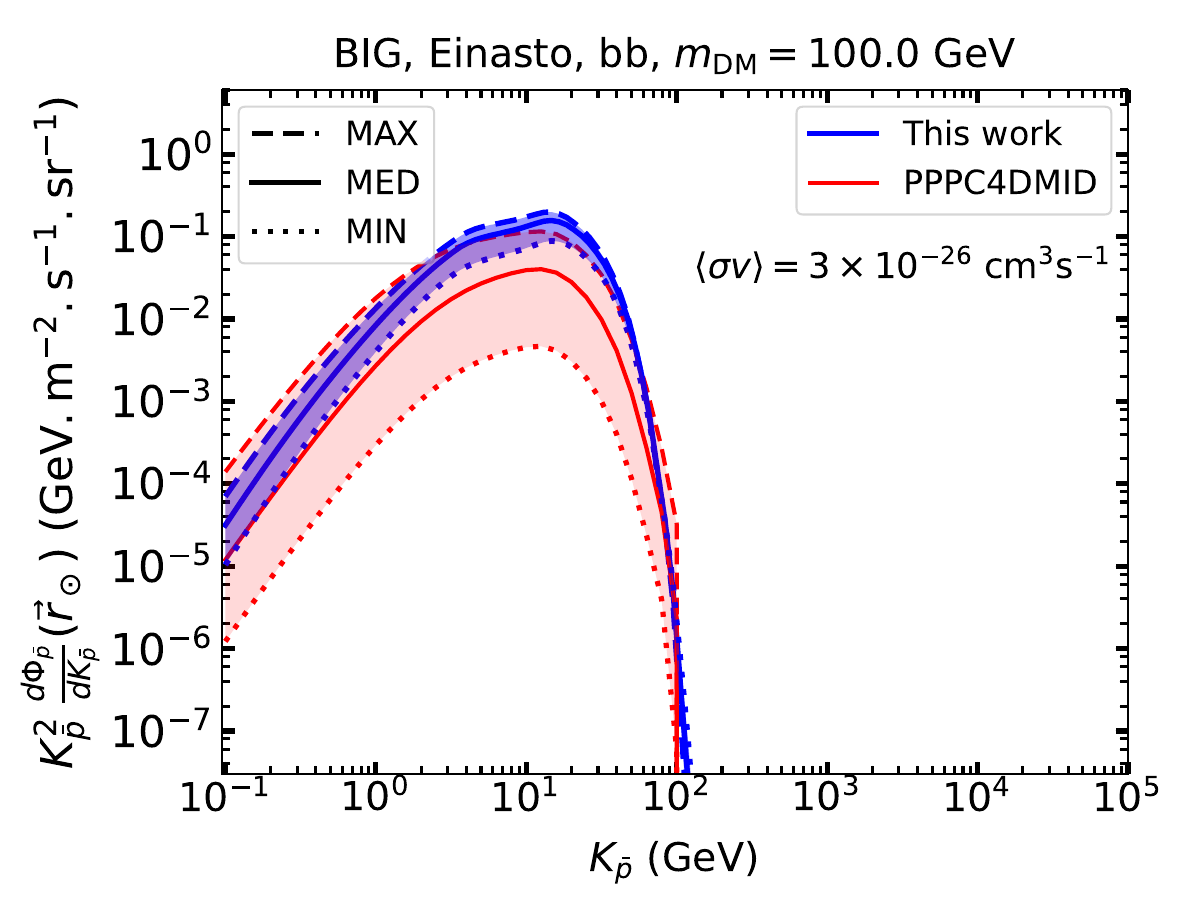} & \hspace{-6mm}
\includegraphics[width=0.36\textwidth]{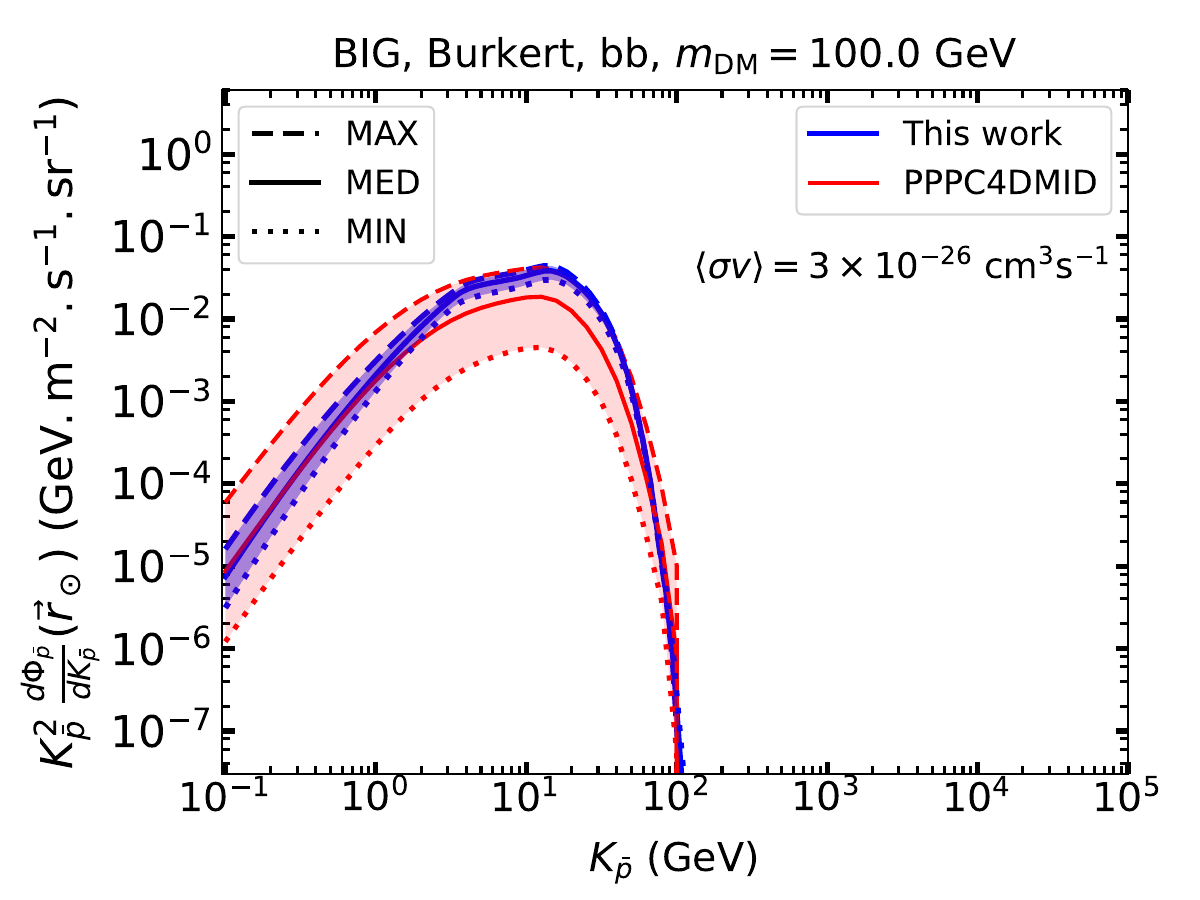}  \hspace{-6mm}
\end{tabular}

\hspace{-10mm}
\begin{tabular}{ccc}
\includegraphics[width=0.36\textwidth]{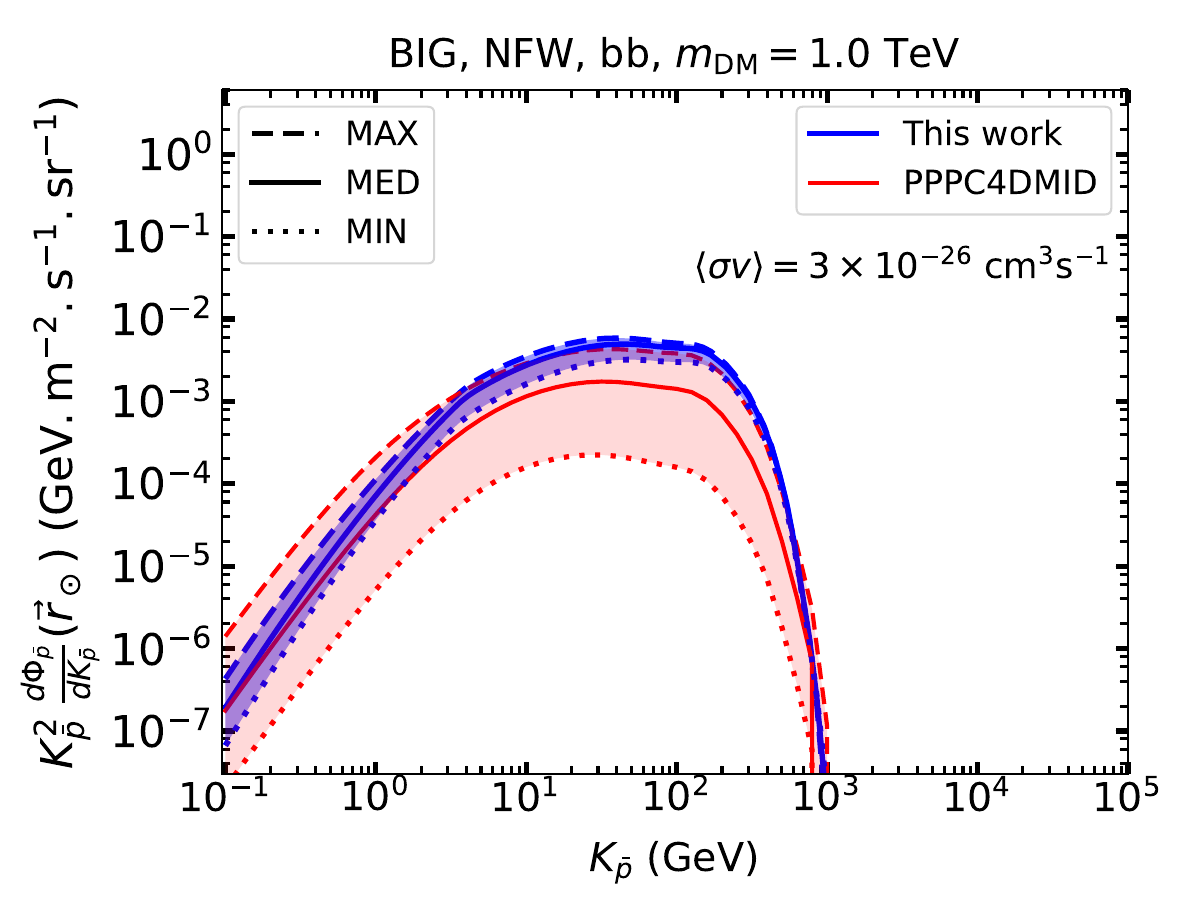} & \hspace{-6mm}
\includegraphics[width=0.36\textwidth]{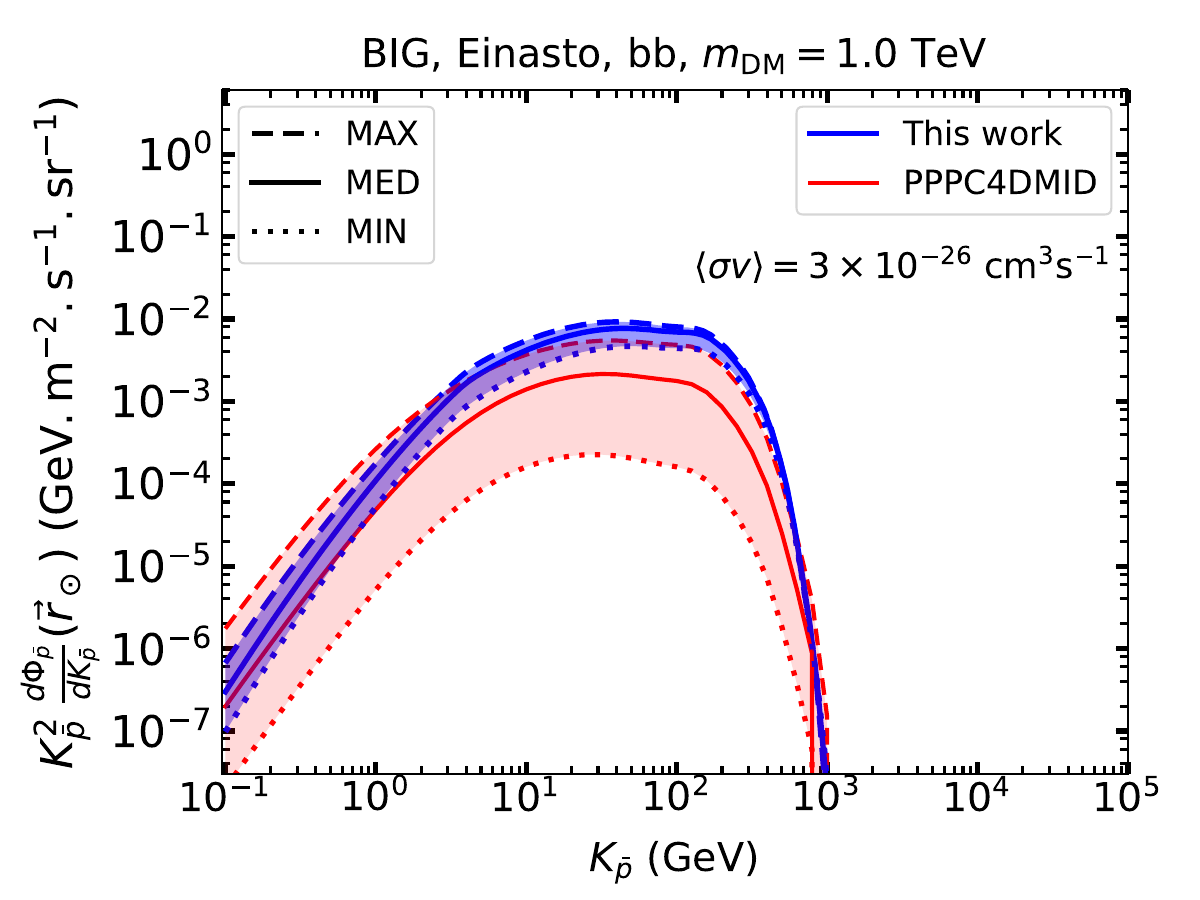} & \hspace{-6mm}
\includegraphics[width=0.36\textwidth]{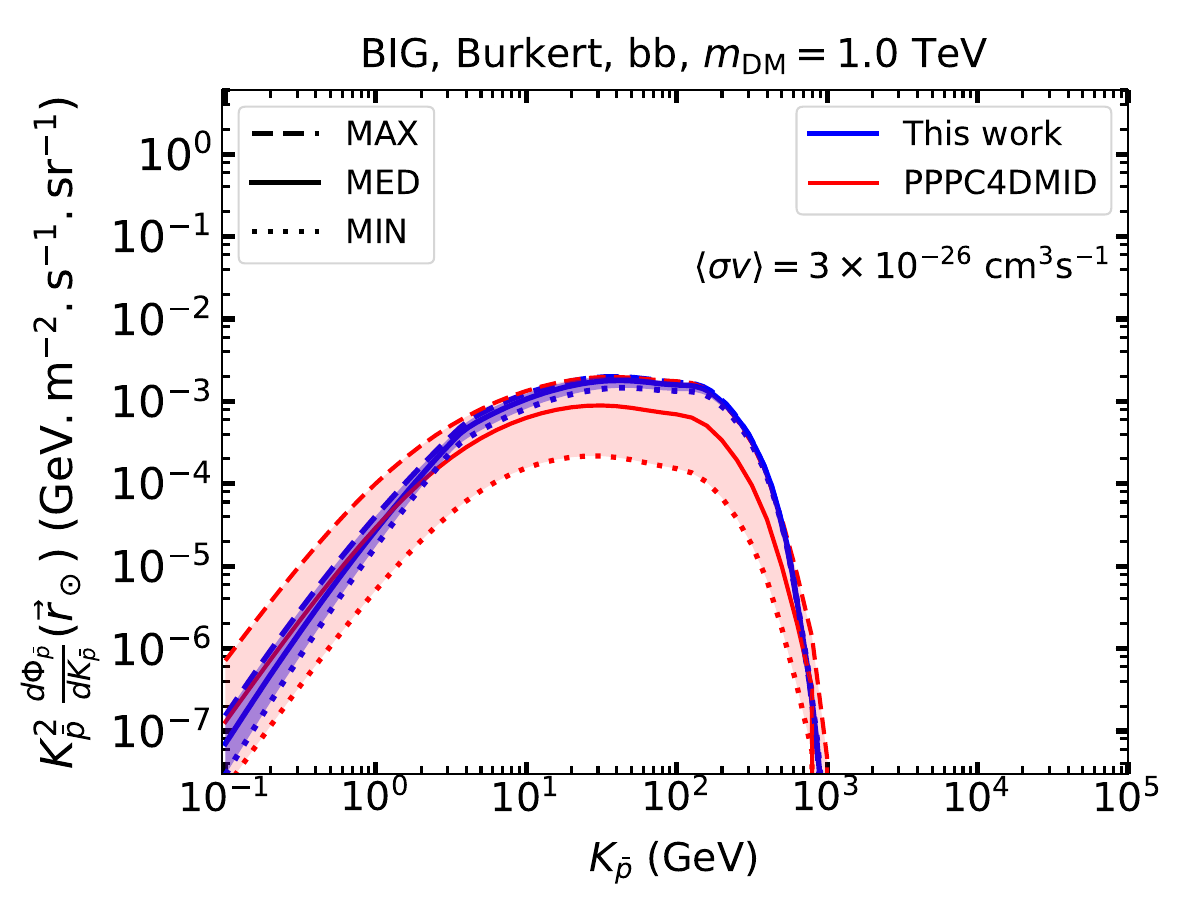}  \hspace{-6mm}
\end{tabular}

\hspace{-10mm}
\begin{tabular}{ccc}
\includegraphics[width=0.36\textwidth]{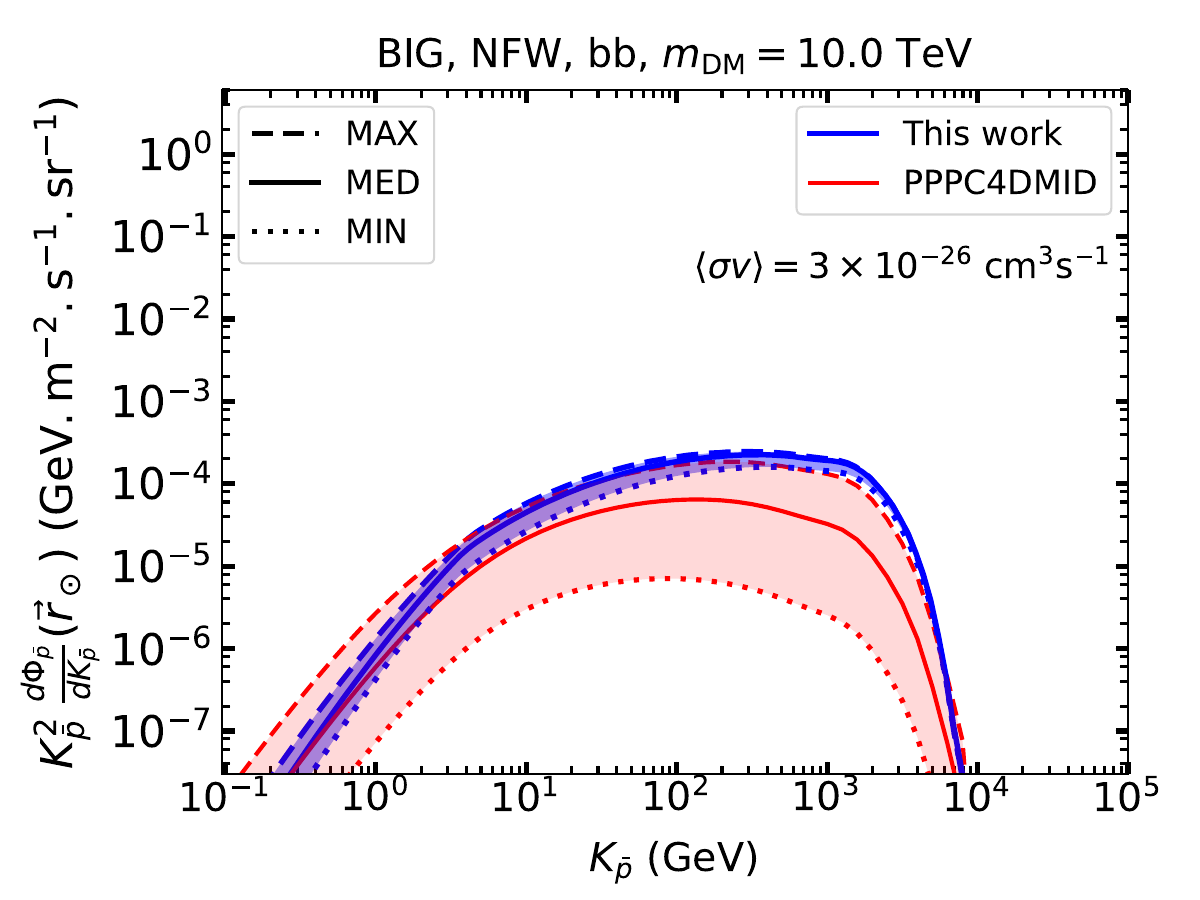} & \hspace{-6mm}
\includegraphics[width=0.36\textwidth]{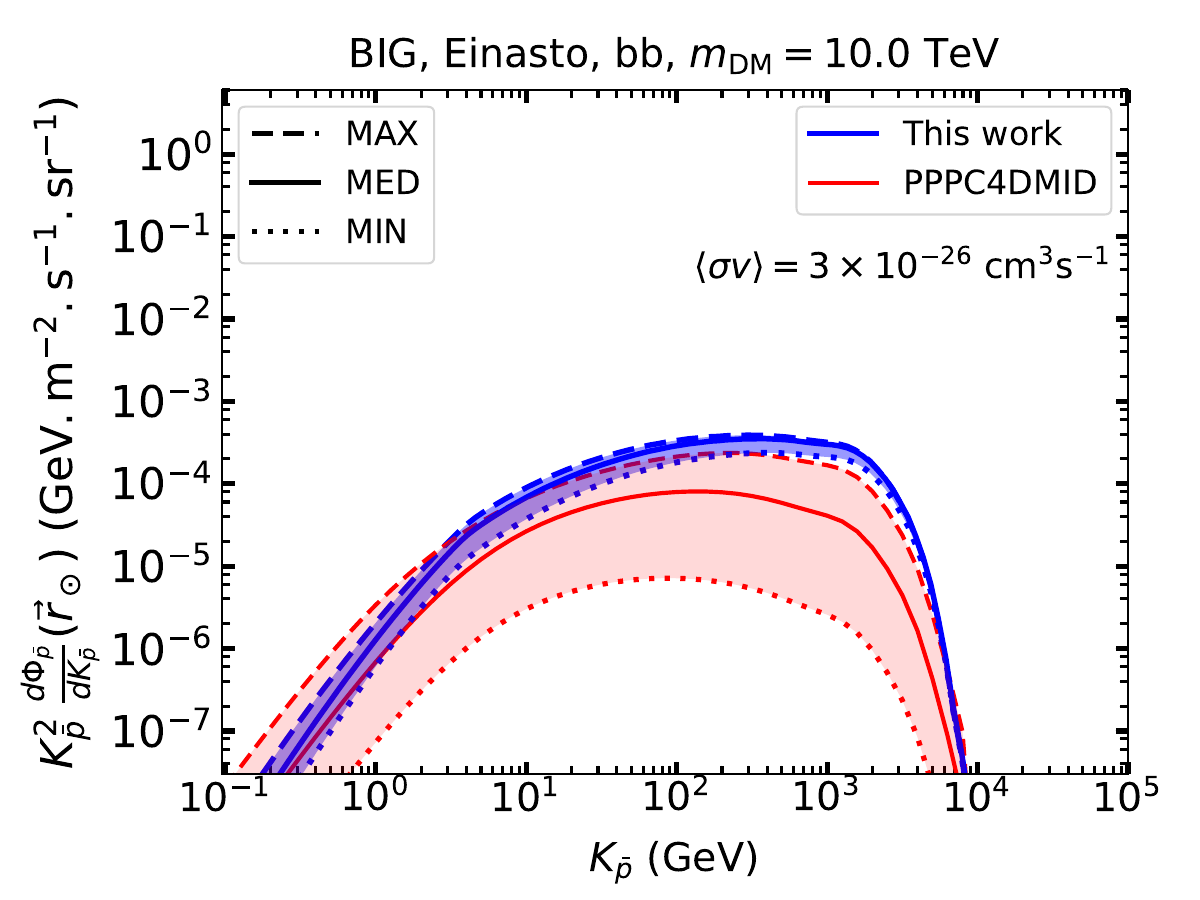} & \hspace{-6mm}
\includegraphics[width=0.36\textwidth]{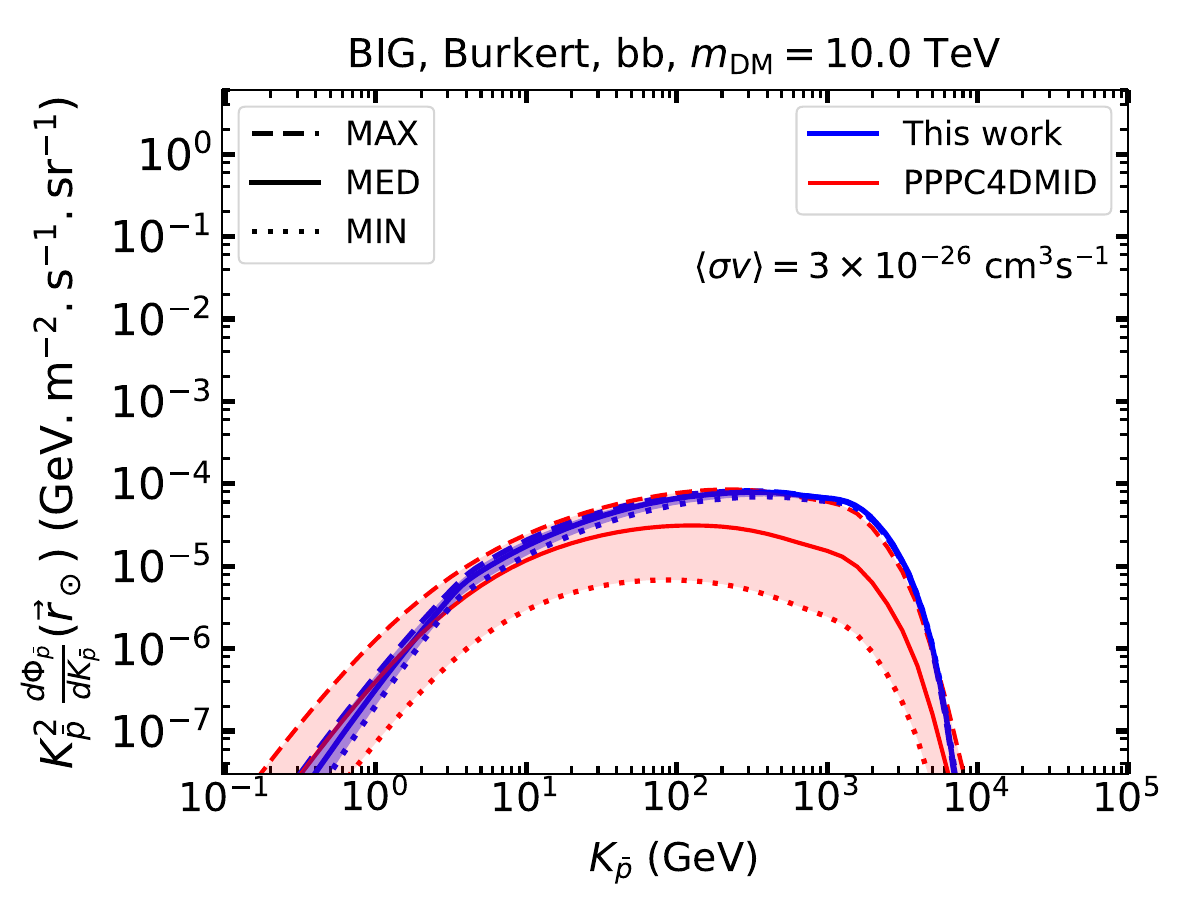}  \hspace{-6mm}
\end{tabular}

\hspace{-10mm}
\begin{tabular}{ccc}
\includegraphics[width=0.36\textwidth]{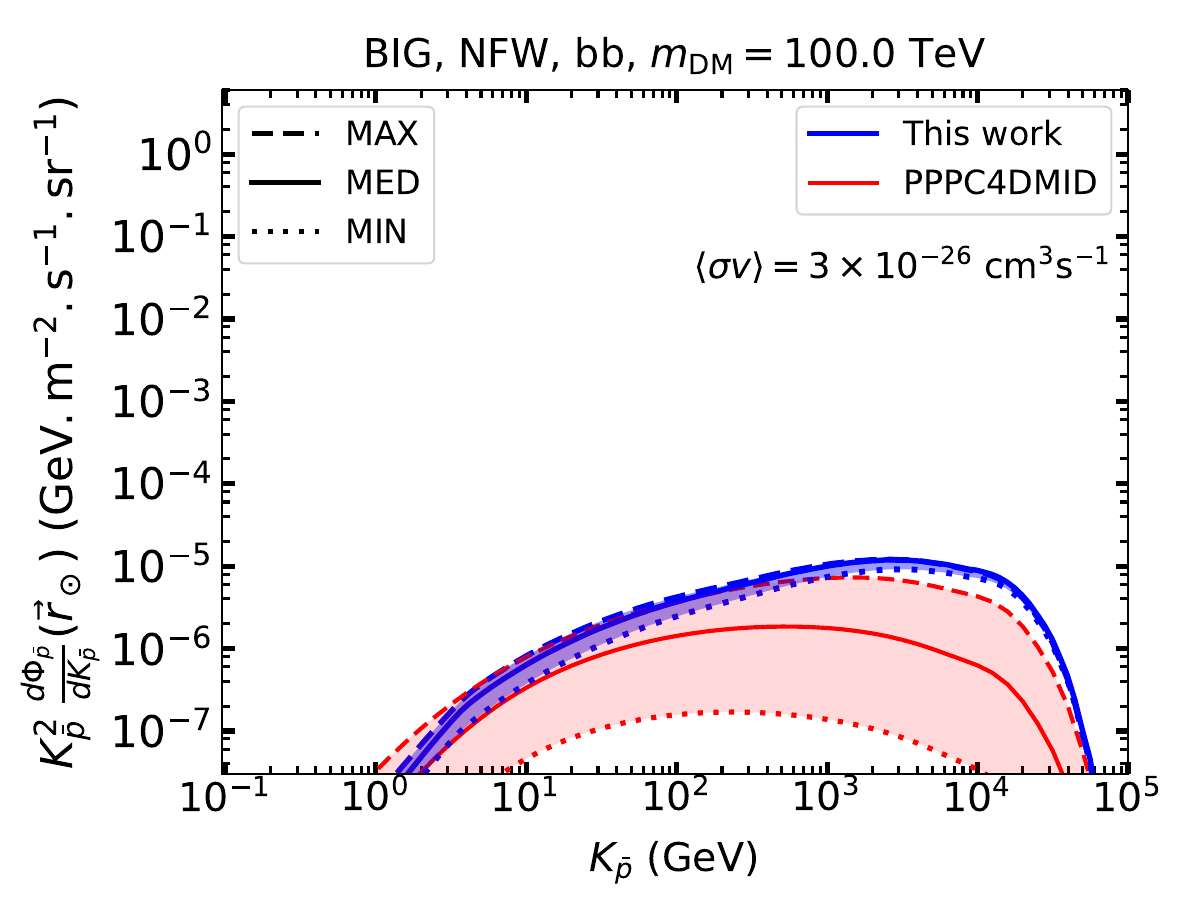} &
\hspace{-6mm}
\includegraphics[width=0.36\textwidth]{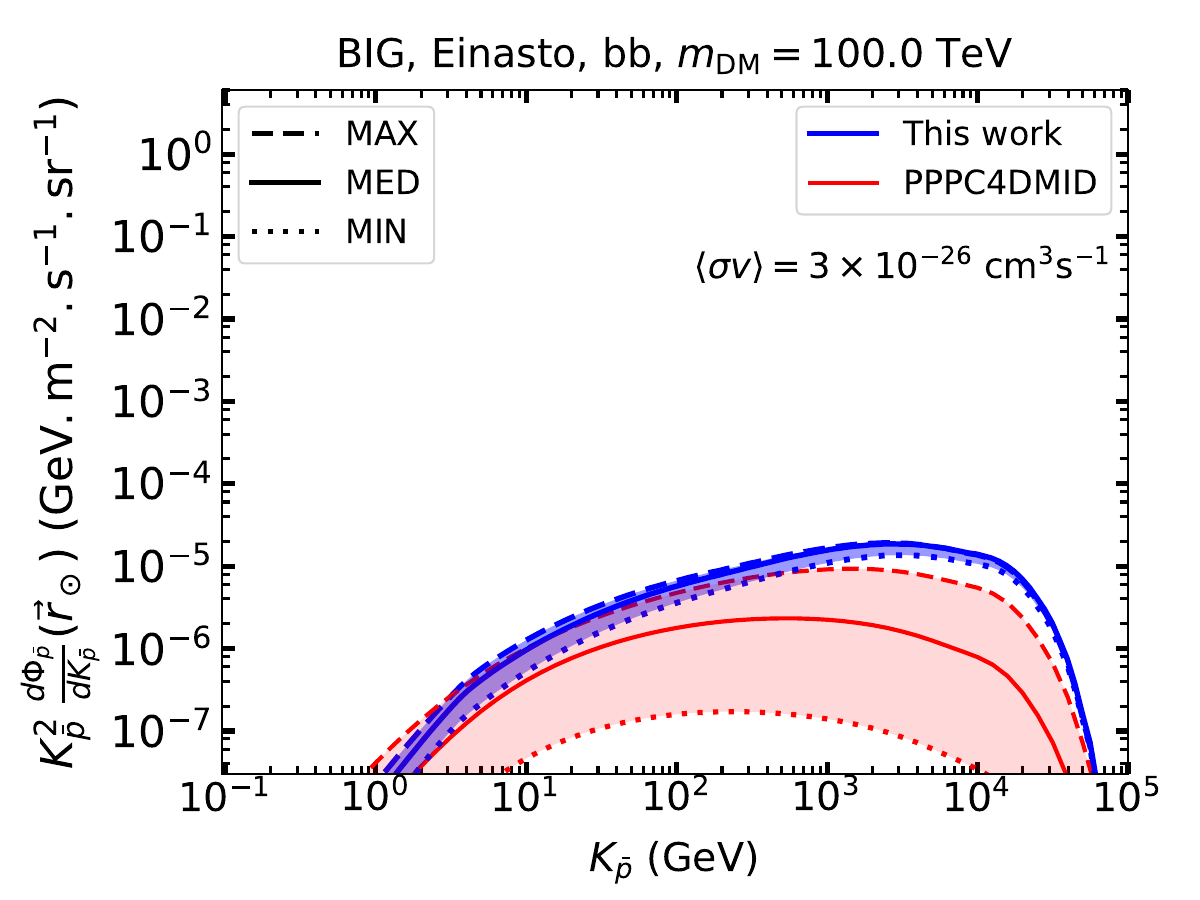} &
\hspace{-6mm}
\includegraphics[width=0.36\textwidth]{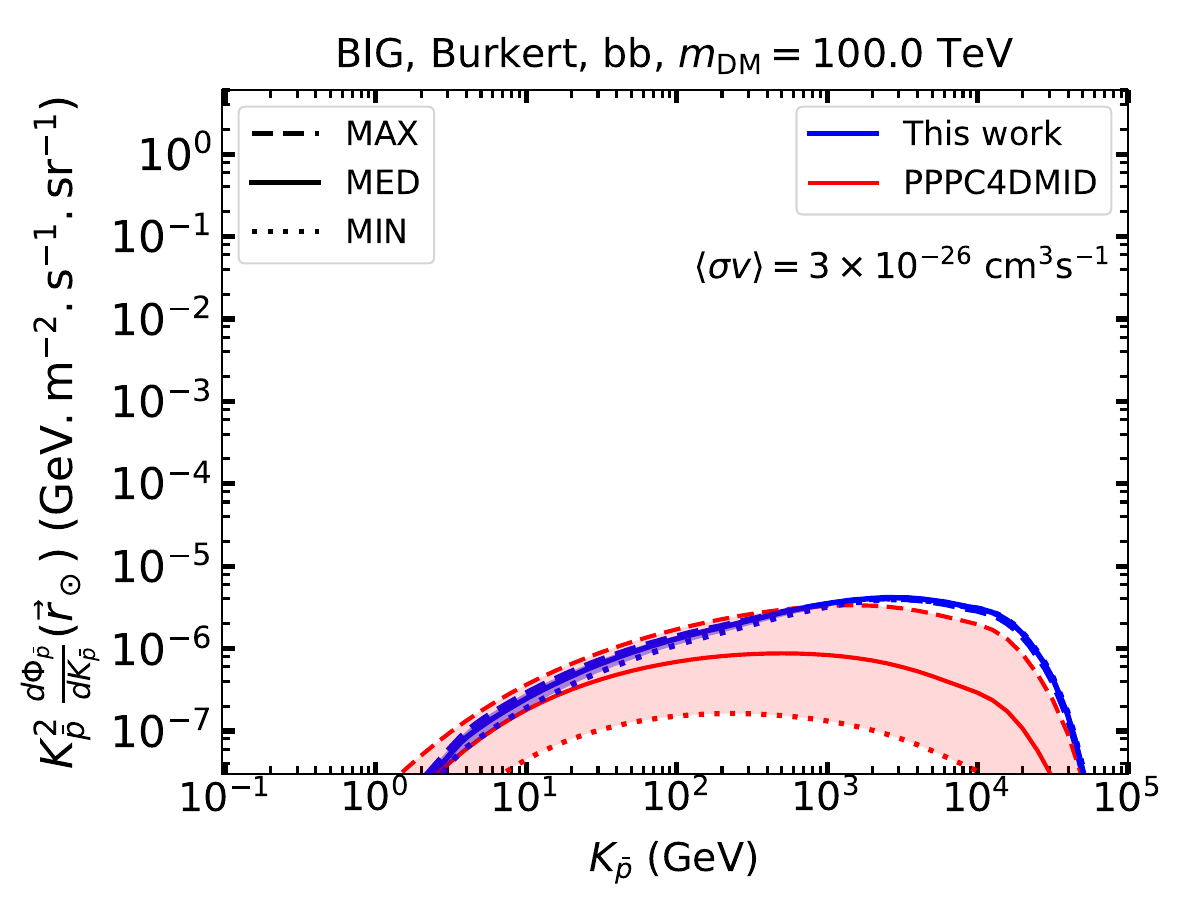}
\hspace{-6mm}
\end{tabular}

\caption{{\it Antiproton fluxes} (IS) from ${\rm DM \, DM} \rightarrow \bm{b\bar{b}}$ annihilations, under the {\sc Big} propagation scheme, compared to the previous \texttt{PPPC4DMID} results. The rows and columns are as in Fig.~\ref{fig:pbar_SLIM_bb_Ann}.}
\label{fig:pbar_BIG_bb_Ann}
\vspace{2cm}
\end{figure}

\begin{figure*}[!ht]
\vspace{1cm}
\hspace{-10mm}
\begin{tabular}{ccc}
\includegraphics[width=0.36\textwidth]{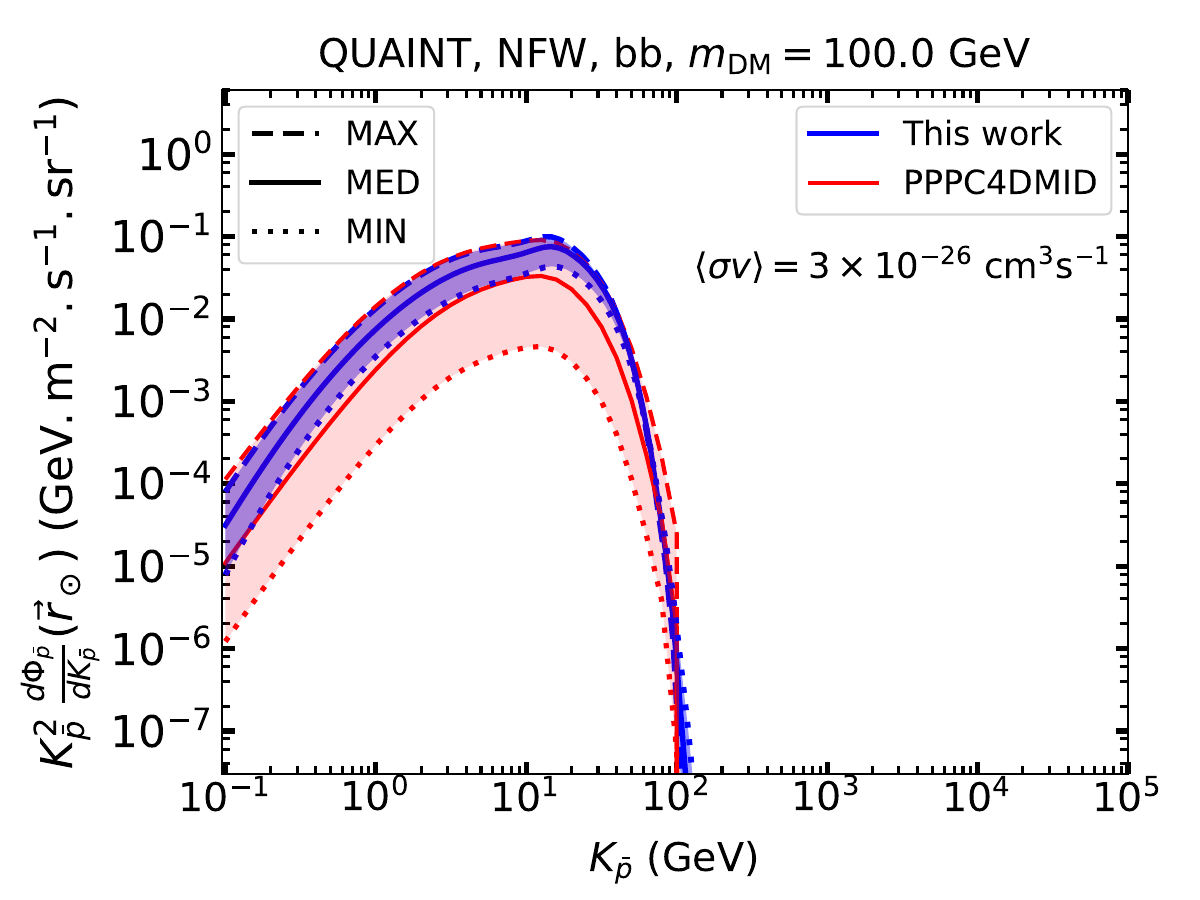} & \hspace{-6mm}
\includegraphics[width=0.36\textwidth]{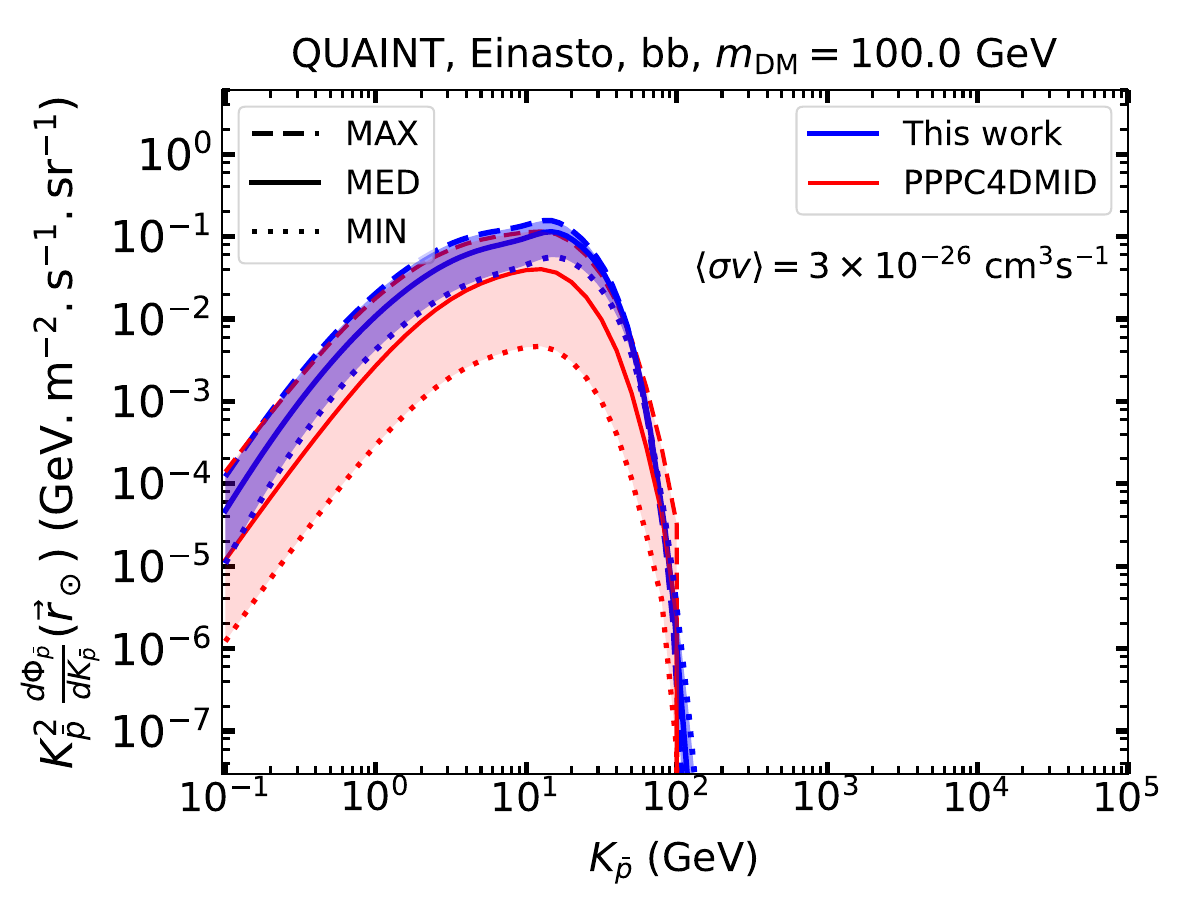} & \hspace{-6mm}
\includegraphics[width=0.36\textwidth]{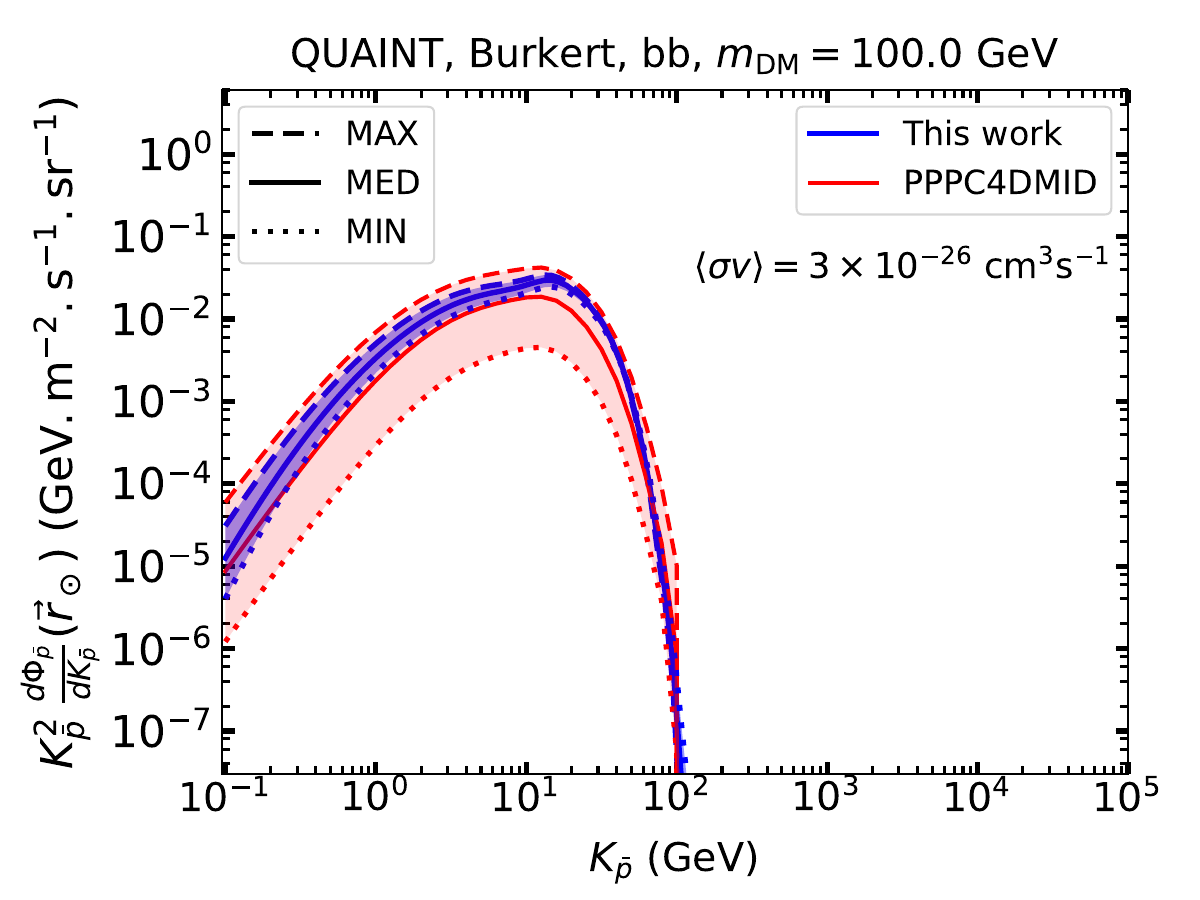}
\end{tabular}

\hspace{-10mm}
\begin{tabular}{ccc}
\includegraphics[width=0.36\textwidth]{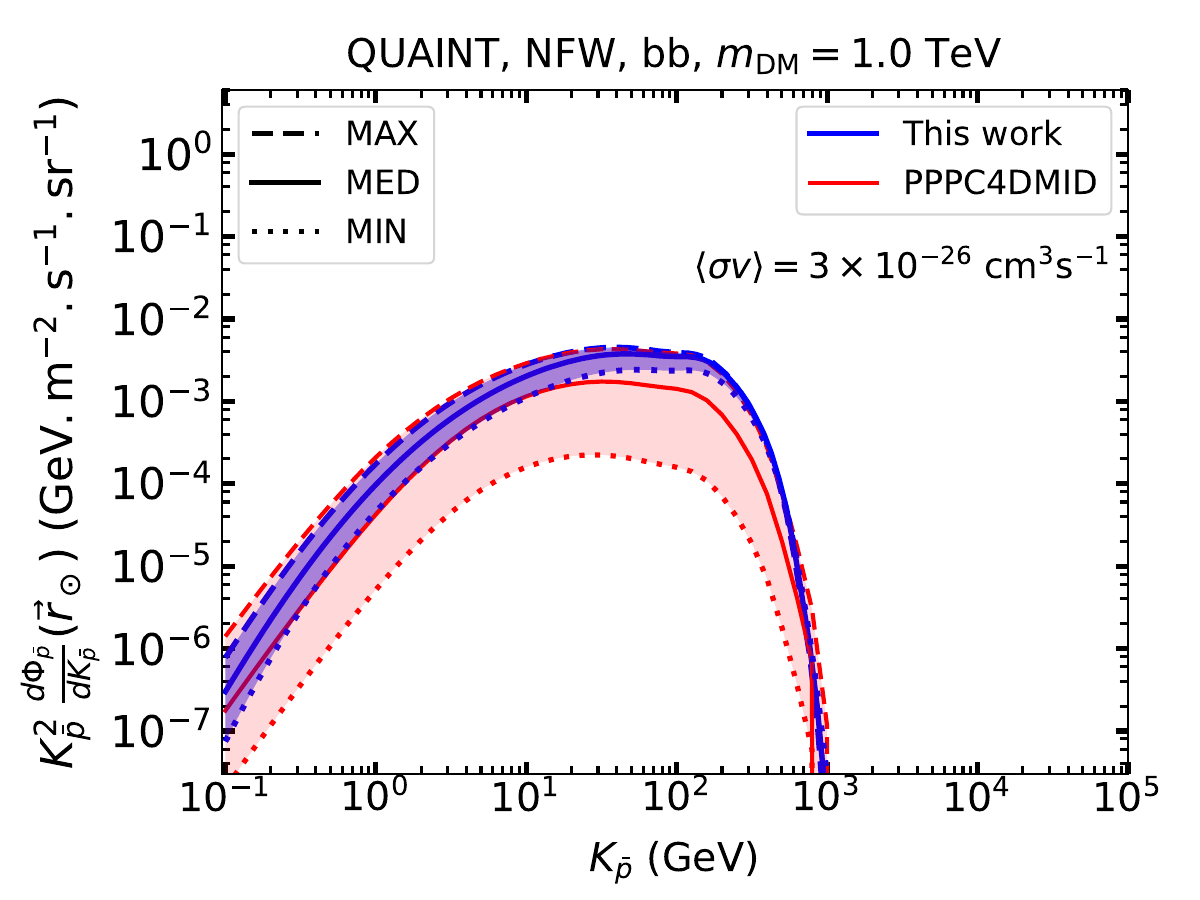} & \hspace{-6mm}
\includegraphics[width=0.36\textwidth]{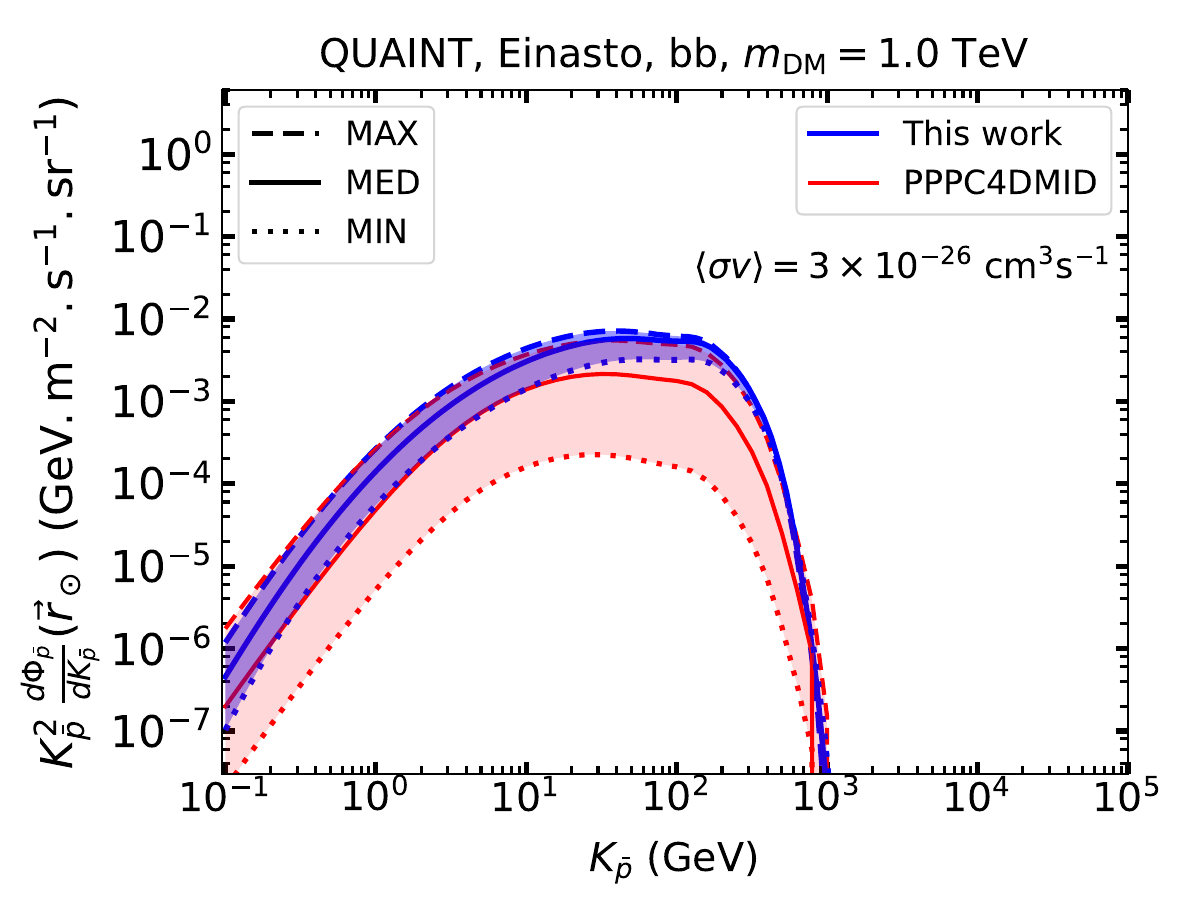} & \hspace{-6mm}
\includegraphics[width=0.36\textwidth]{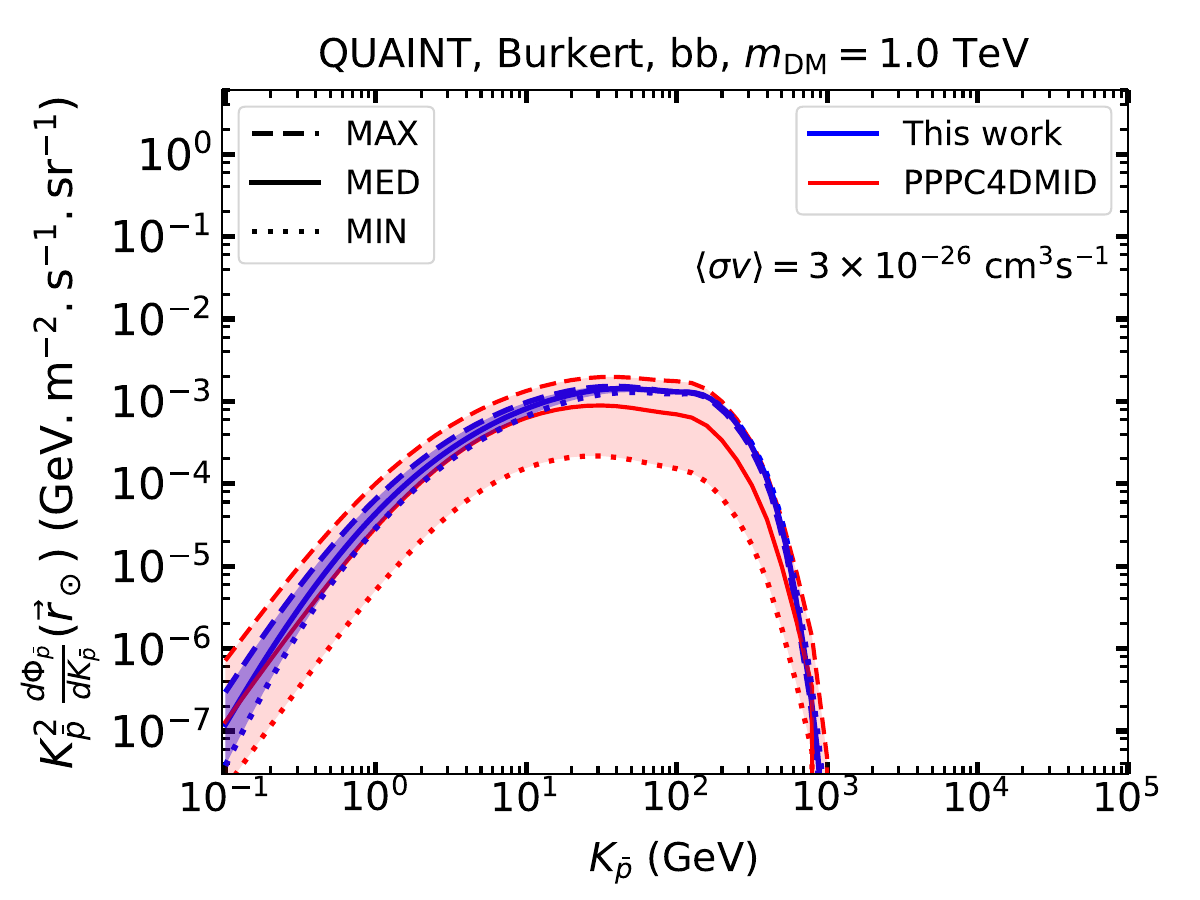}
\end{tabular}

\hspace{-10mm}
\begin{tabular}{ccc}
\includegraphics[width=0.36\textwidth]{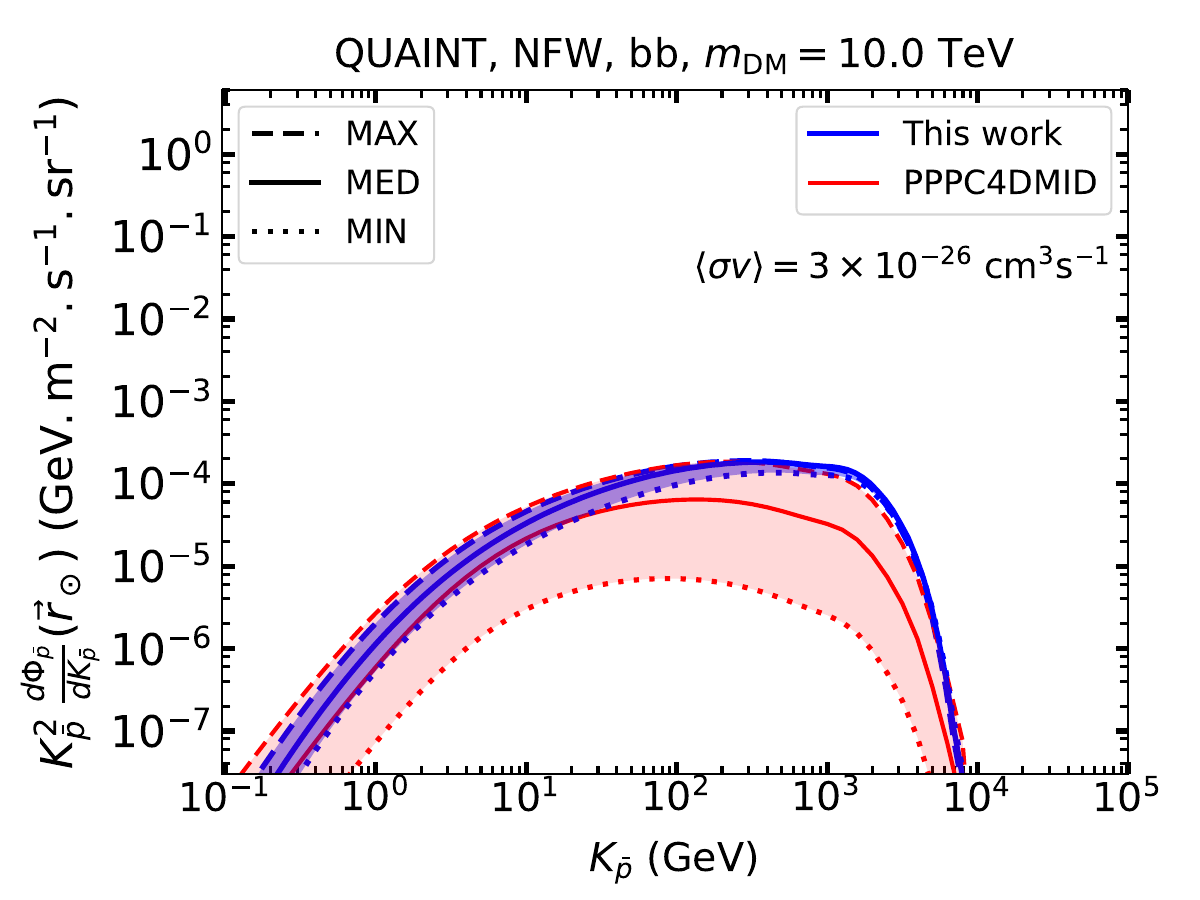} & \hspace{-6mm}
\includegraphics[width=0.36\textwidth]{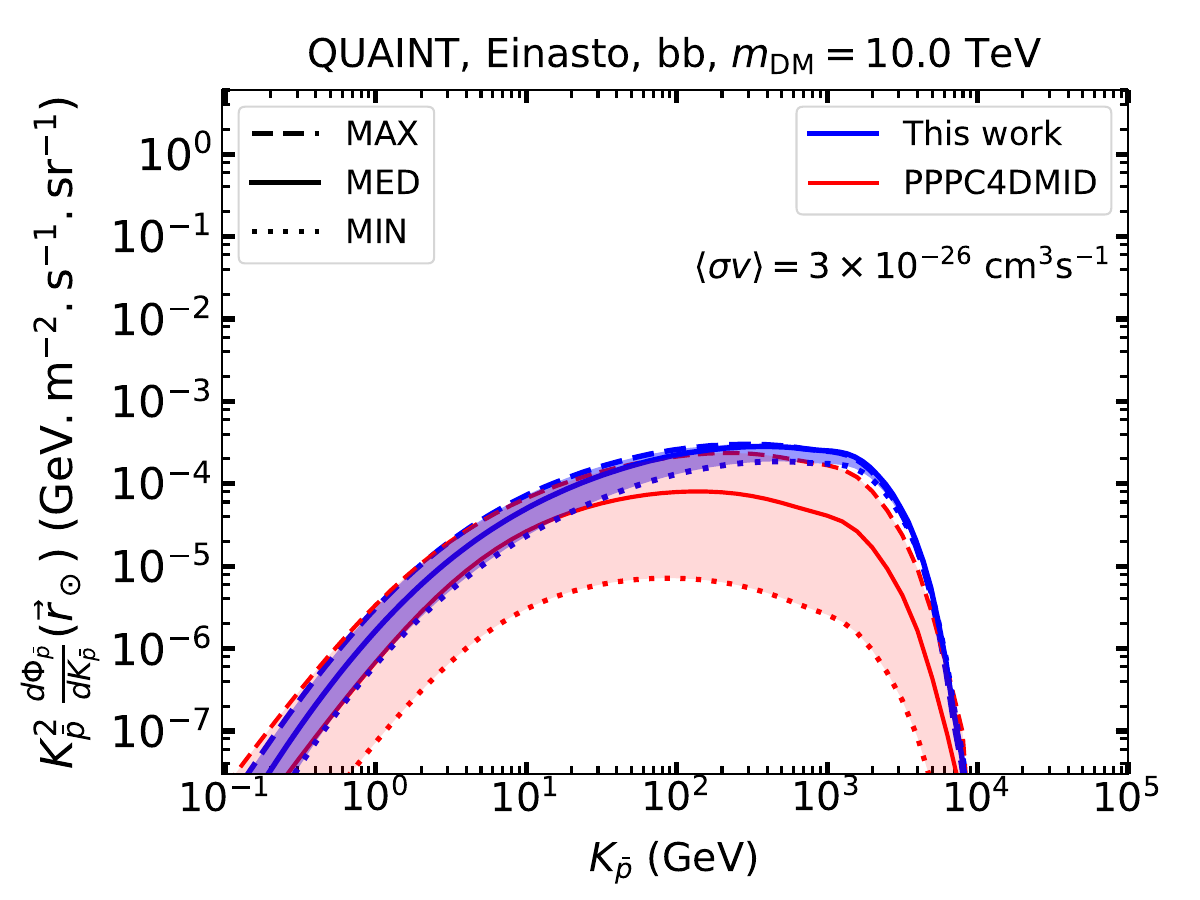} & \hspace{-6mm}
\includegraphics[width=0.36\textwidth]{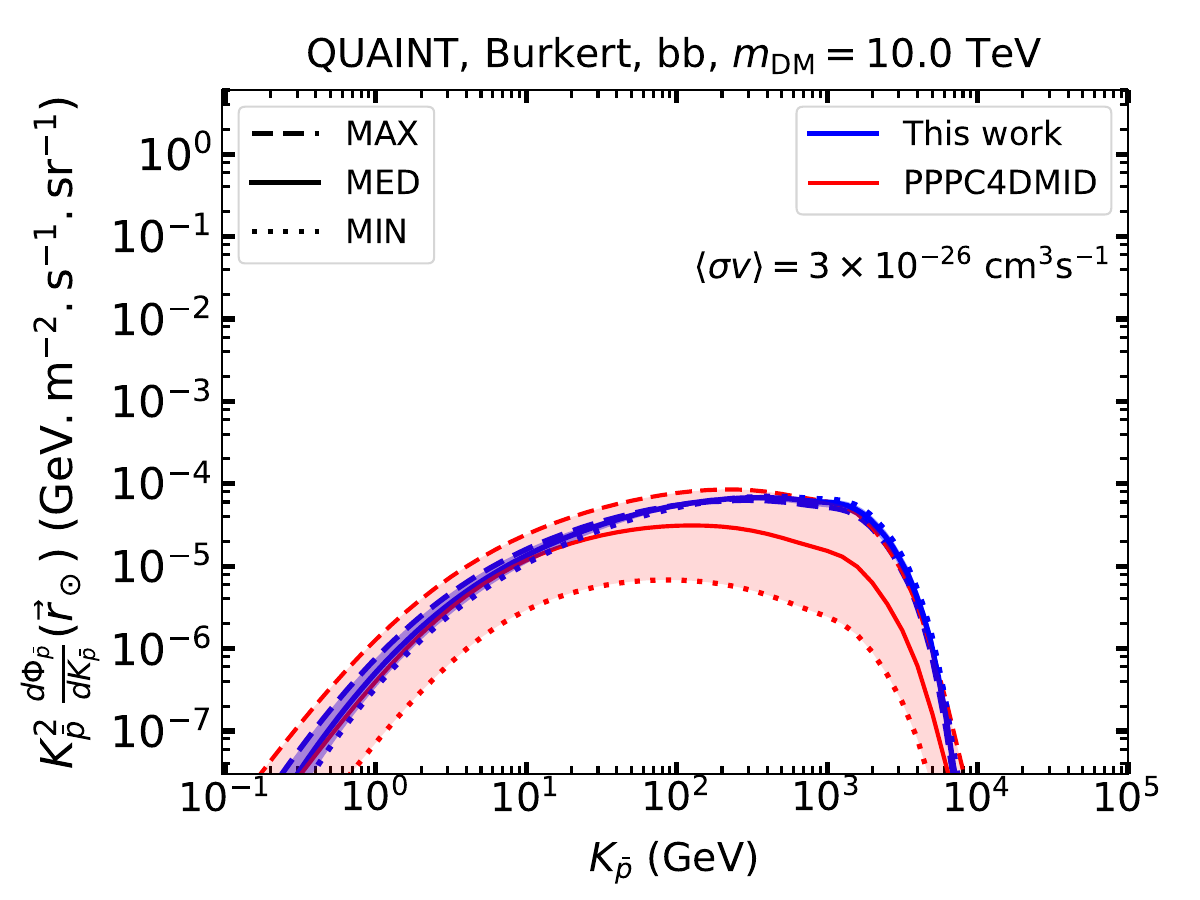}
\end{tabular}

\hspace{-10mm}
\begin{tabular}{ccc}
\includegraphics[width=0.36\textwidth]{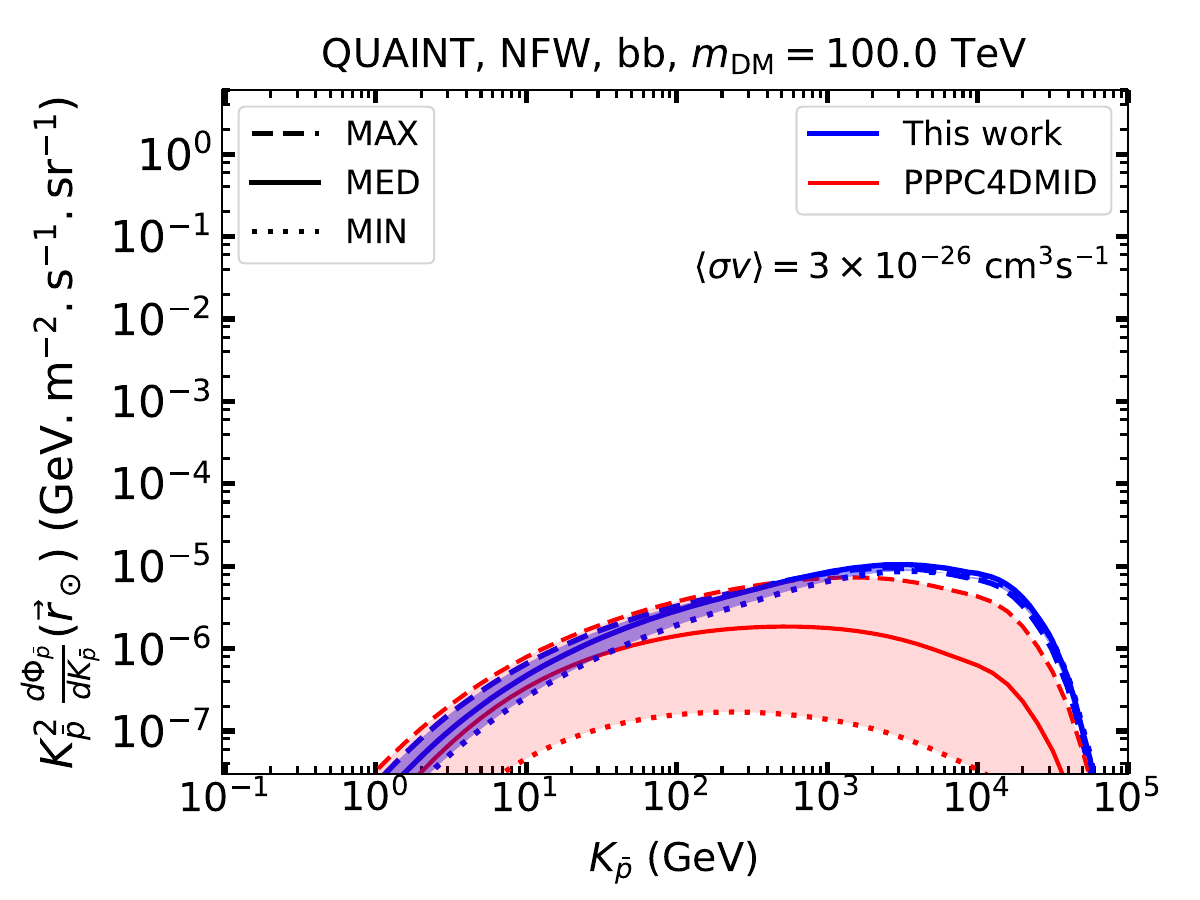}& \hspace{-6mm}
\includegraphics[width=0.36\textwidth]{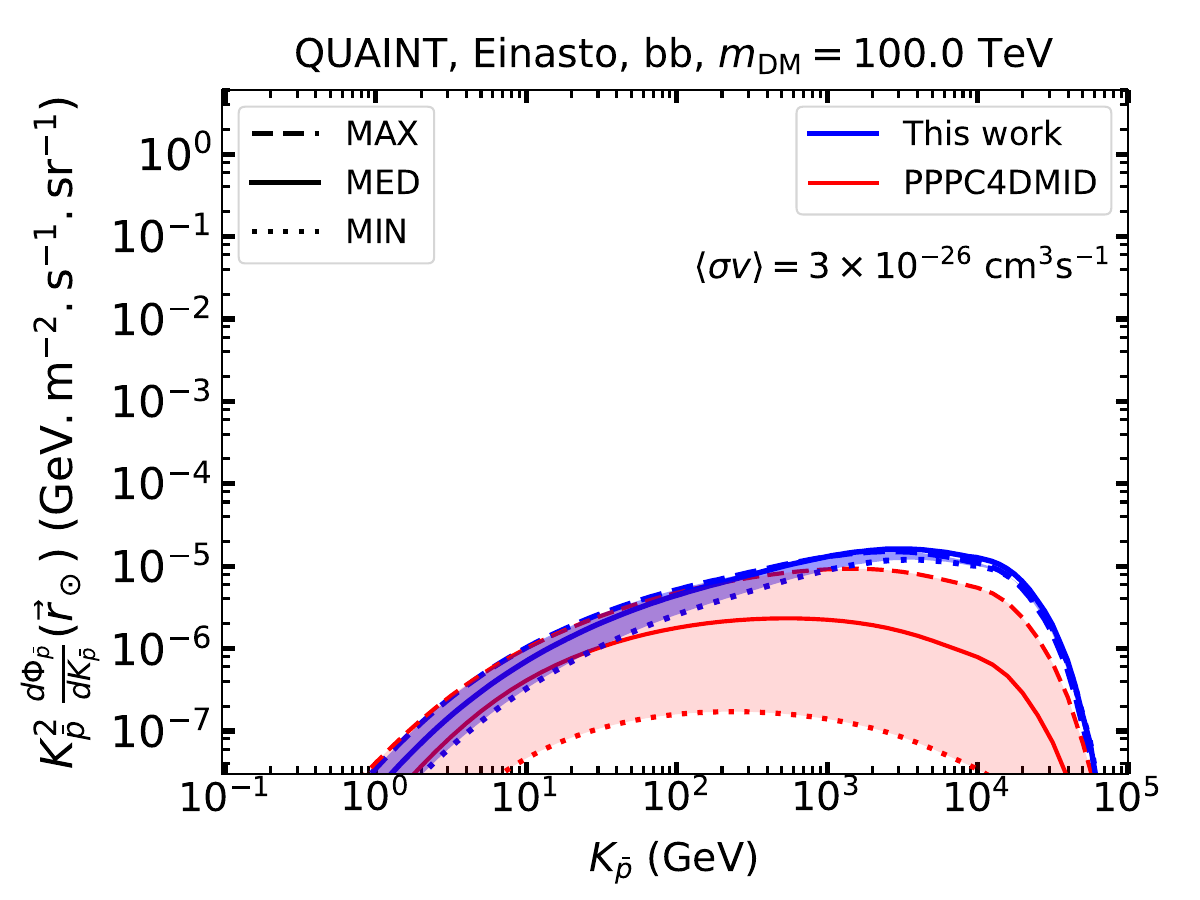} & \hspace{-6mm}
\includegraphics[width=0.36\textwidth]{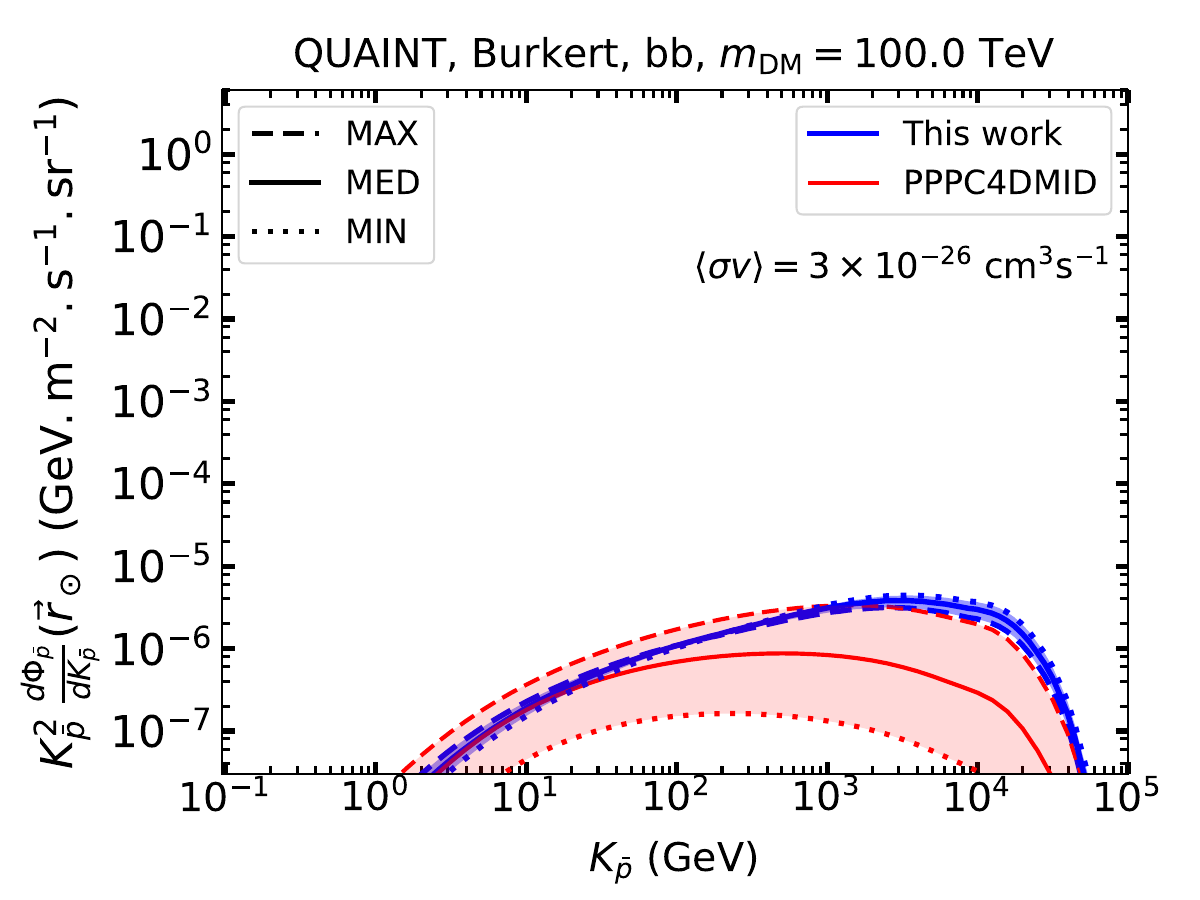}
\end{tabular}

\caption{{\it Antiproton fluxes} (IS) from ${\rm DM \, DM} \rightarrow \bm{b\bar{b}}$ annihilations, under the {\sc Quaint} propagation scheme, compared to the previous \texttt{PPPC4DMID} results.
The rows and columns are as in Fig.~\ref{fig:pbar_SLIM_bb_Ann}.}
\label{fig:pbar_QUAINT_bb_Ann}
\vspace{2cm}
\end{figure*}

\begin{figure*}[!ht]
\vspace{1cm}
\hspace{-10mm}
\begin{tabular}{ccc}
\includegraphics[width=0.36\textwidth]{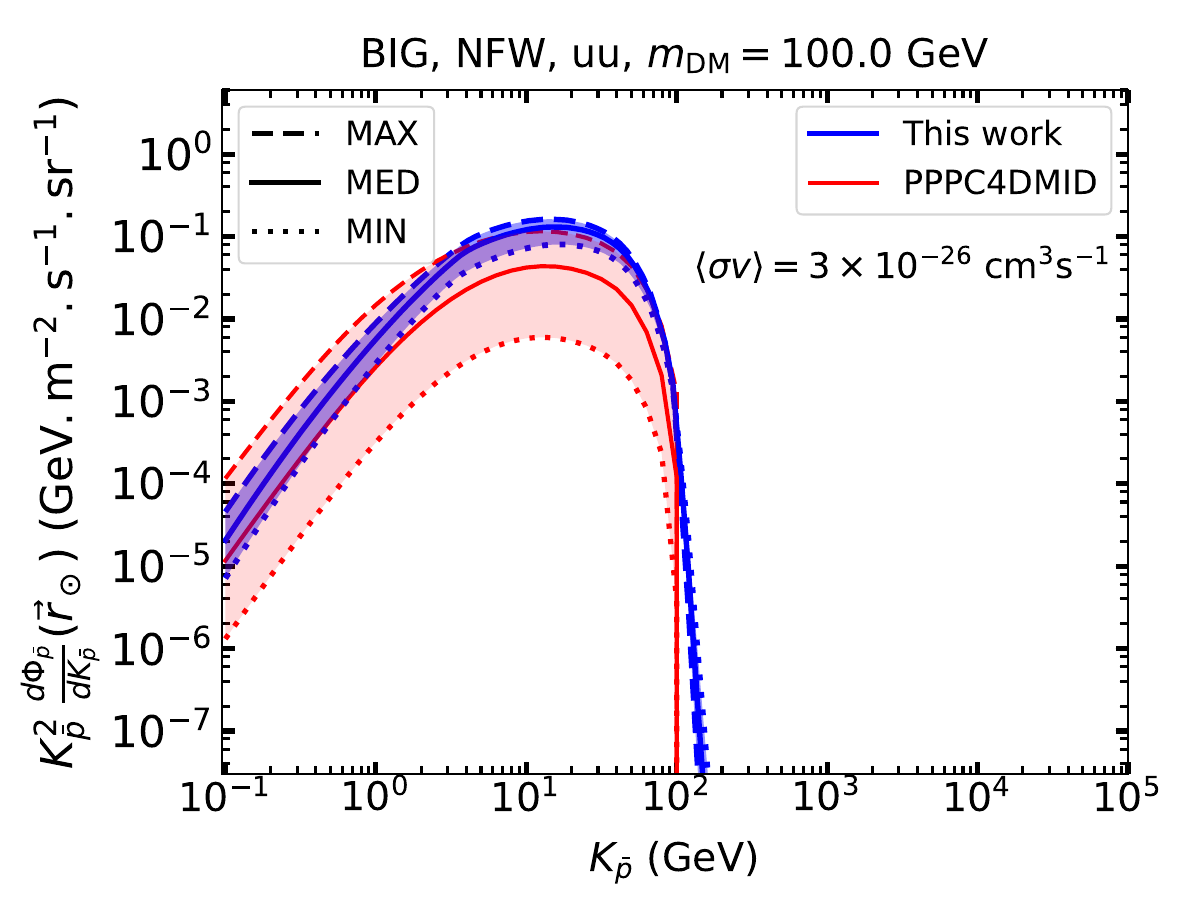} & \hspace{-6mm}
\includegraphics[width=0.36\textwidth]{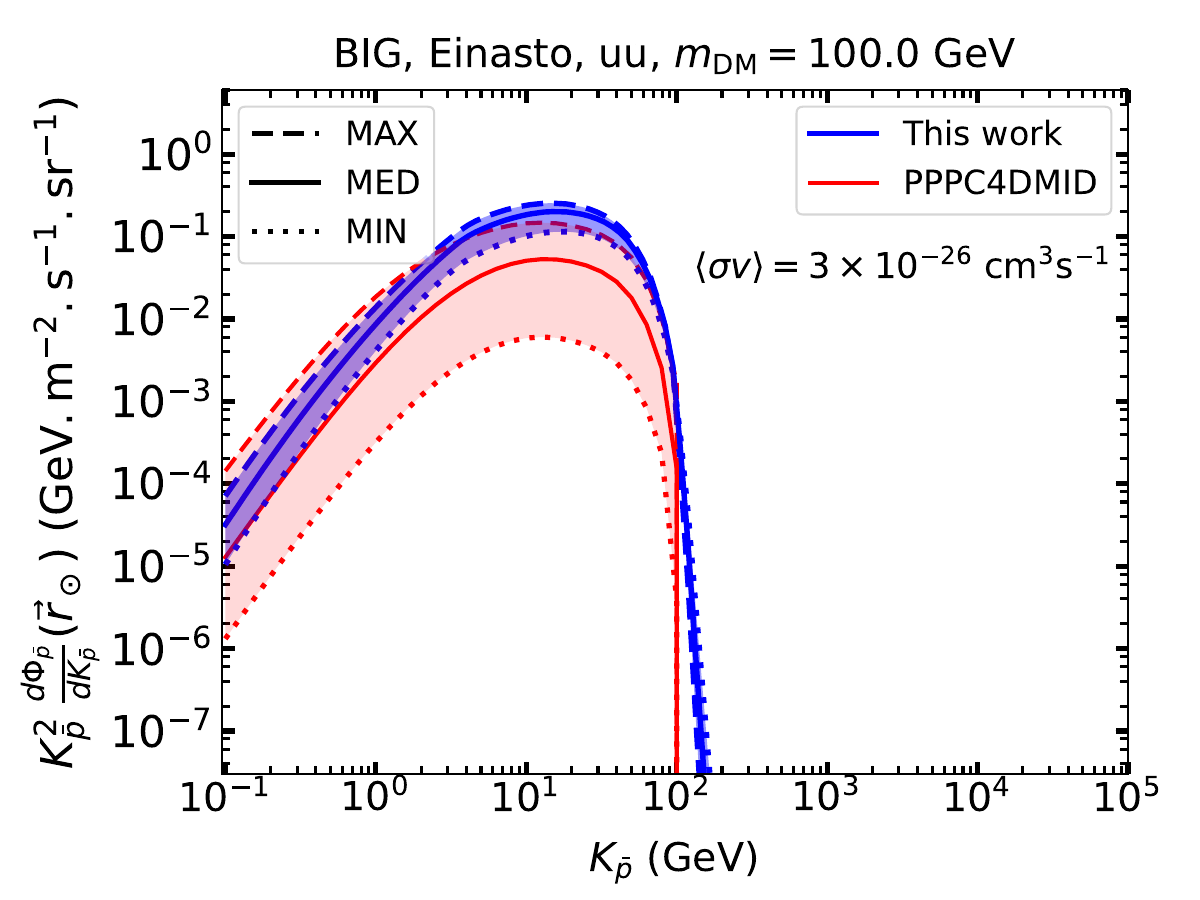} & \hspace{-6mm}
\includegraphics[width=0.36\textwidth]{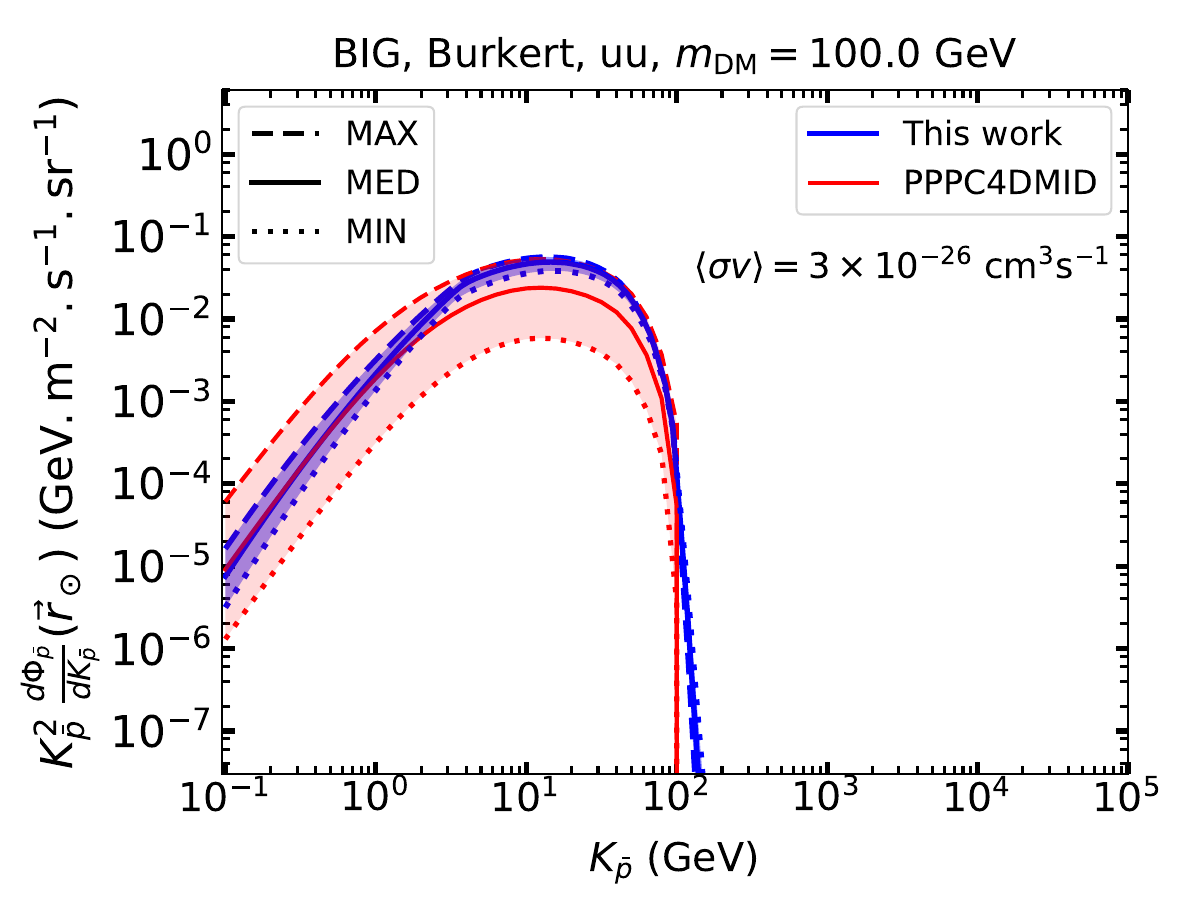}
\end{tabular}

\hspace{-10mm}
\begin{tabular}{ccc}
\includegraphics[width=0.36\textwidth]{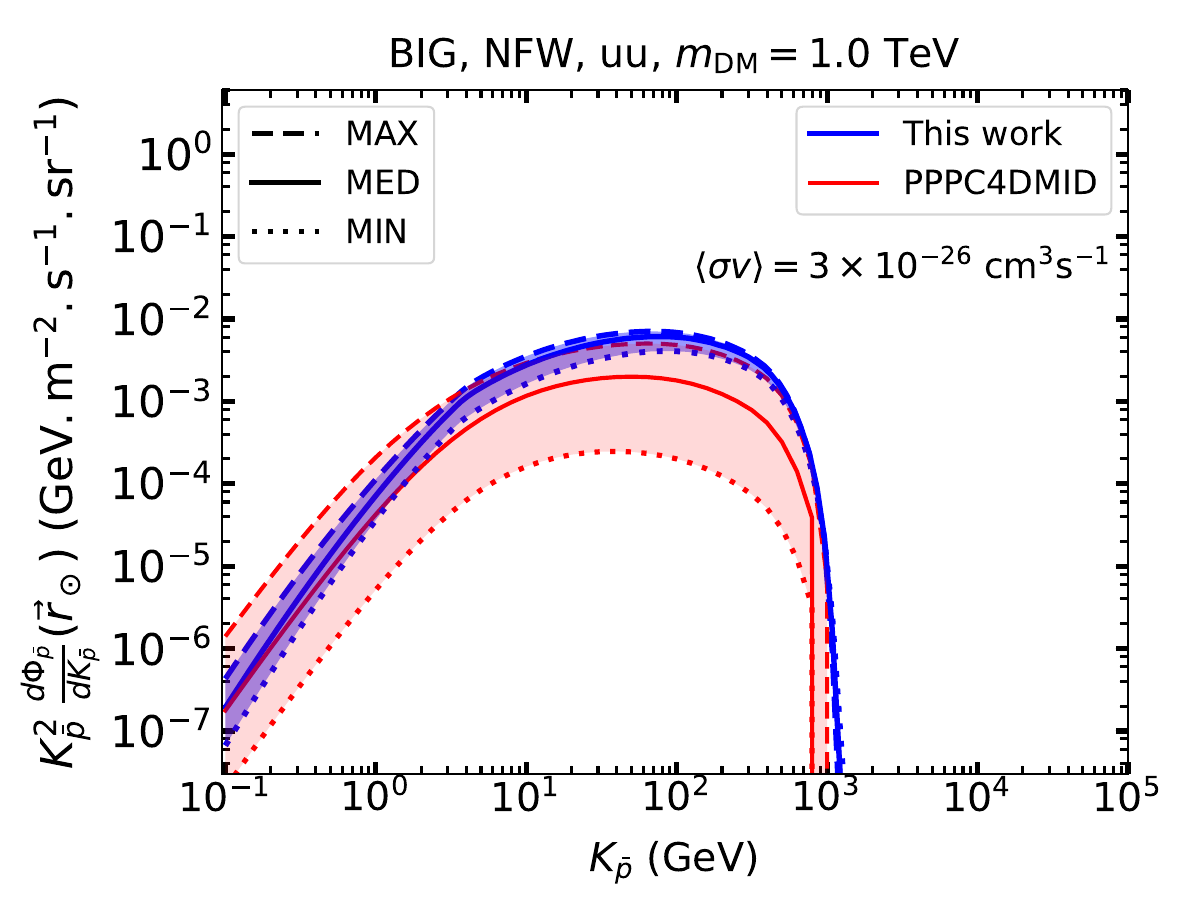} & \hspace{-6mm}
\includegraphics[width=0.36\textwidth]{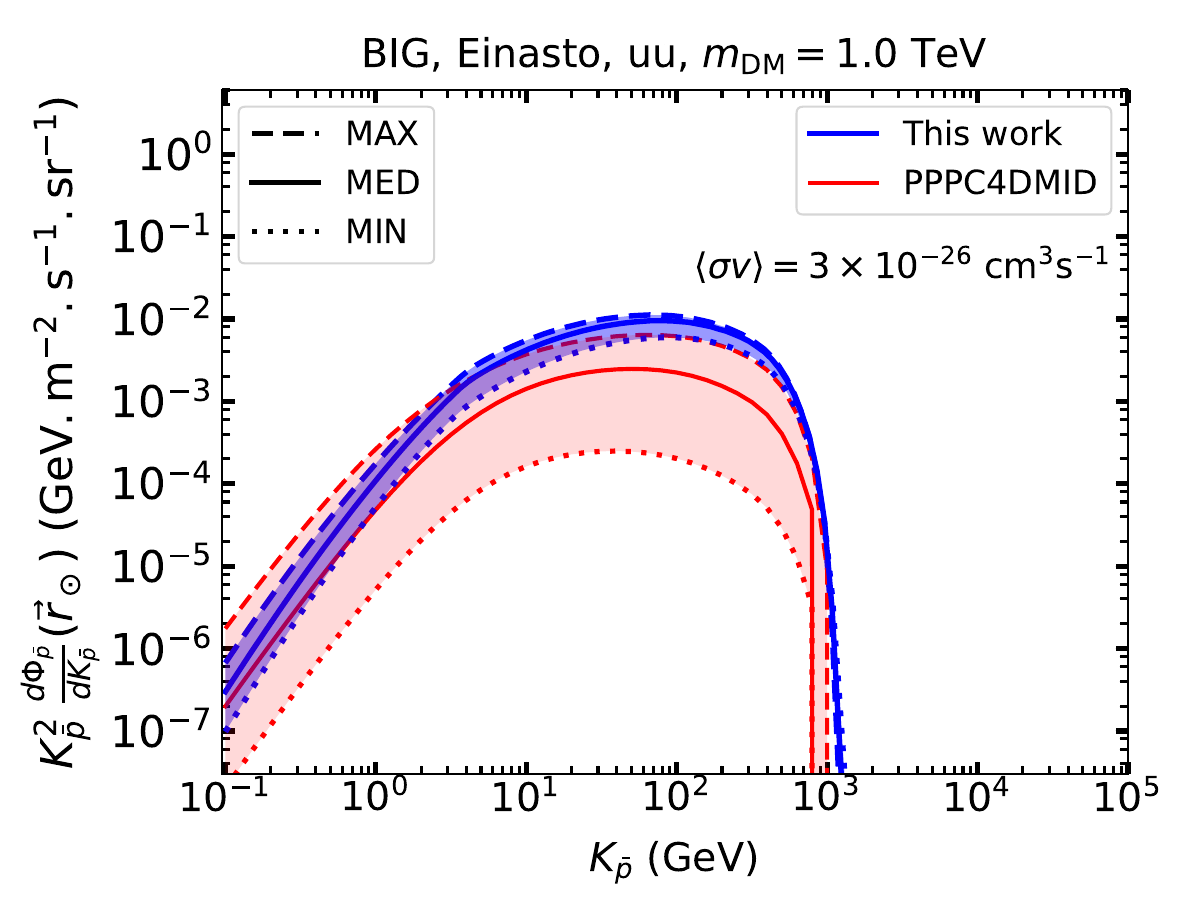} & \hspace{-6mm}
\includegraphics[width=0.36\textwidth]{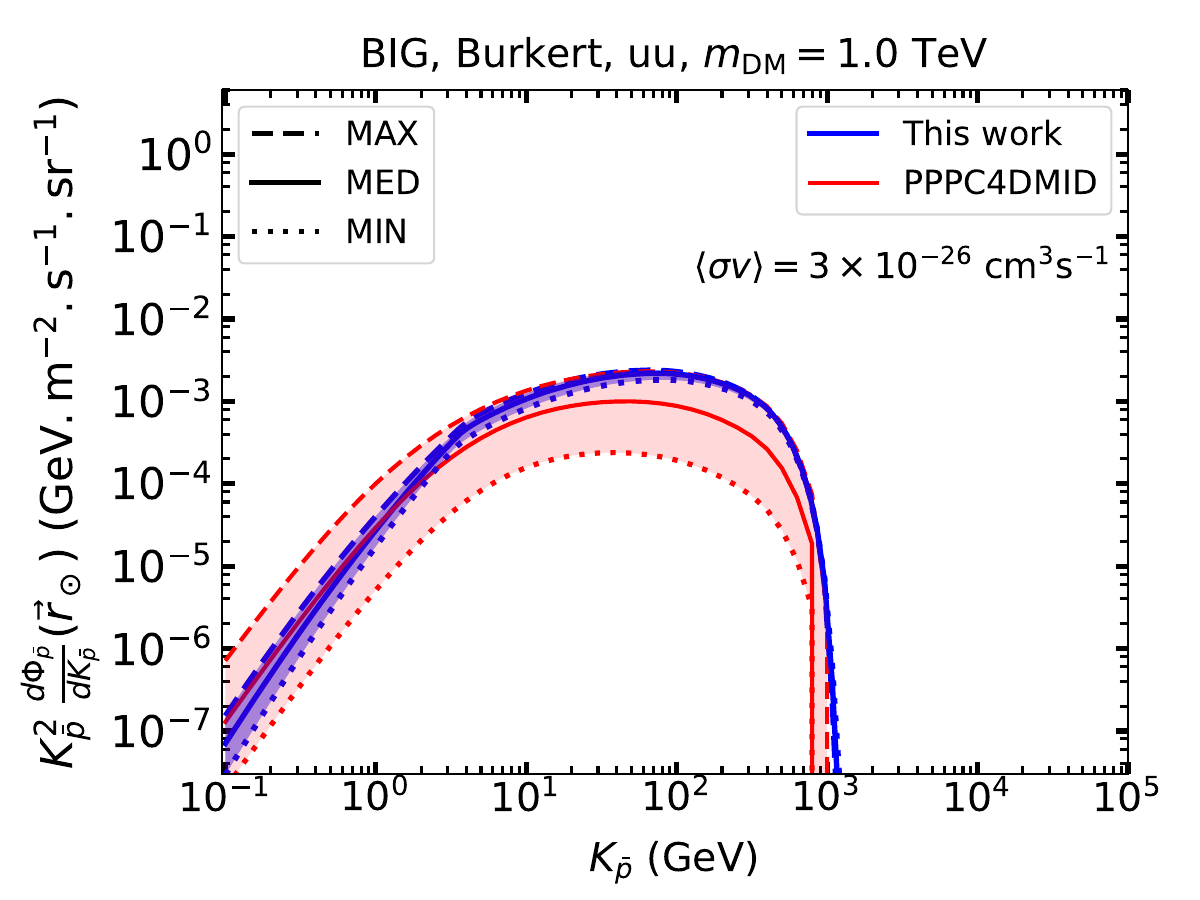}
\end{tabular}

\hspace{-10mm}
\begin{tabular}{ccc}
\includegraphics[width=0.36\textwidth]{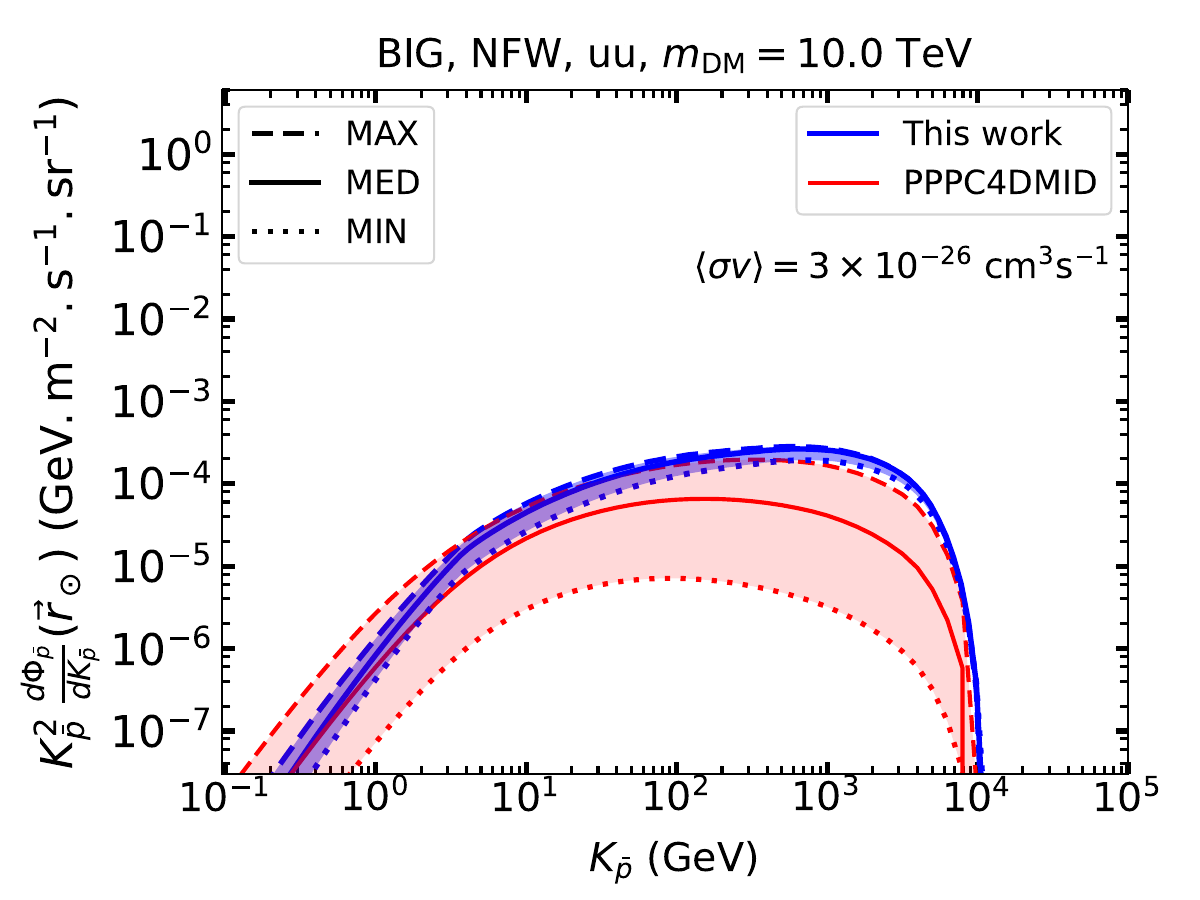} & \hspace{-6mm}
\includegraphics[width=0.36\textwidth]{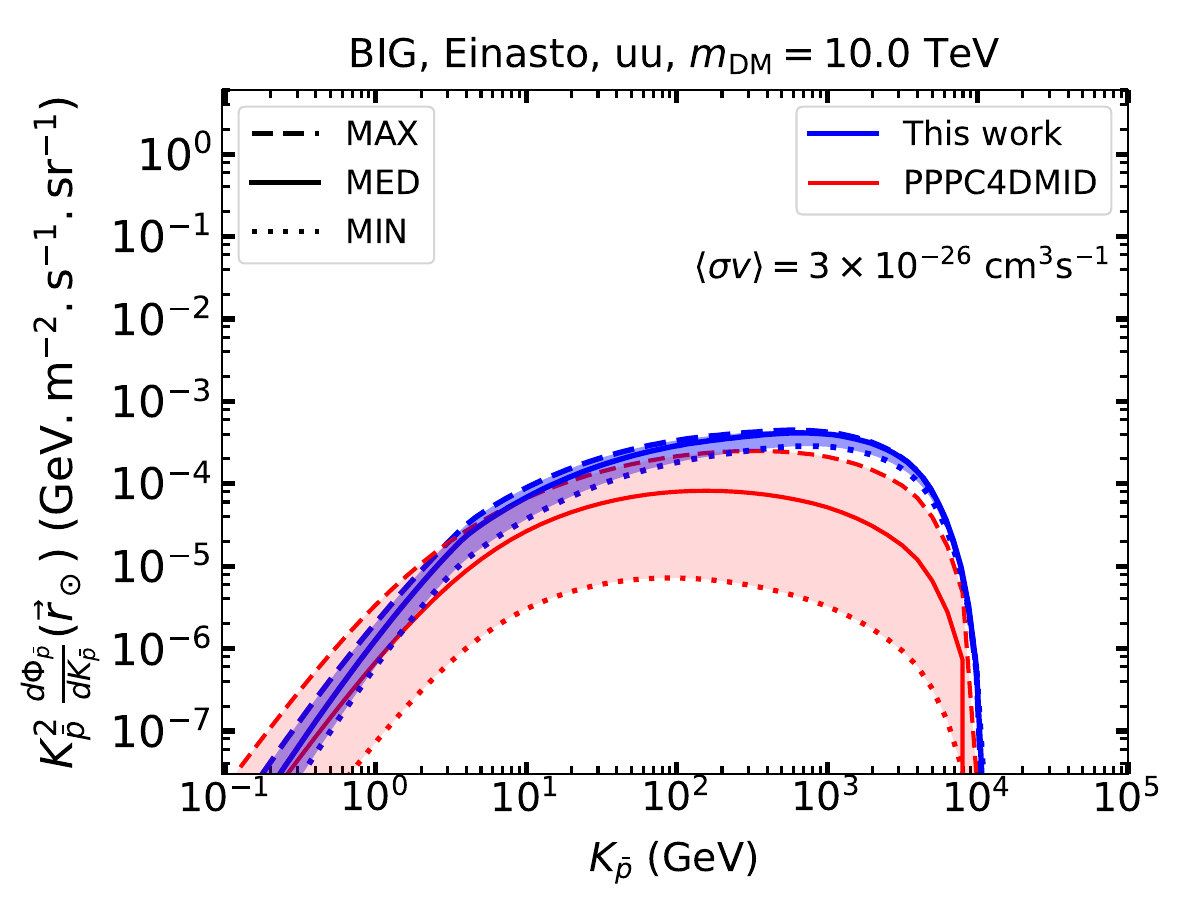} & \hspace{-6mm}
\includegraphics[width=0.36\textwidth]{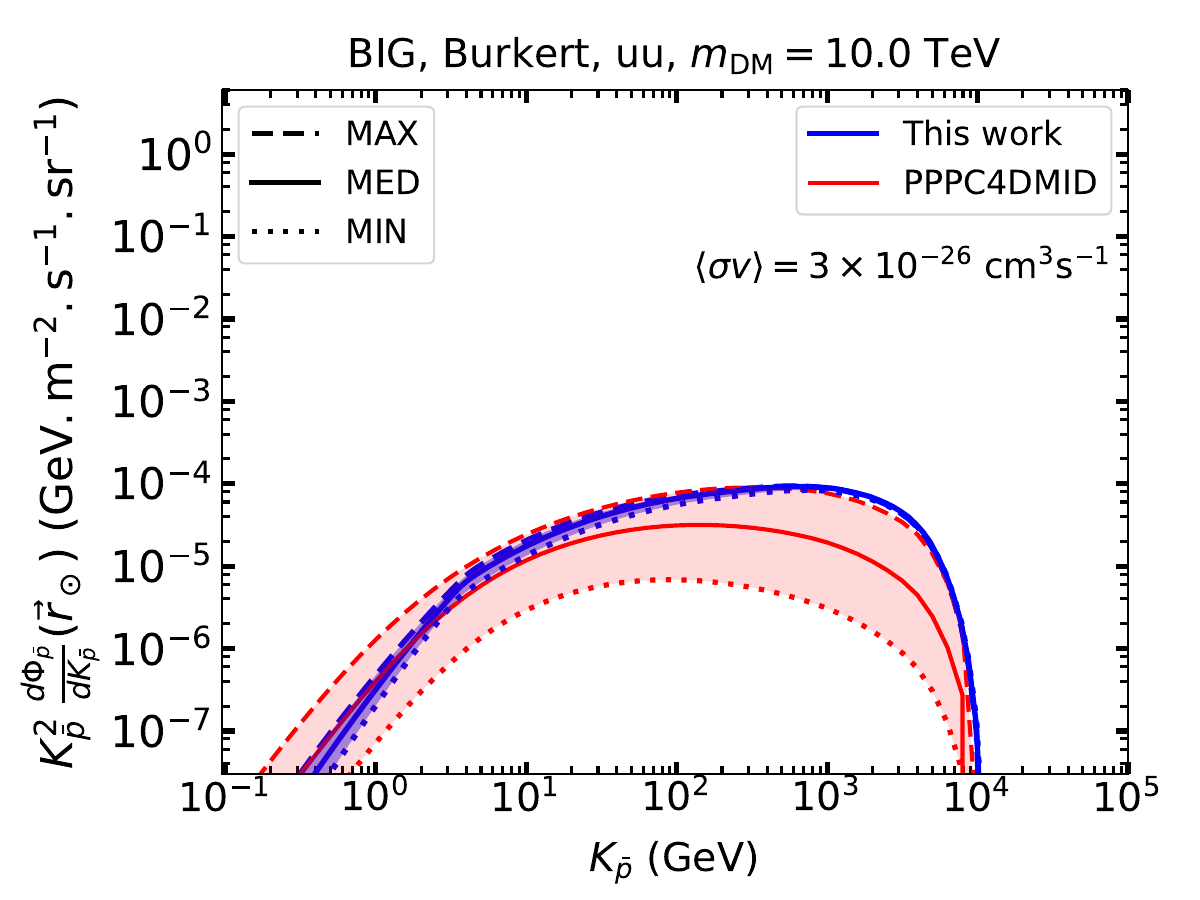}
\end{tabular}

\hspace{-10mm}
\begin{tabular}{ccc}
\includegraphics[width=0.36\textwidth]{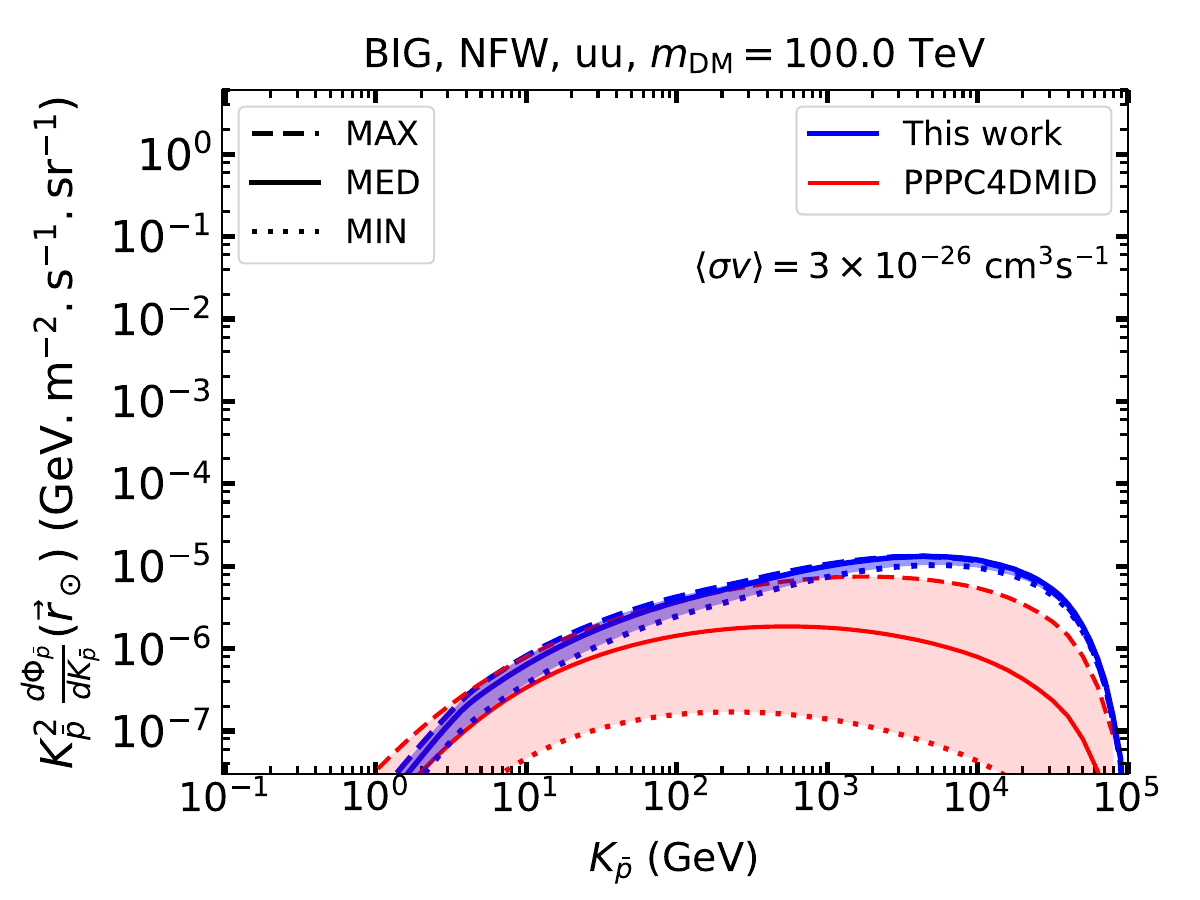} & \hspace{-6mm}
\includegraphics[width=0.36\textwidth]{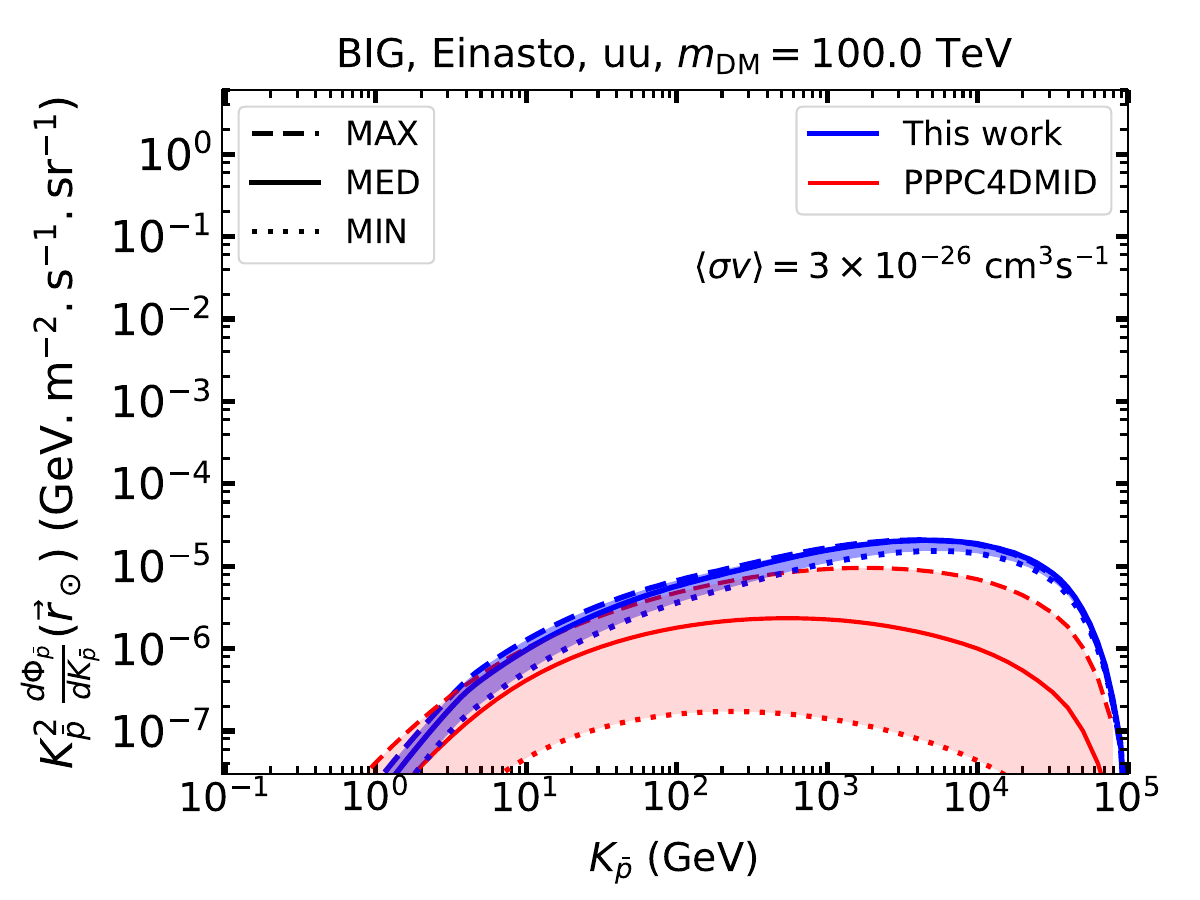} & \hspace{-6mm}
\includegraphics[width=0.36\textwidth]{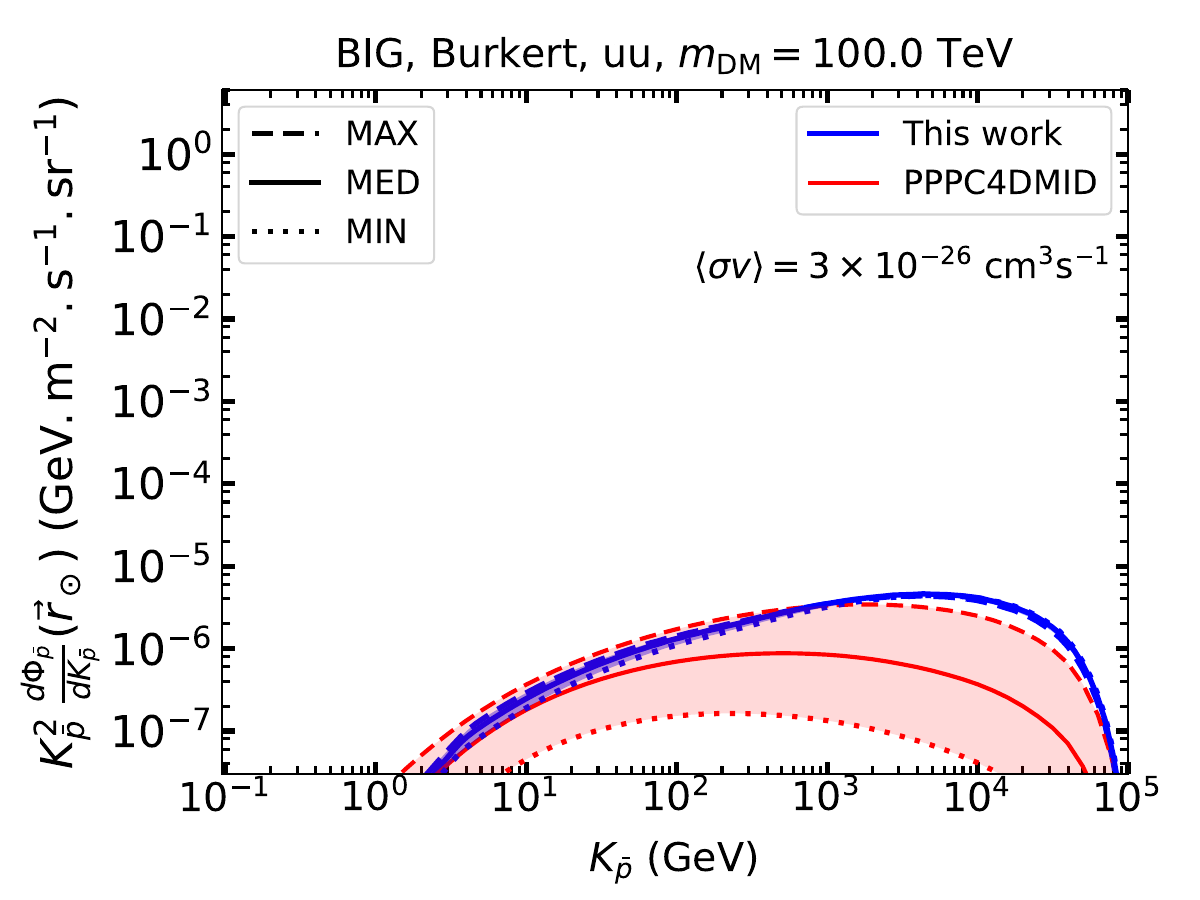}
\end{tabular}

\caption{{\it Antiproton fluxes} (IS) from ${\rm DM \, DM} \rightarrow \bm{u\bar{u}}$ annihilations, under the {\sc Big} propagation scheme, compared to the previous \texttt{PPPC4DMID} results (for $\bm{q\bar{q}}$).
The rows and columns are as in Fig.~\ref{fig:pbar_SLIM_bb_Ann}.}
\label{fig:pbar_BIG_uu_Ann}
\vspace{2cm}
\end{figure*}

\begin{figure*}[!ht]
\vspace{1cm}
\hspace{-10mm}
\begin{tabular}{ccc}
\includegraphics[width=0.36\textwidth]{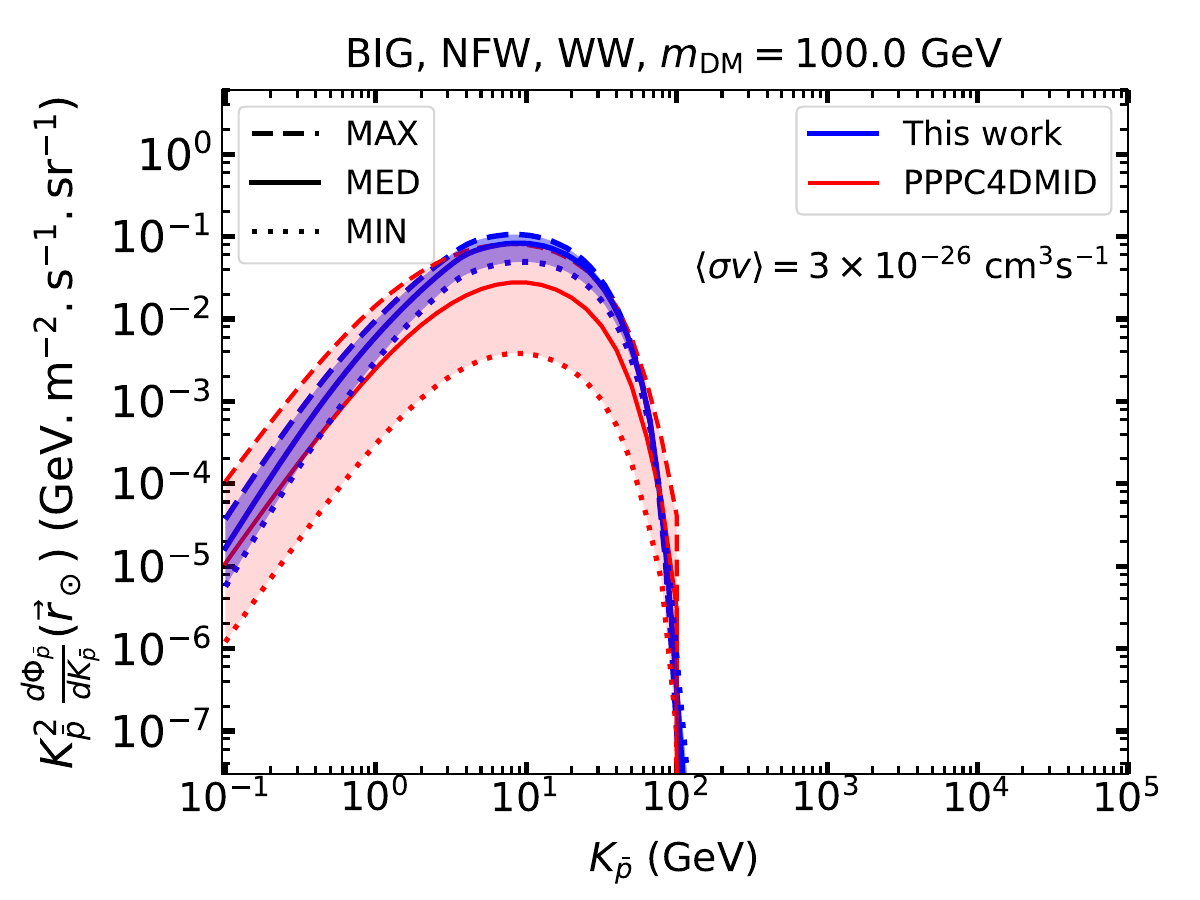} & \hspace{-6mm}
\includegraphics[width=0.36\textwidth]{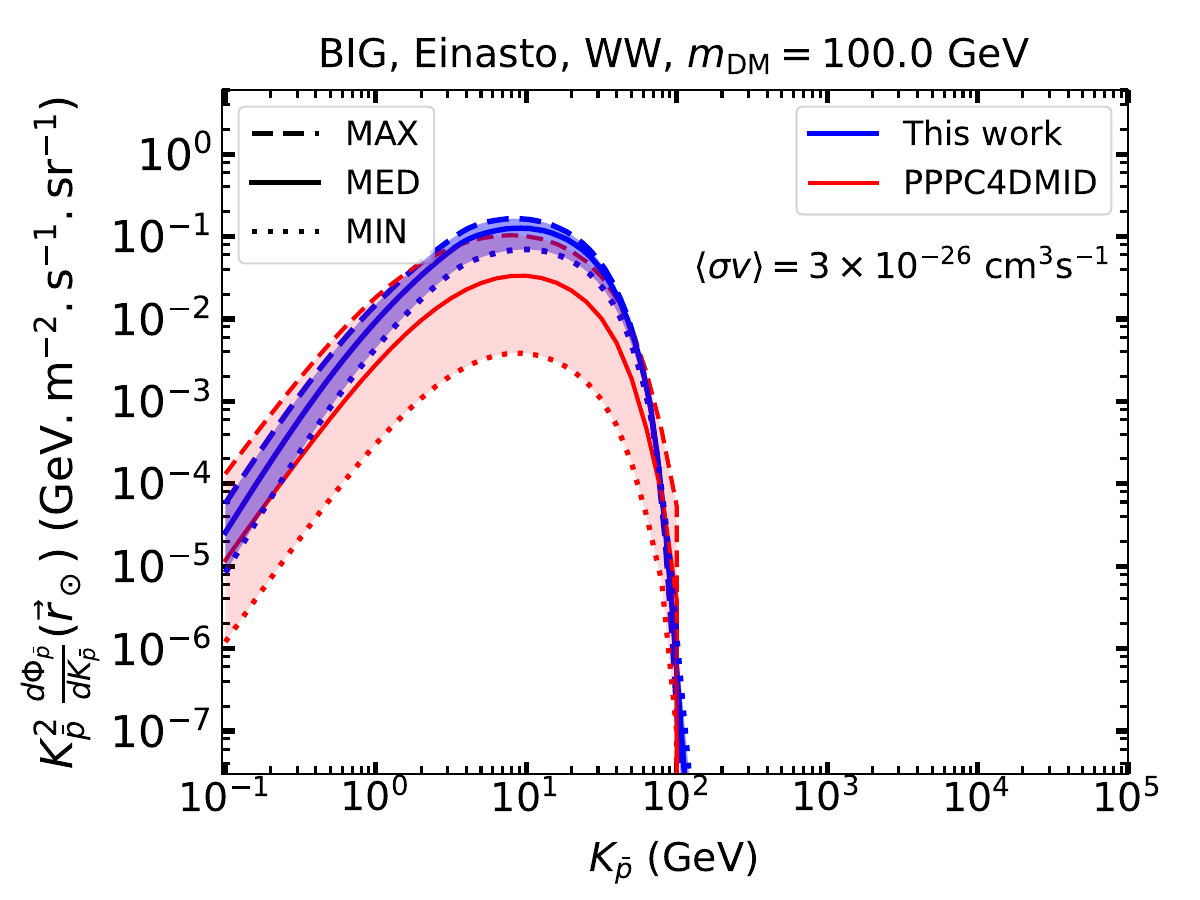} & \hspace{-6mm}
\includegraphics[width=0.36\textwidth]{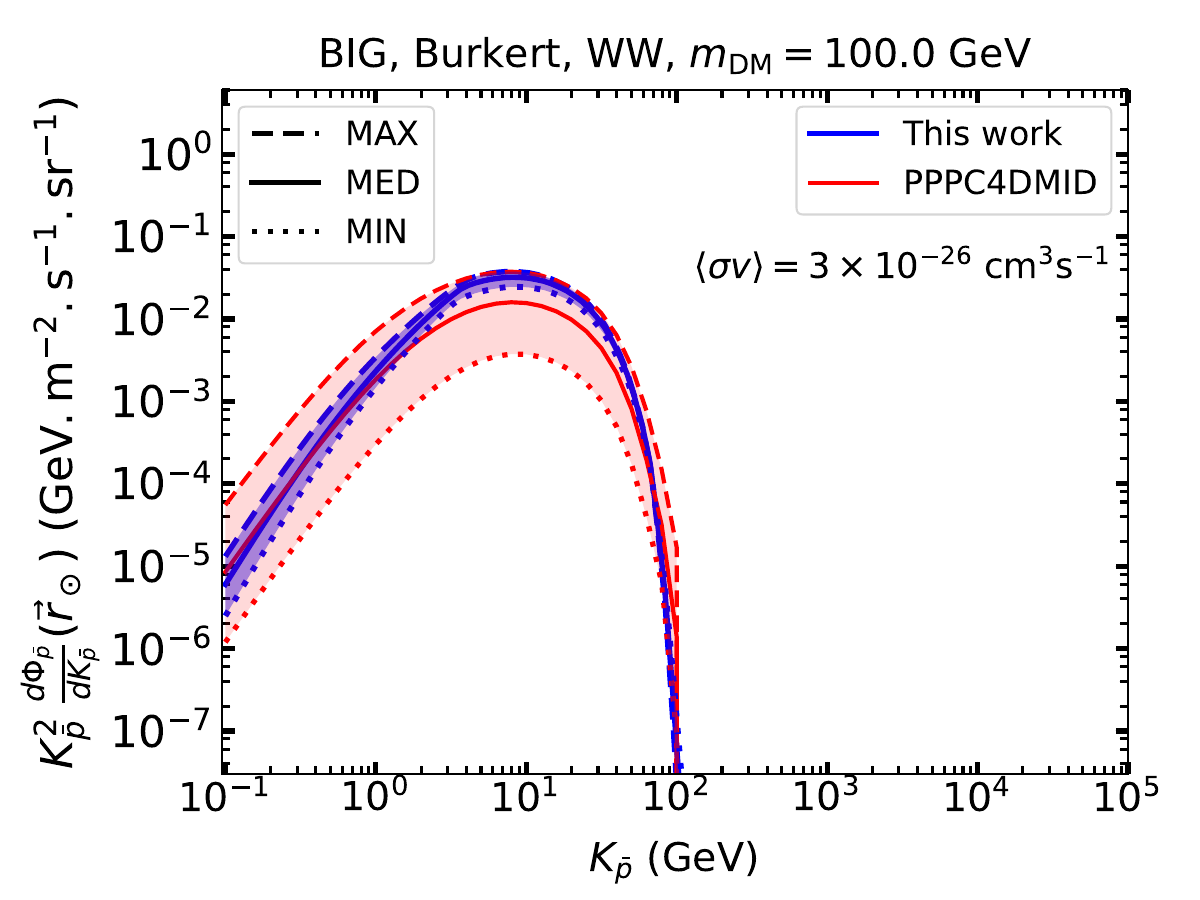}
\end{tabular}

\hspace{-10mm}
\begin{tabular}{ccc}
\includegraphics[width=0.36\textwidth]{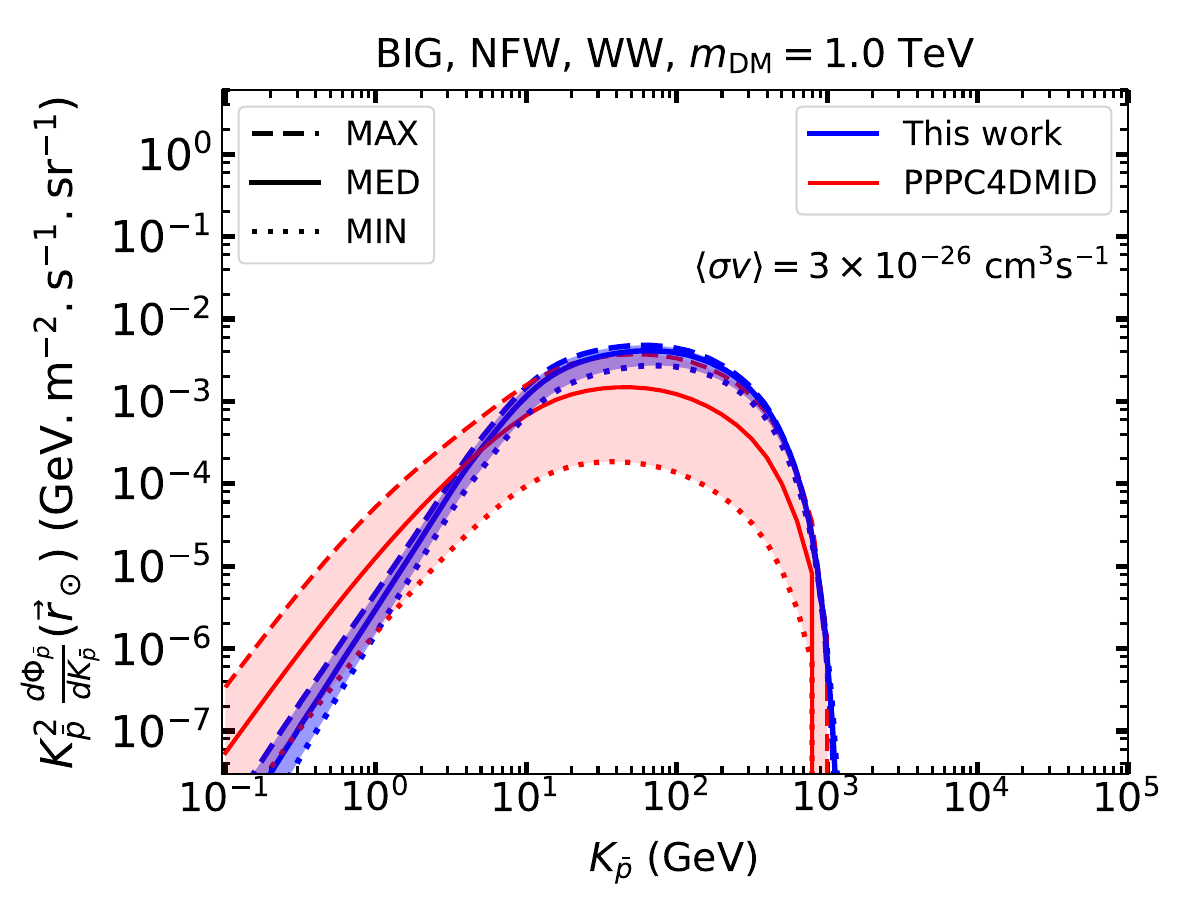} & \hspace{-6mm}
\includegraphics[width=0.36\textwidth]{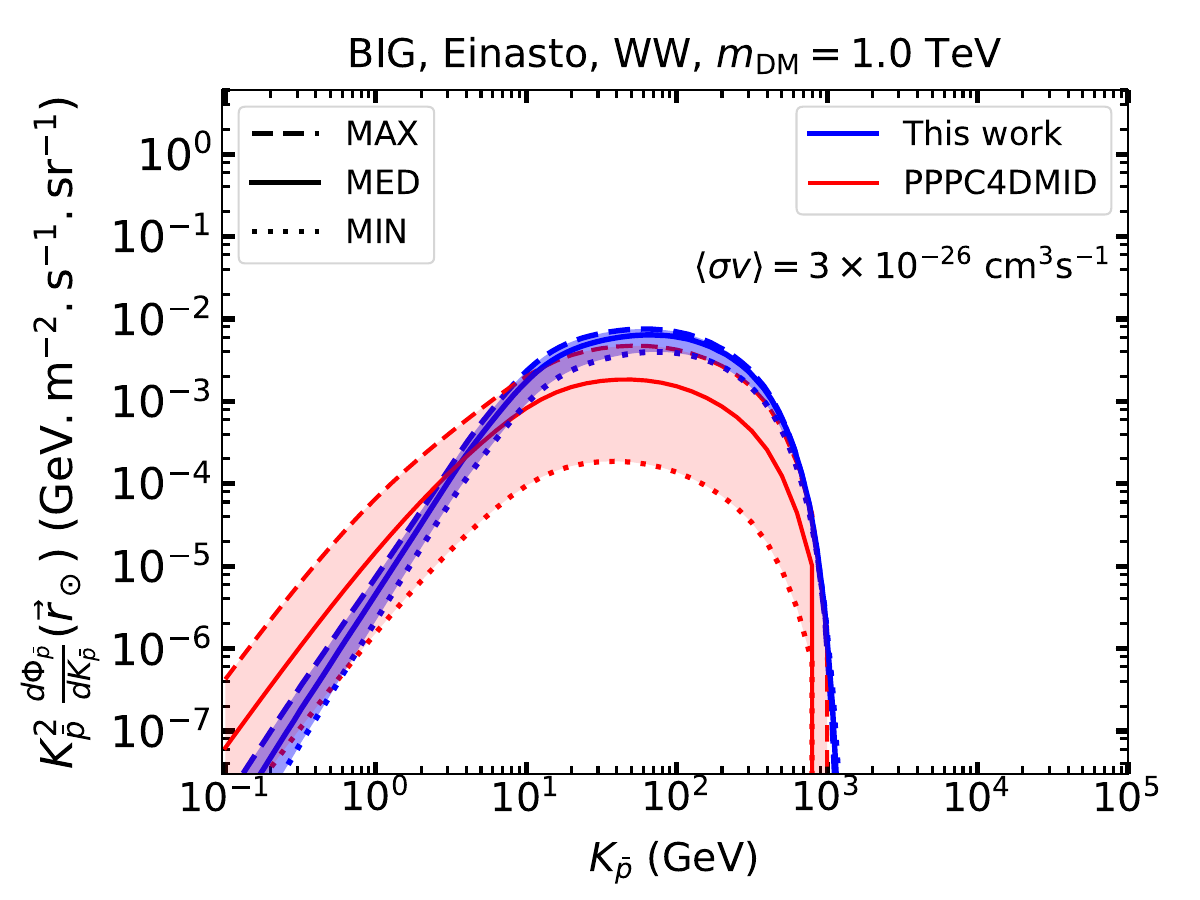} & \hspace{-6mm}
\includegraphics[width=0.36\textwidth]{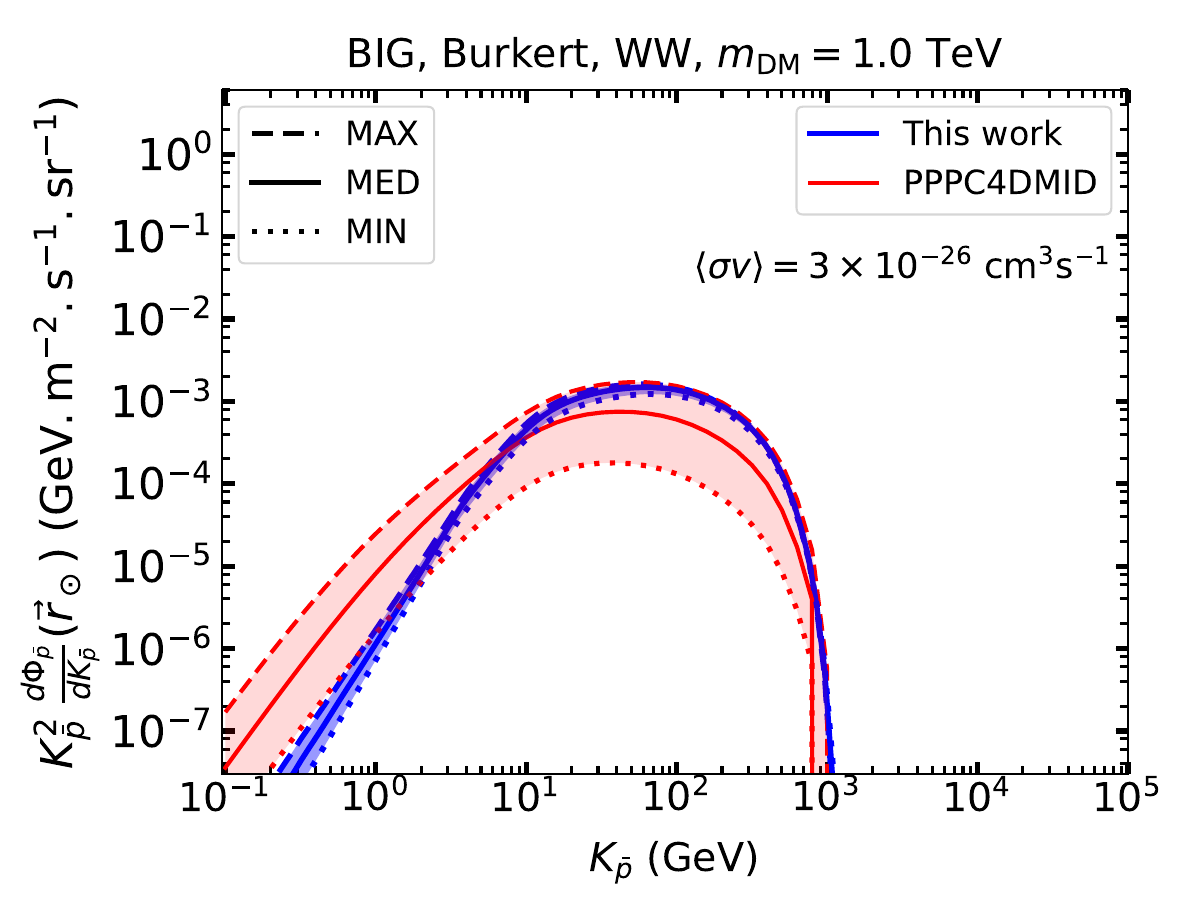}
\end{tabular}

\hspace{-10mm}
\begin{tabular}{ccc}
\includegraphics[width=0.36\textwidth]{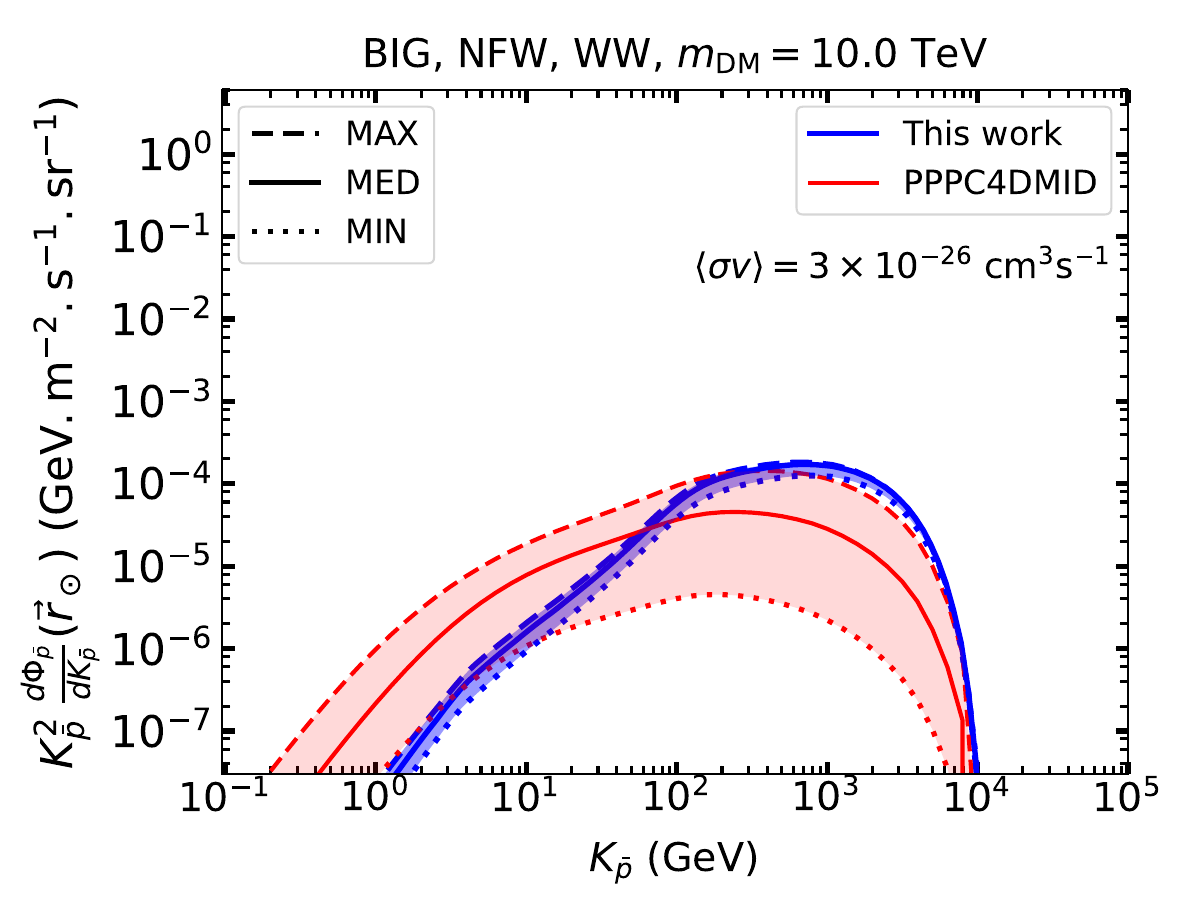} & \hspace{-6mm}
\includegraphics[width=0.36\textwidth]{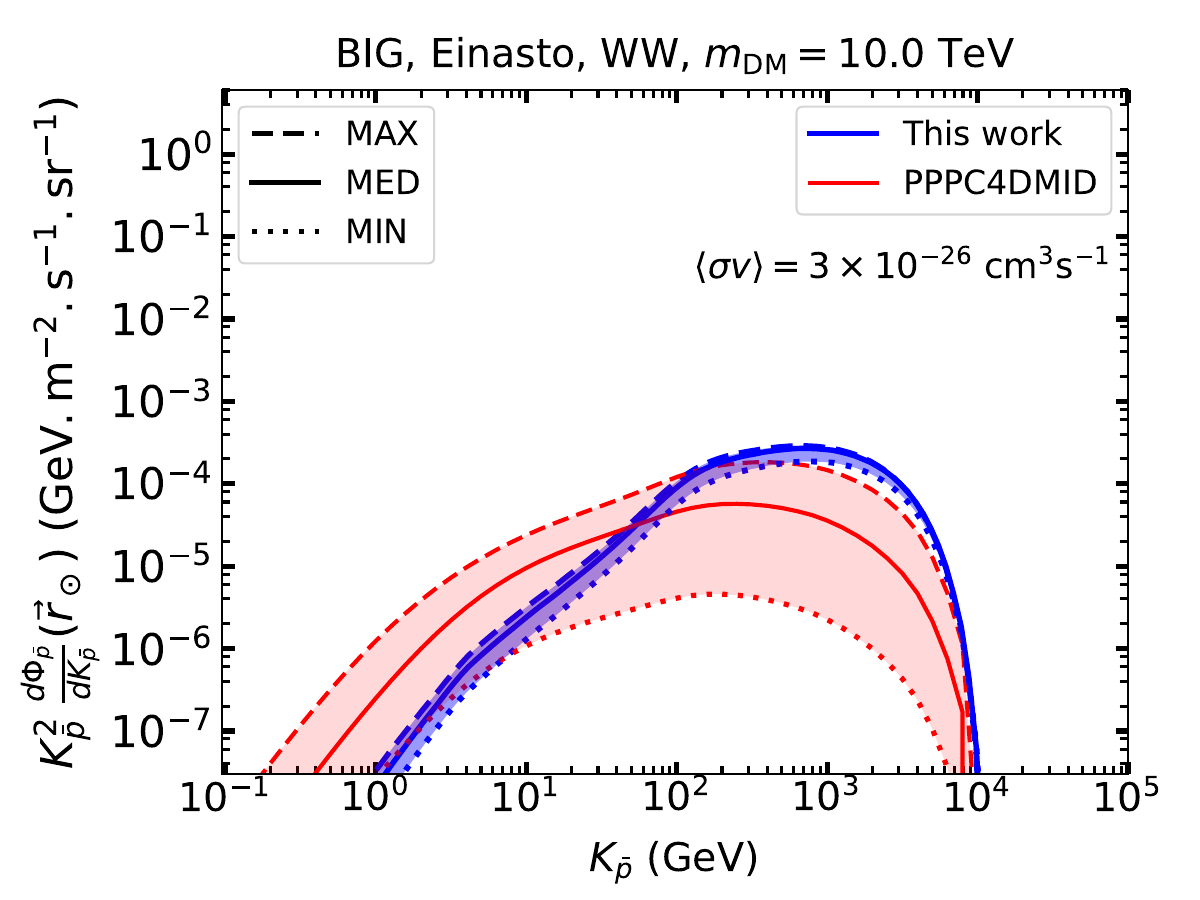} & \hspace{-6mm}
\includegraphics[width=0.36\textwidth]{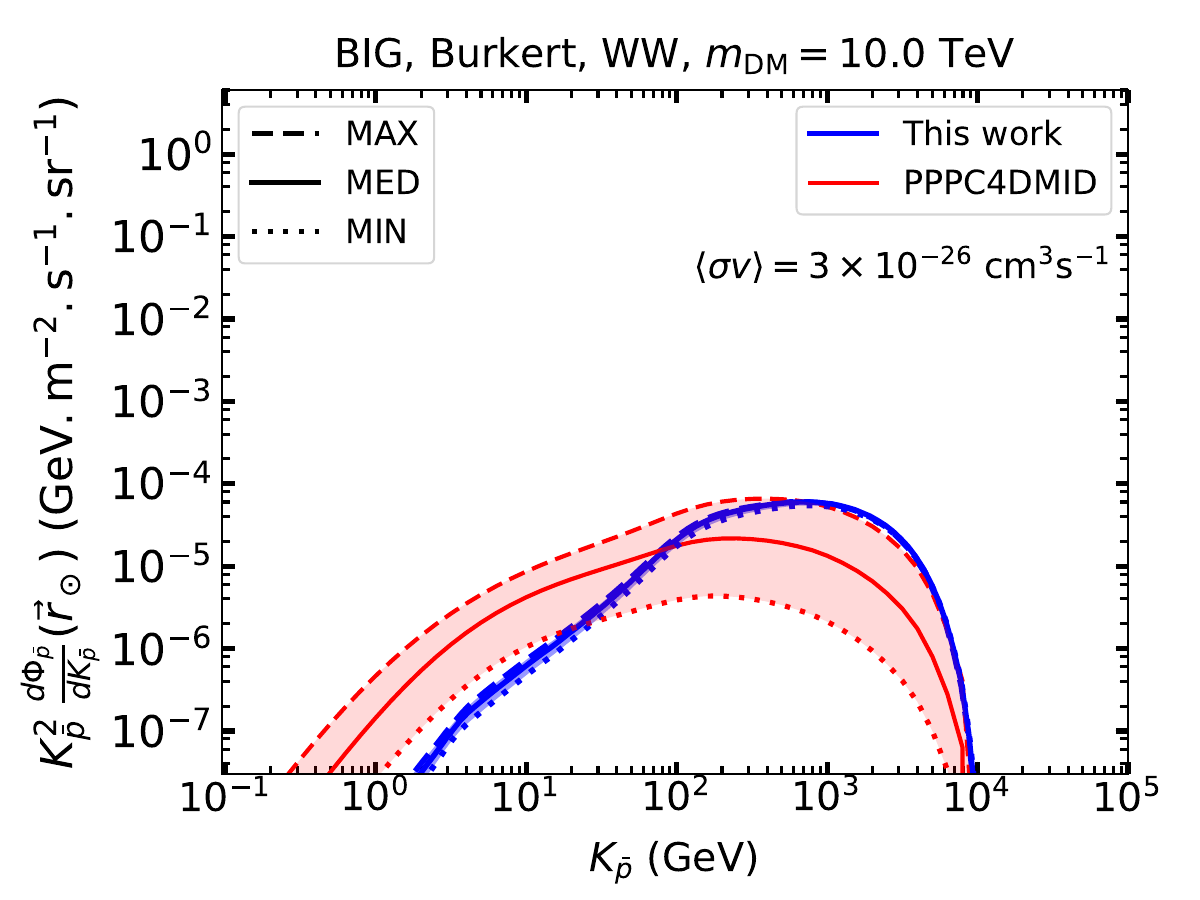}
\end{tabular}

\hspace{-10mm}
\begin{tabular}{ccc}
\includegraphics[width=0.36\textwidth]{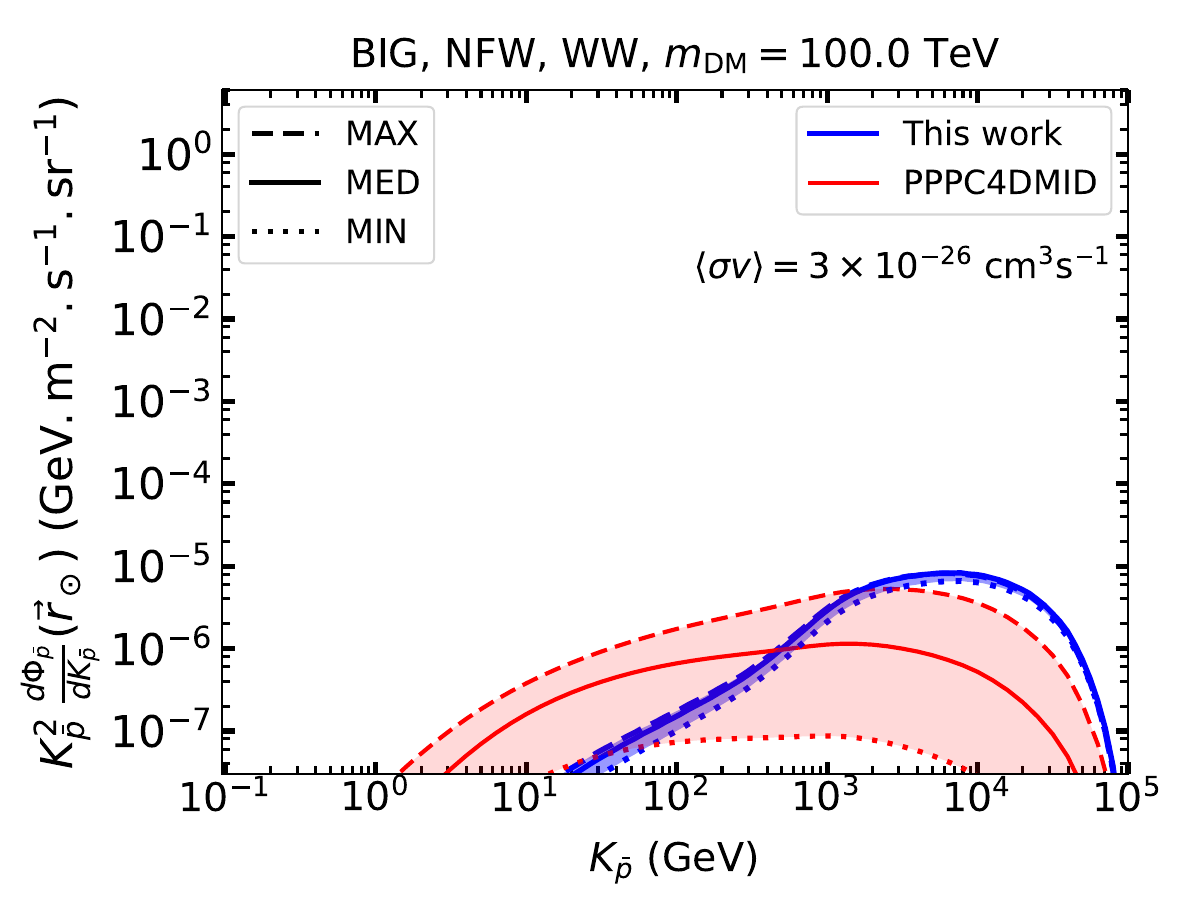} & \hspace{-6mm}
\includegraphics[width=0.36\textwidth]{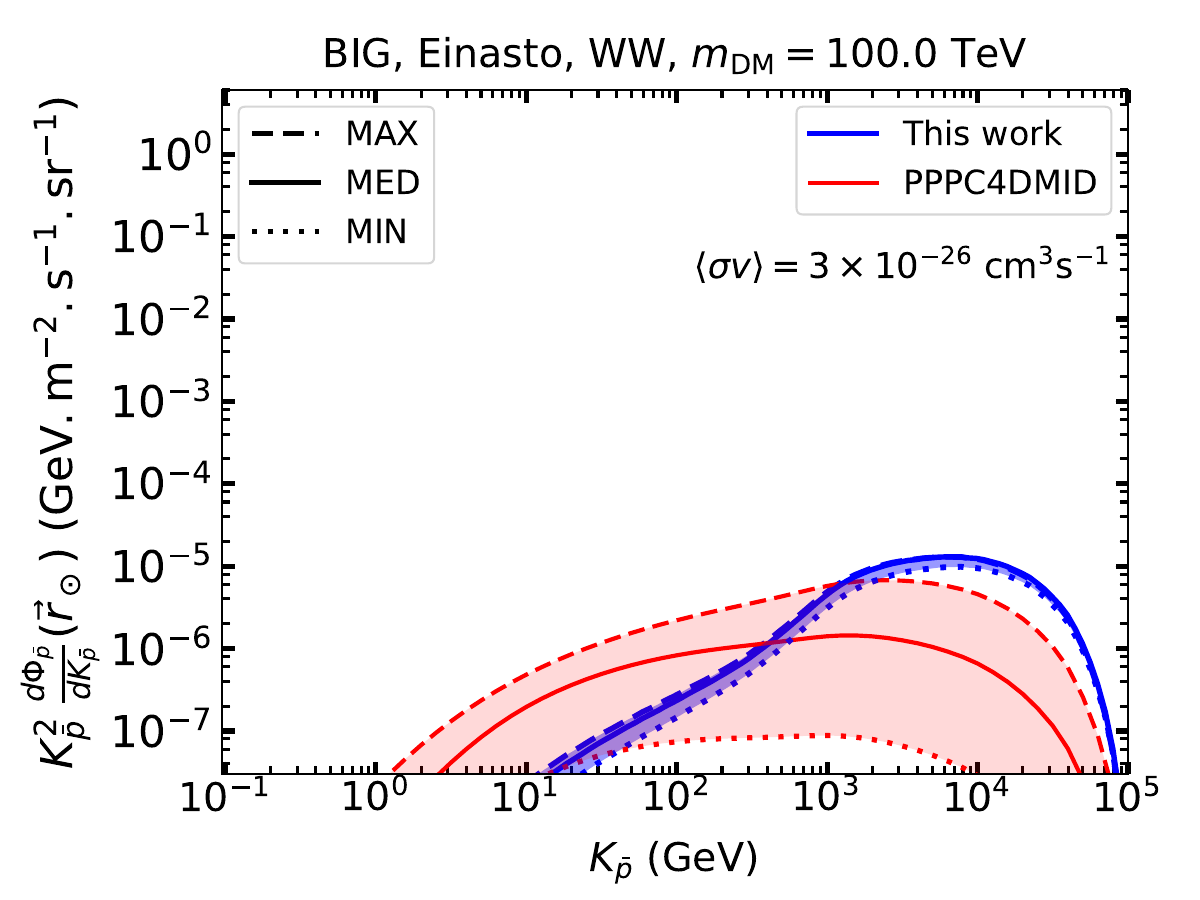} & \hspace{-6mm}
\includegraphics[width=0.36\textwidth]{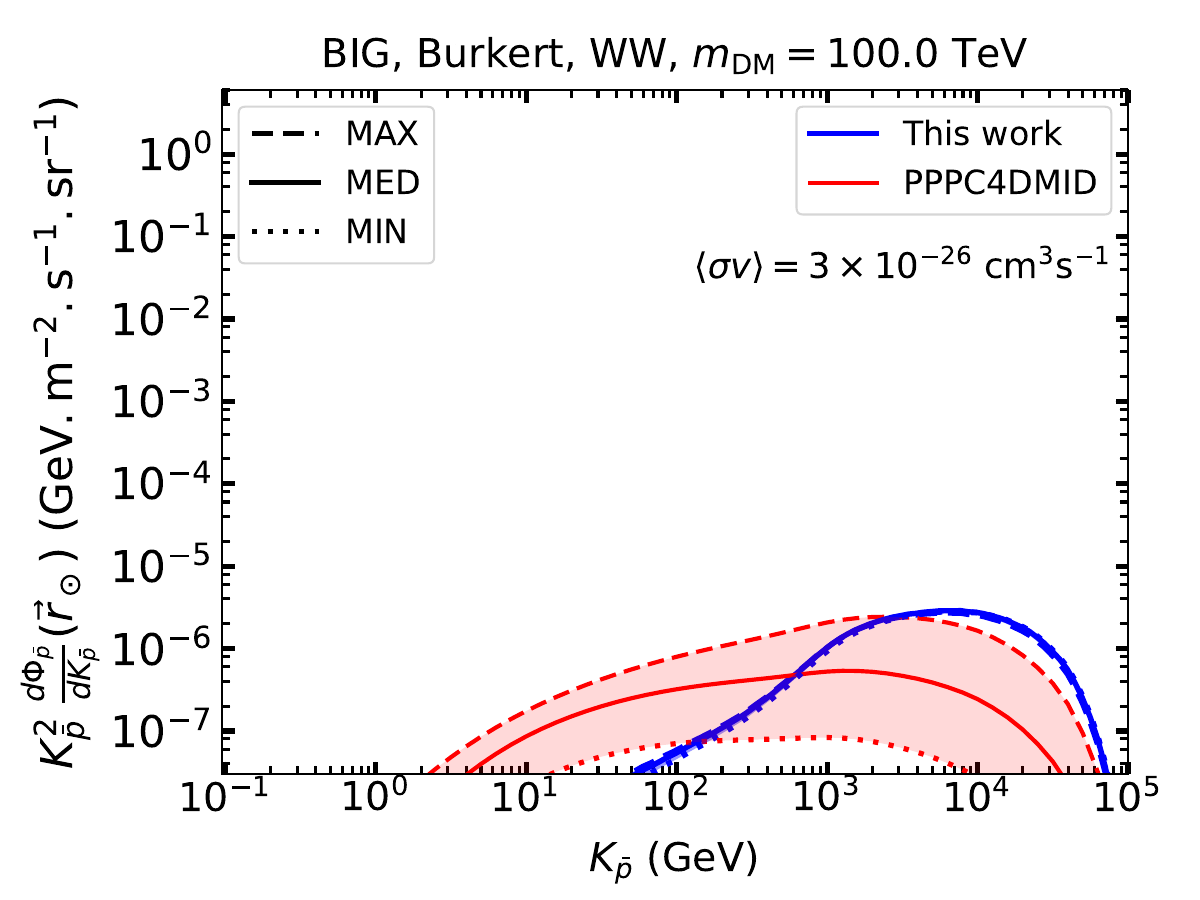}
\end{tabular}

\caption{{\it Antiproton fluxes} (IS) from ${\rm DM \, DM} \rightarrow \bm{W^+W^-}$ annihilations, under the {\sc Big} propagation scheme, compared to the previous \texttt{PPPC4DMID} results.
The rows and columns are as in Fig.~\ref{fig:pbar_SLIM_bb_Ann}.}
\label{fig:pbar_BIG_WW_Ann}
\vspace{2cm}
\end{figure*}

\begin{figure*}[!ht]
\vspace{1cm}
\hspace{-10mm}
\begin{tabular}{ccc}
\includegraphics[width=0.36\textwidth]{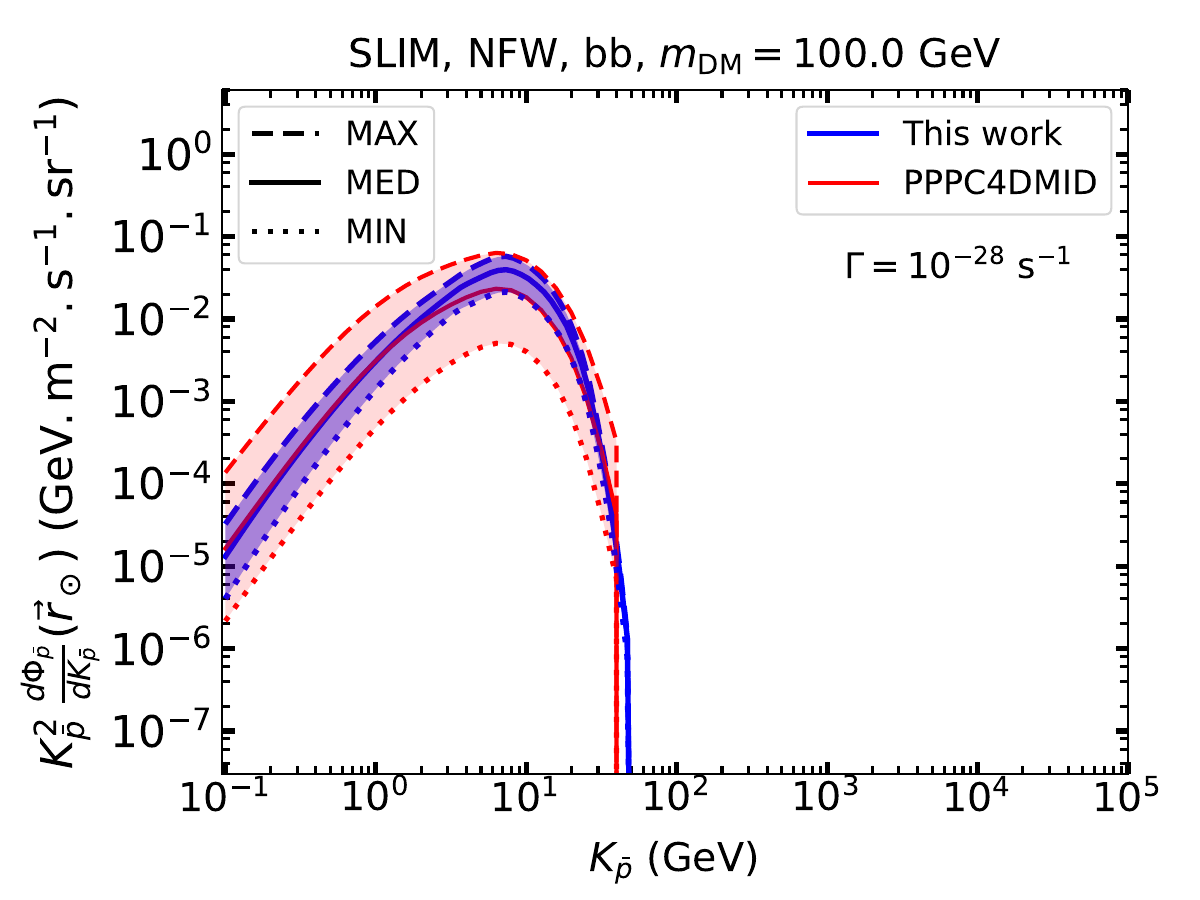} & \hspace{-6mm}
\includegraphics[width=0.36\textwidth]{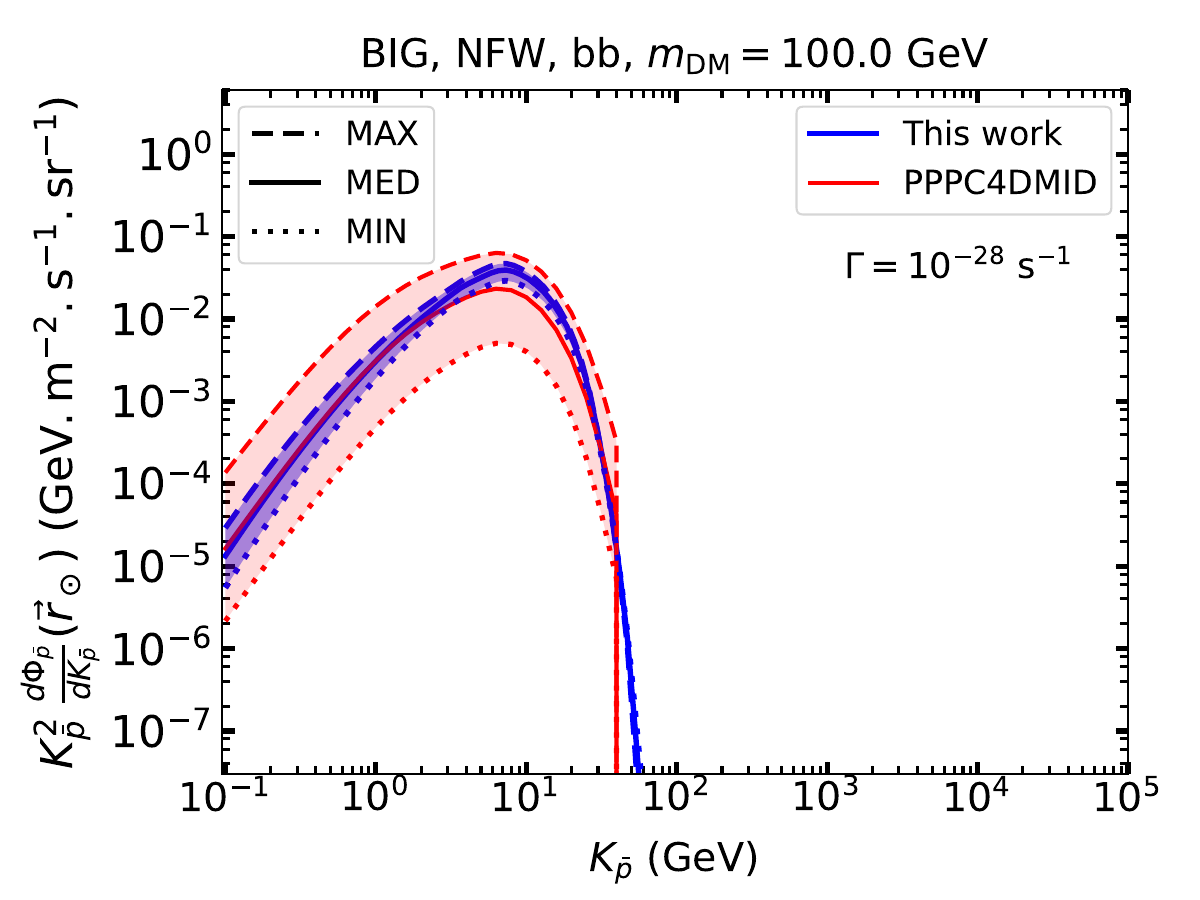} & \hspace{-6mm}
\includegraphics[width=0.36\textwidth]{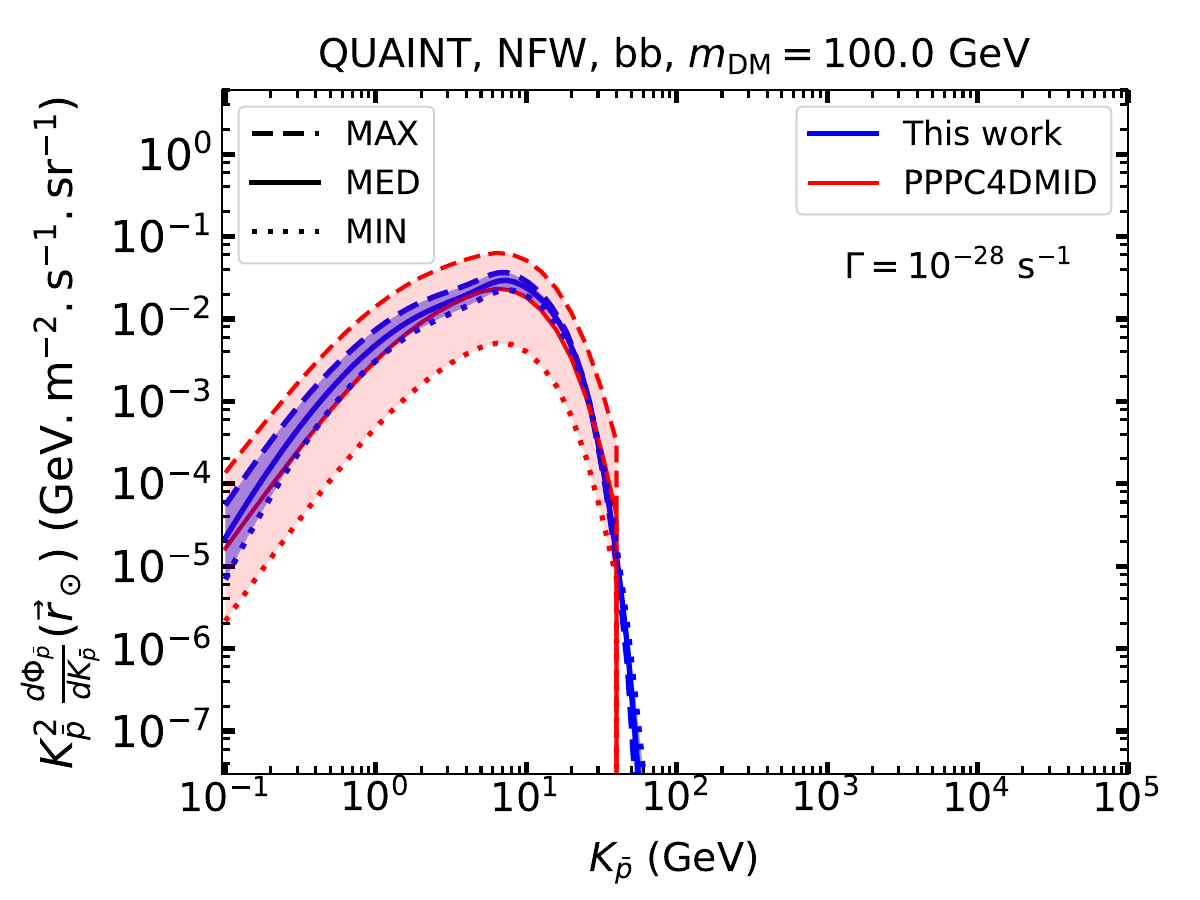}
\end{tabular}

\hspace{-10mm}
\begin{tabular}{ccc}
\includegraphics[width=0.36\textwidth]{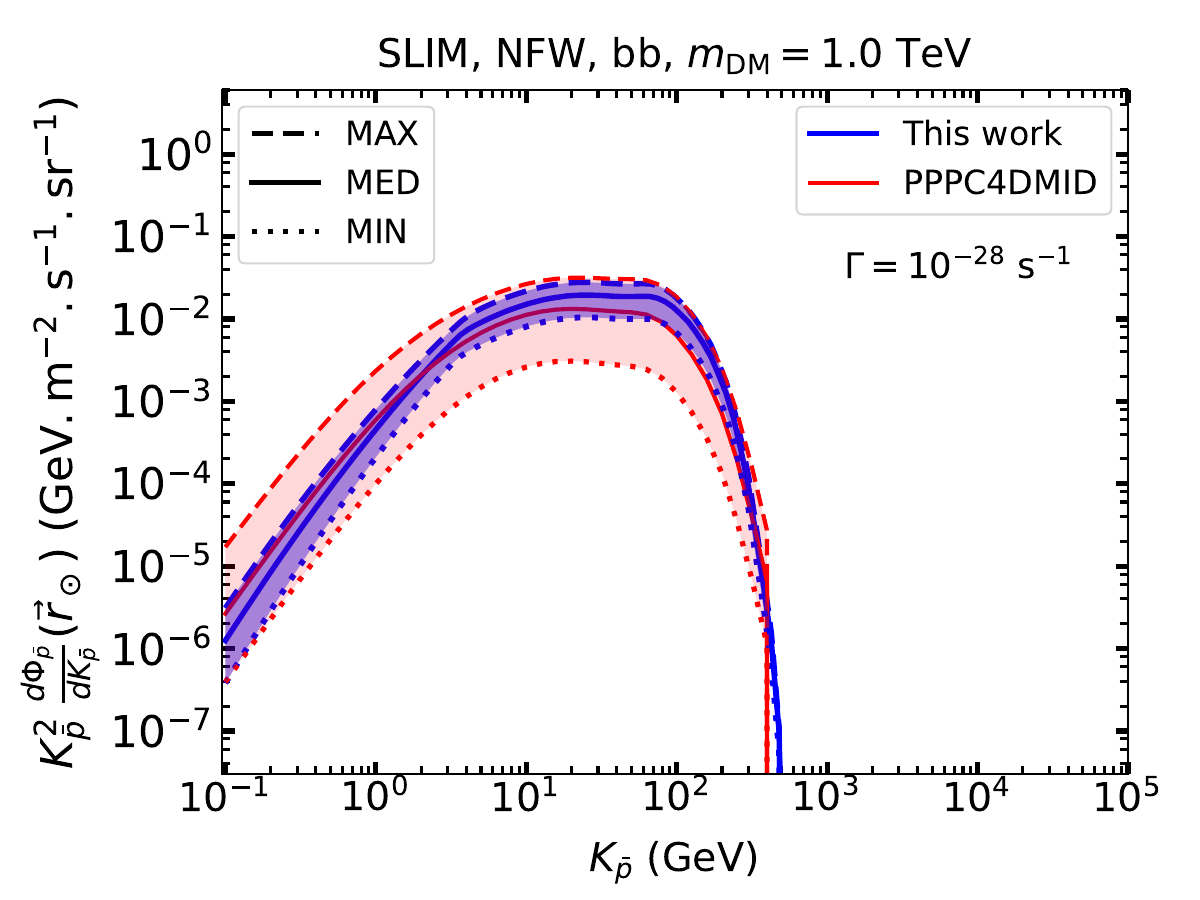} & \hspace{-6mm}
\includegraphics[width=0.36\textwidth]{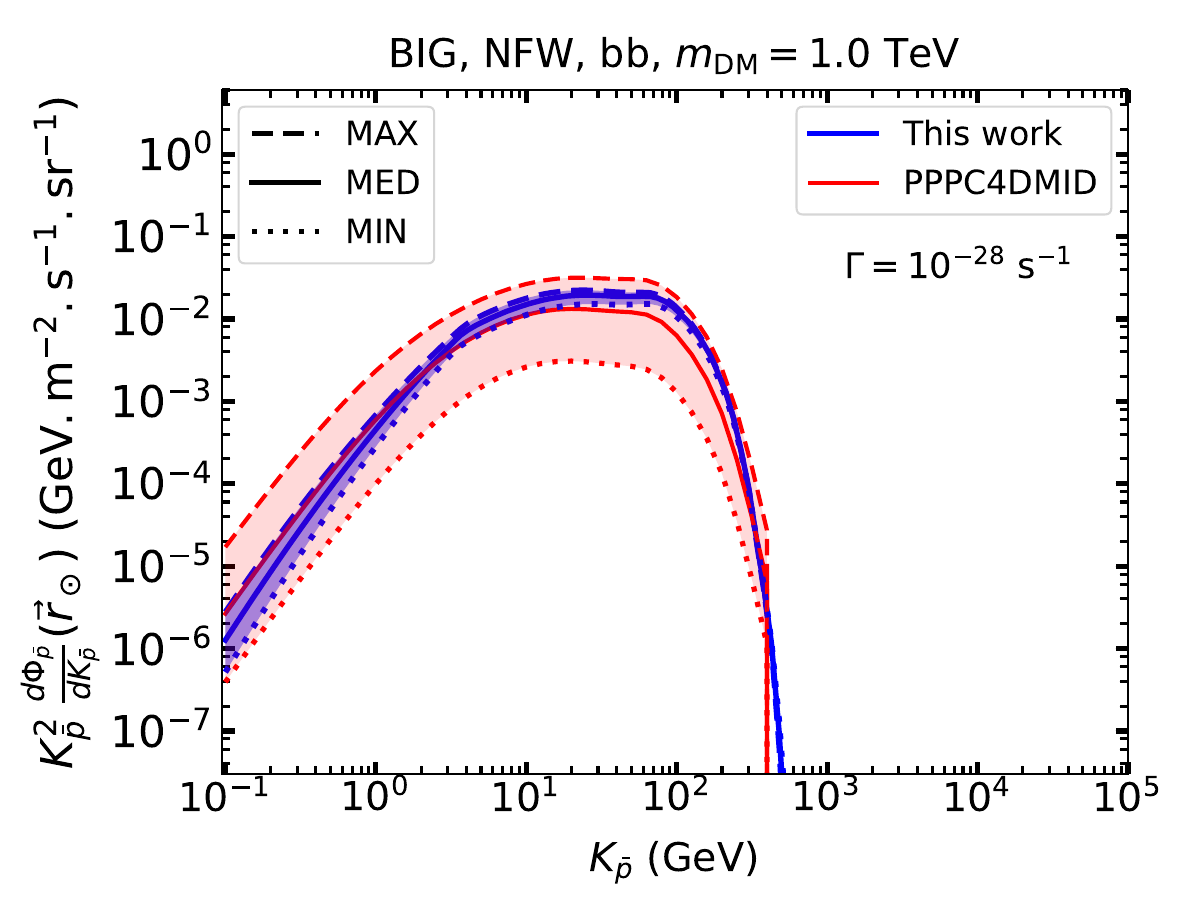} & \hspace{-6mm}
\includegraphics[width=0.36\textwidth]{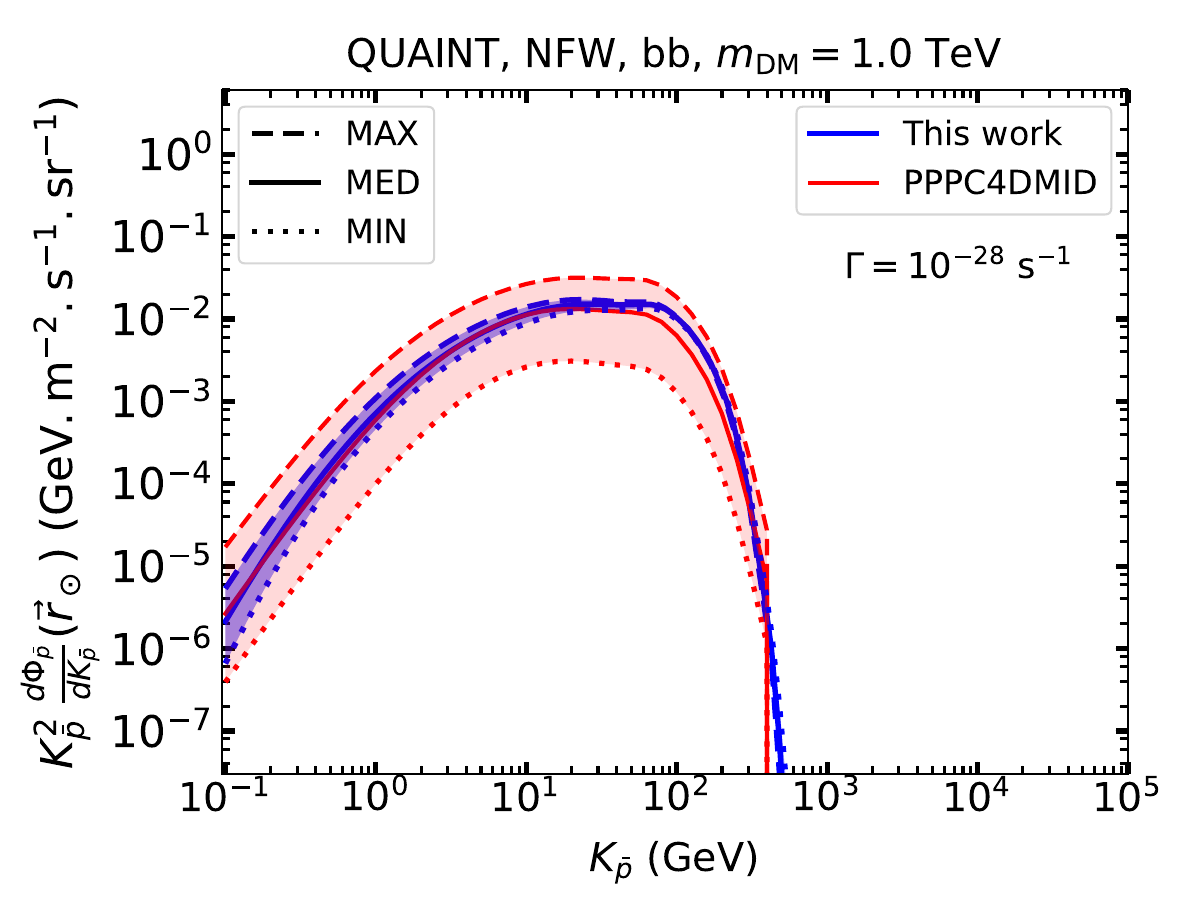}
\end{tabular}

\hspace{-10mm}
\begin{tabular}{ccc}
\includegraphics[width=0.36\textwidth]{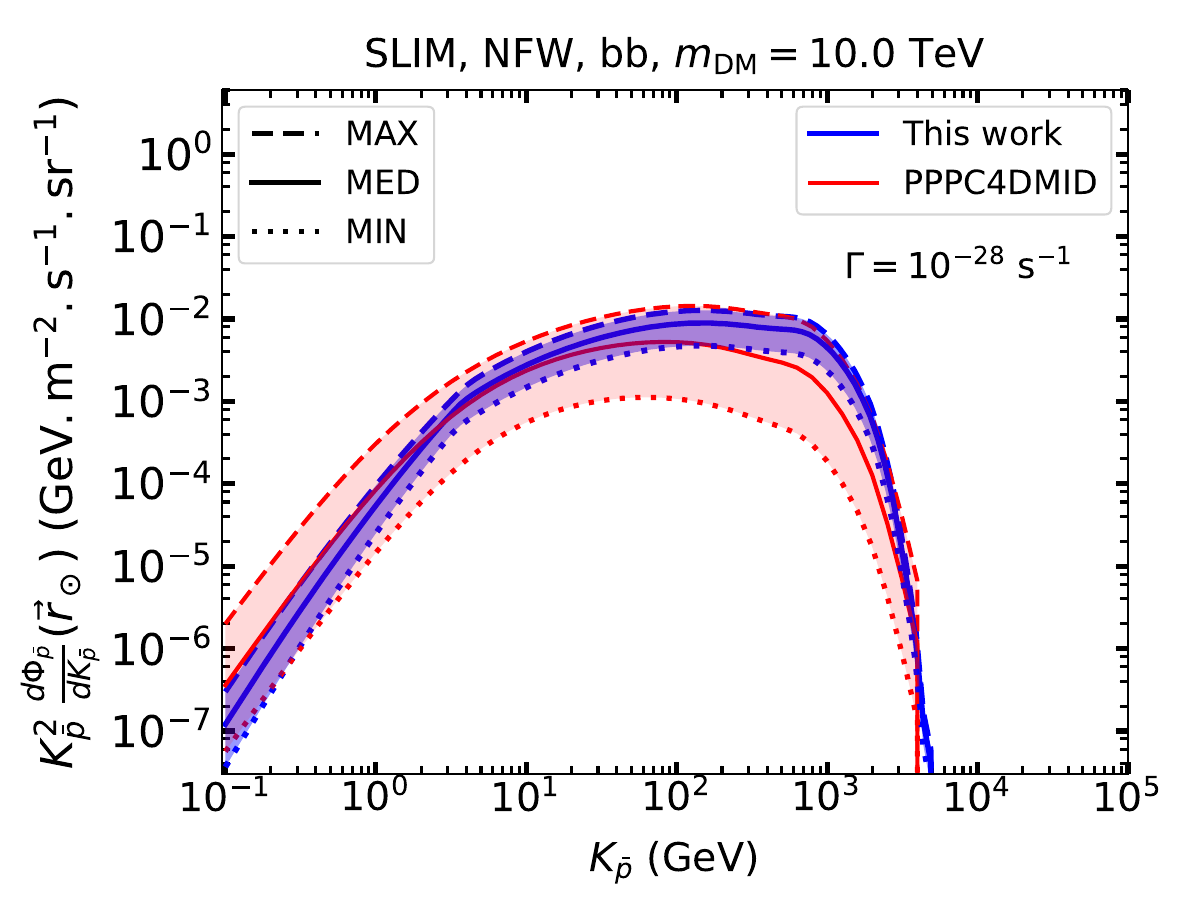} & \hspace{-6mm}
\includegraphics[width=0.36\textwidth]{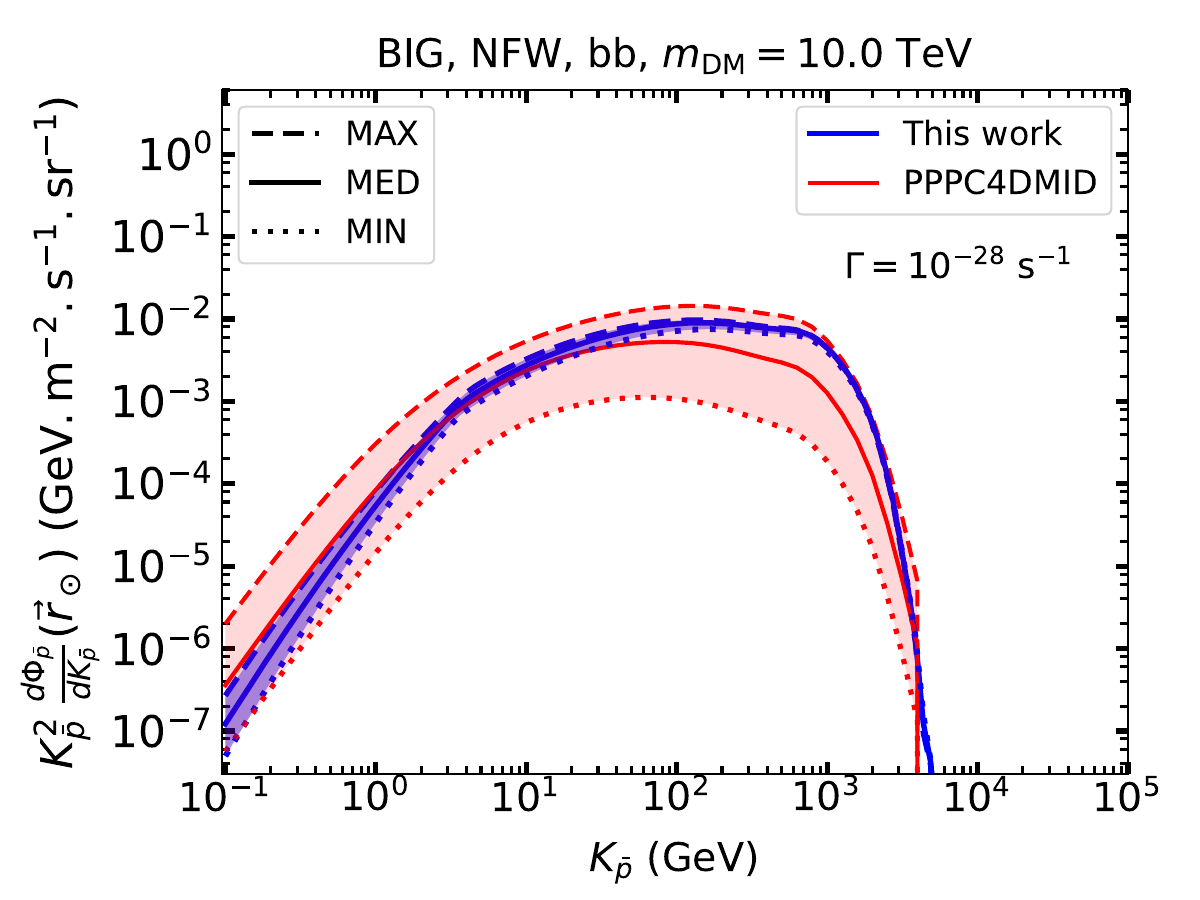} & \hspace{-6mm}
\includegraphics[width=0.36\textwidth]{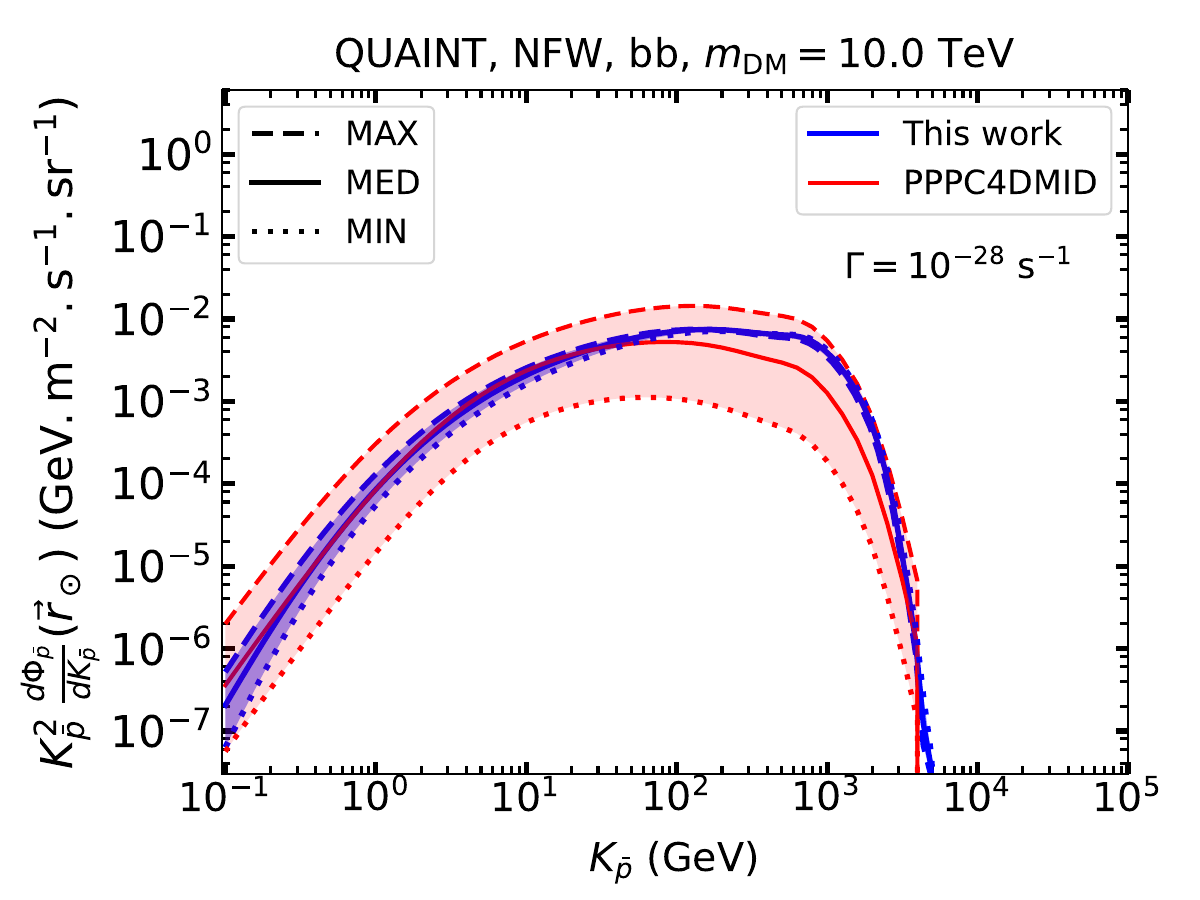}
\end{tabular}

\hspace{-10mm}
\begin{tabular}{ccc}
\includegraphics[width=0.36\textwidth]{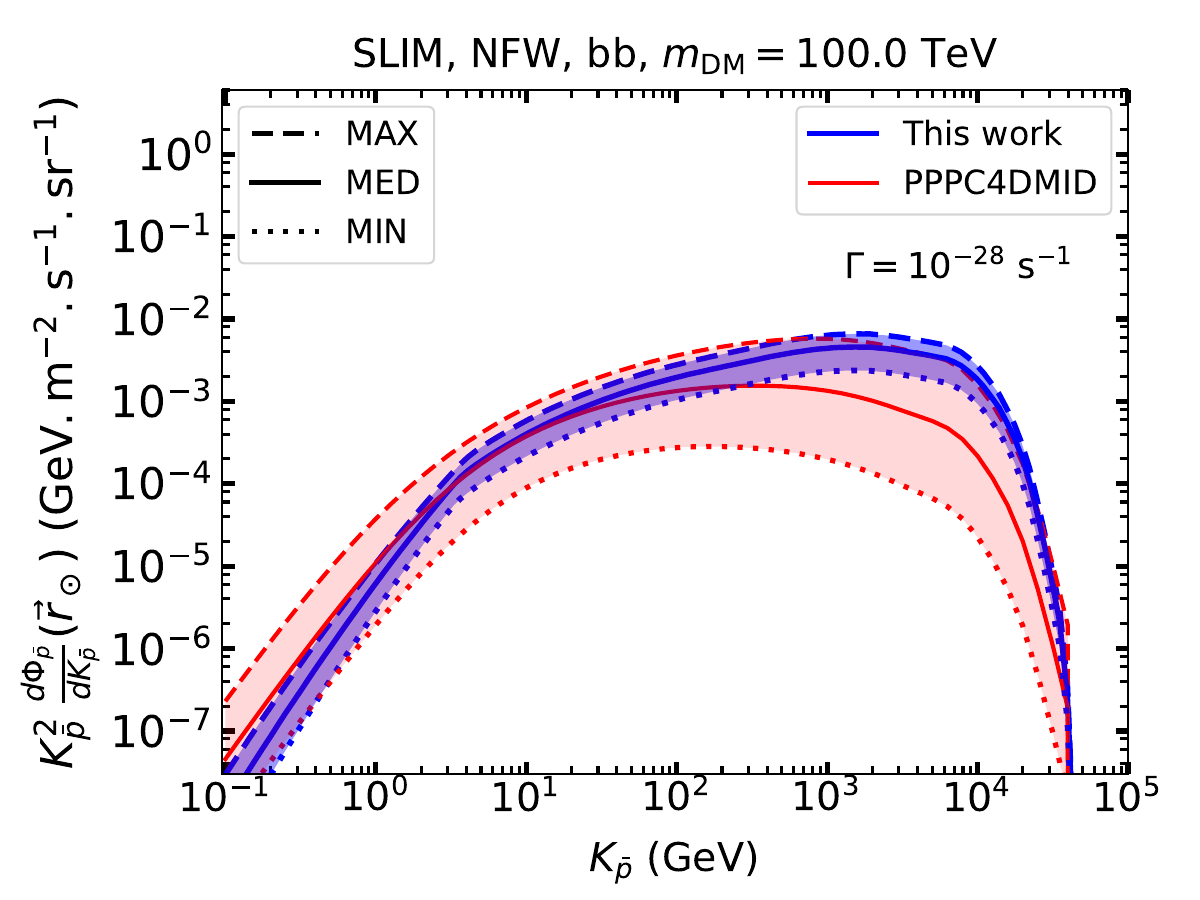} & \hspace{-6mm}
\includegraphics[width=0.36\textwidth]{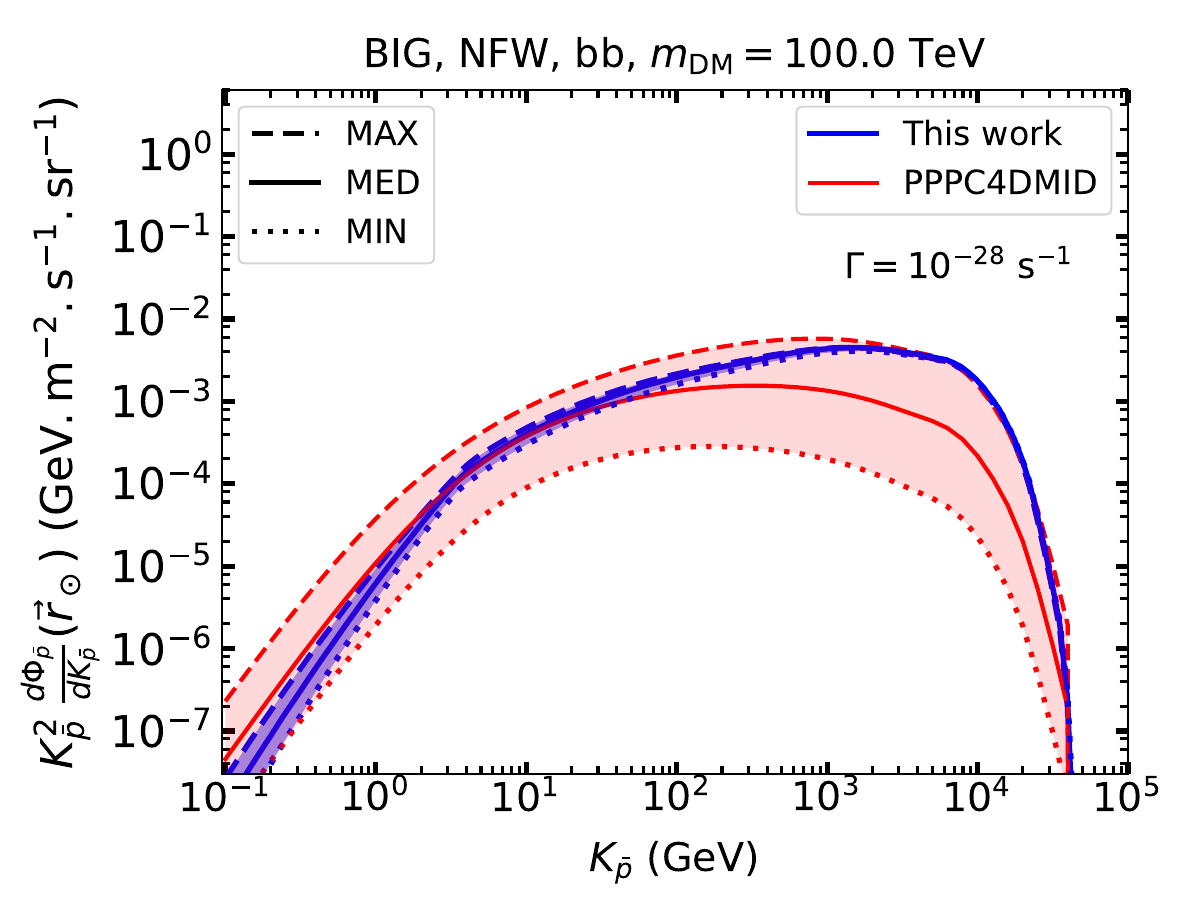} & \hspace{-6mm}
\includegraphics[width=0.36\textwidth]{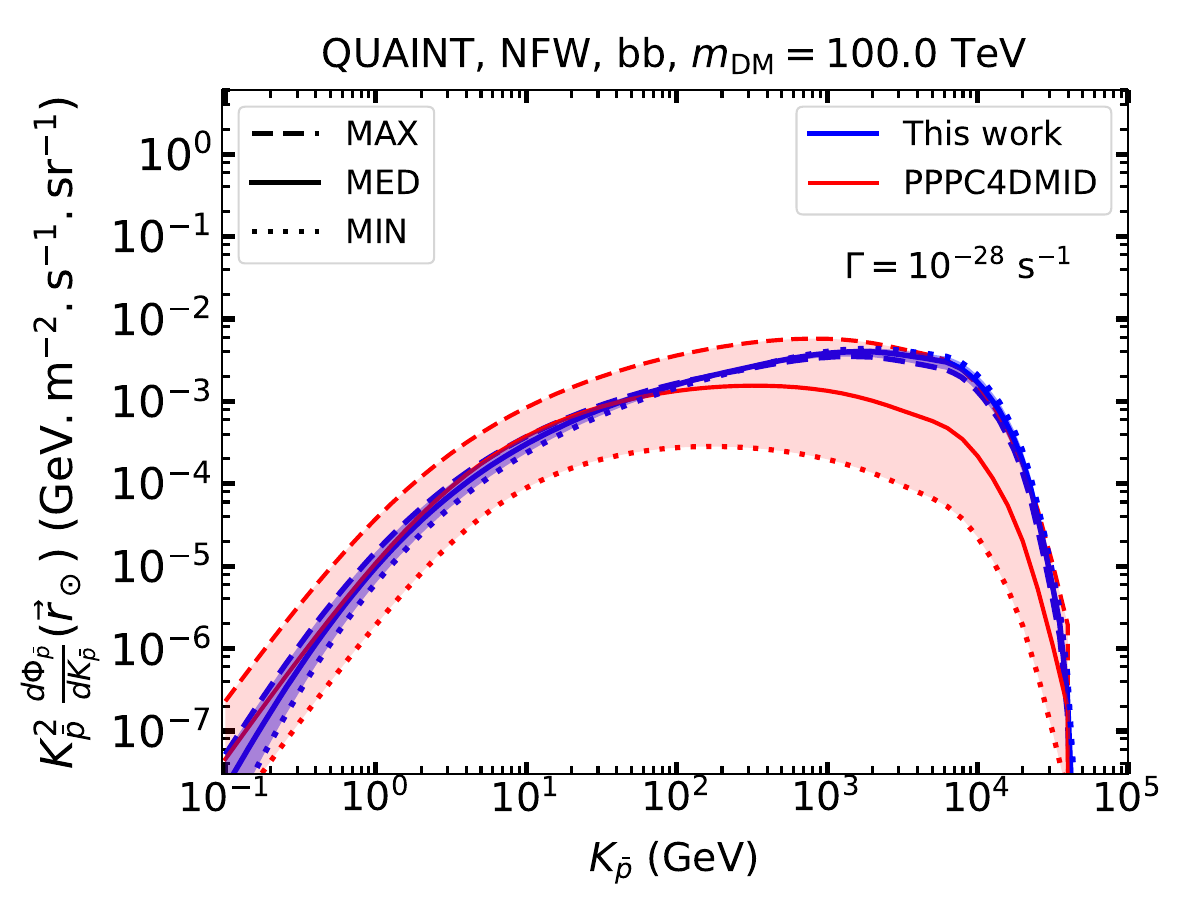}
\end{tabular}

\caption{{\it Antiproton fluxes} (IS) from ${\rm DM} \rightarrow \bm{b\bar{b}}$ {\bf decays}, under the {\sc Slim} (left column), {\sc Big} (middle column)
and {\sc Quaint} (right column) propagation  schemes, compared to the previous \texttt{PPPC4DMID} results.
We fix for definiteness an {\bf NFW} DM profile, and the rows are as in Fig.~\ref{fig:pbar_SLIM_bb_Ann}.
\label{fig:pbar_bb_Dec}}
\vspace{2cm}
\end{figure*}

\begin{figure*}[!ht]
\vspace{1cm}
\hspace{-10mm}
\begin{tabular}{ccc}
\includegraphics[width=0.36\textwidth]{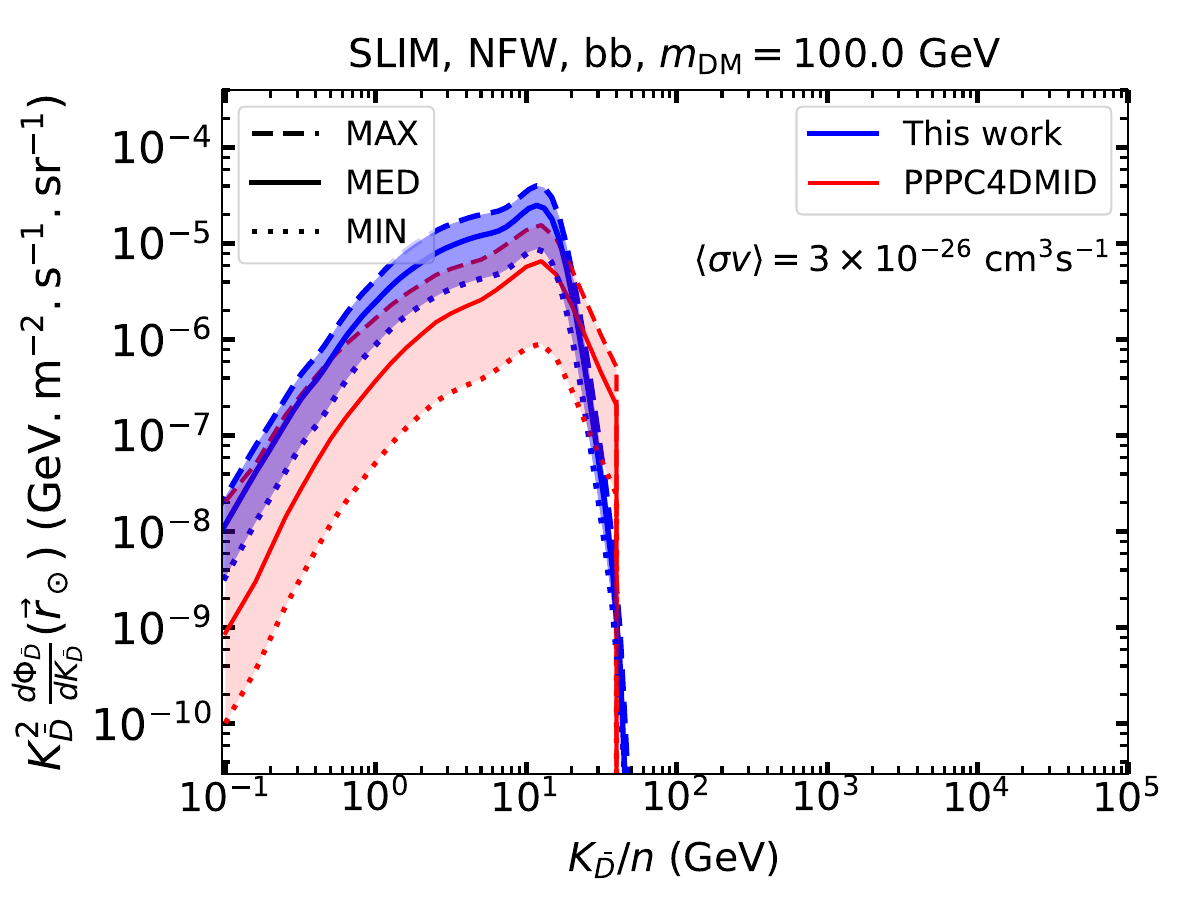} & \hspace{-6mm}
\includegraphics[width=0.36\textwidth]{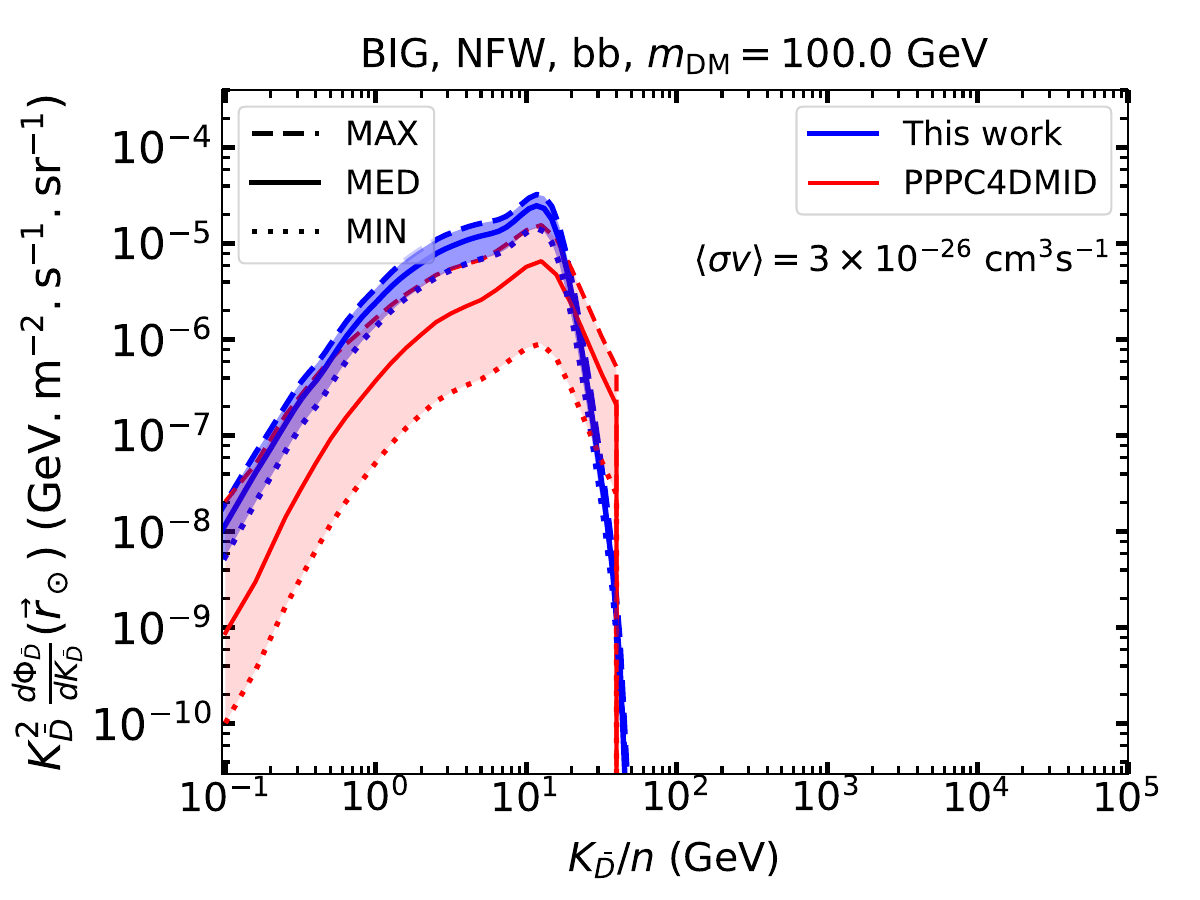} & \hspace{-6mm}
\includegraphics[width=0.36\textwidth]{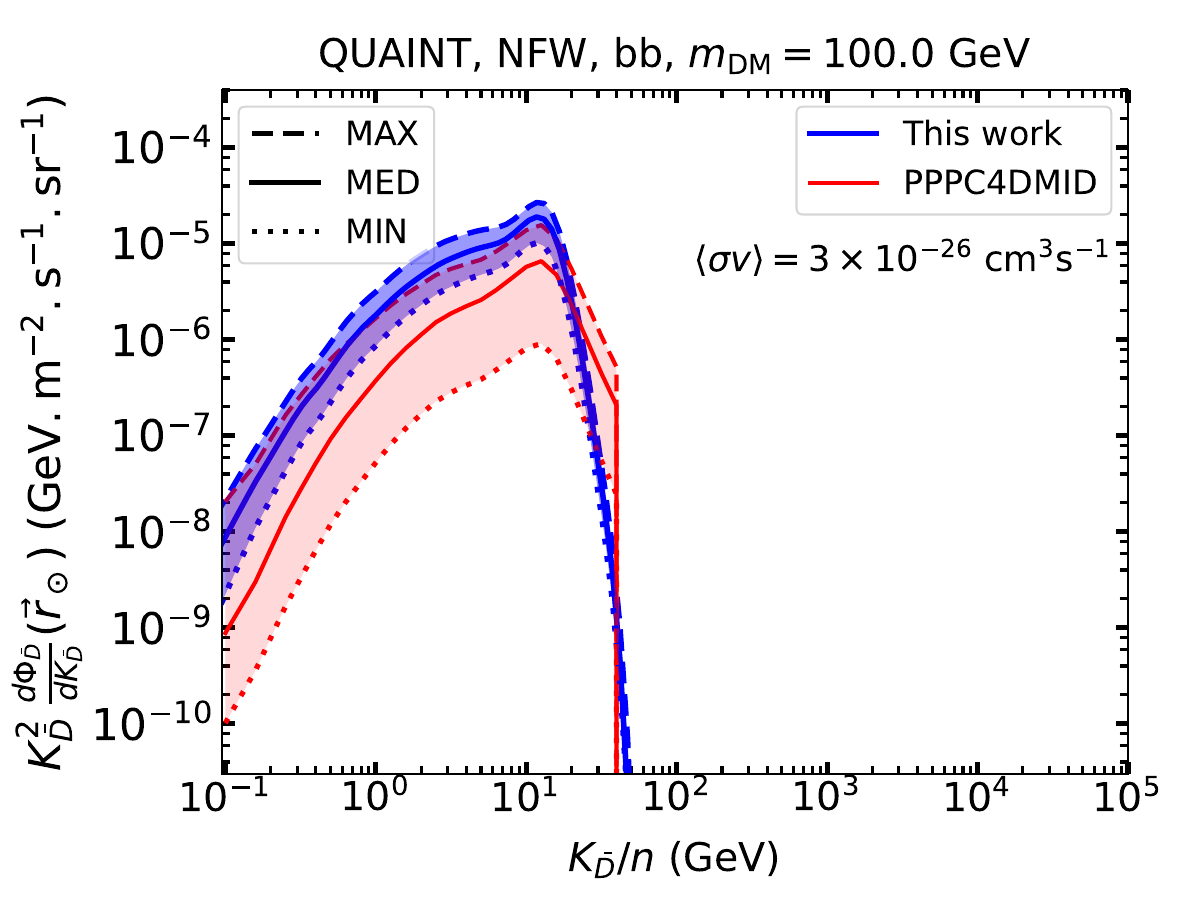}
\end{tabular}

\hspace{-10mm}
\begin{tabular}{ccc}
\includegraphics[width=0.36\textwidth]{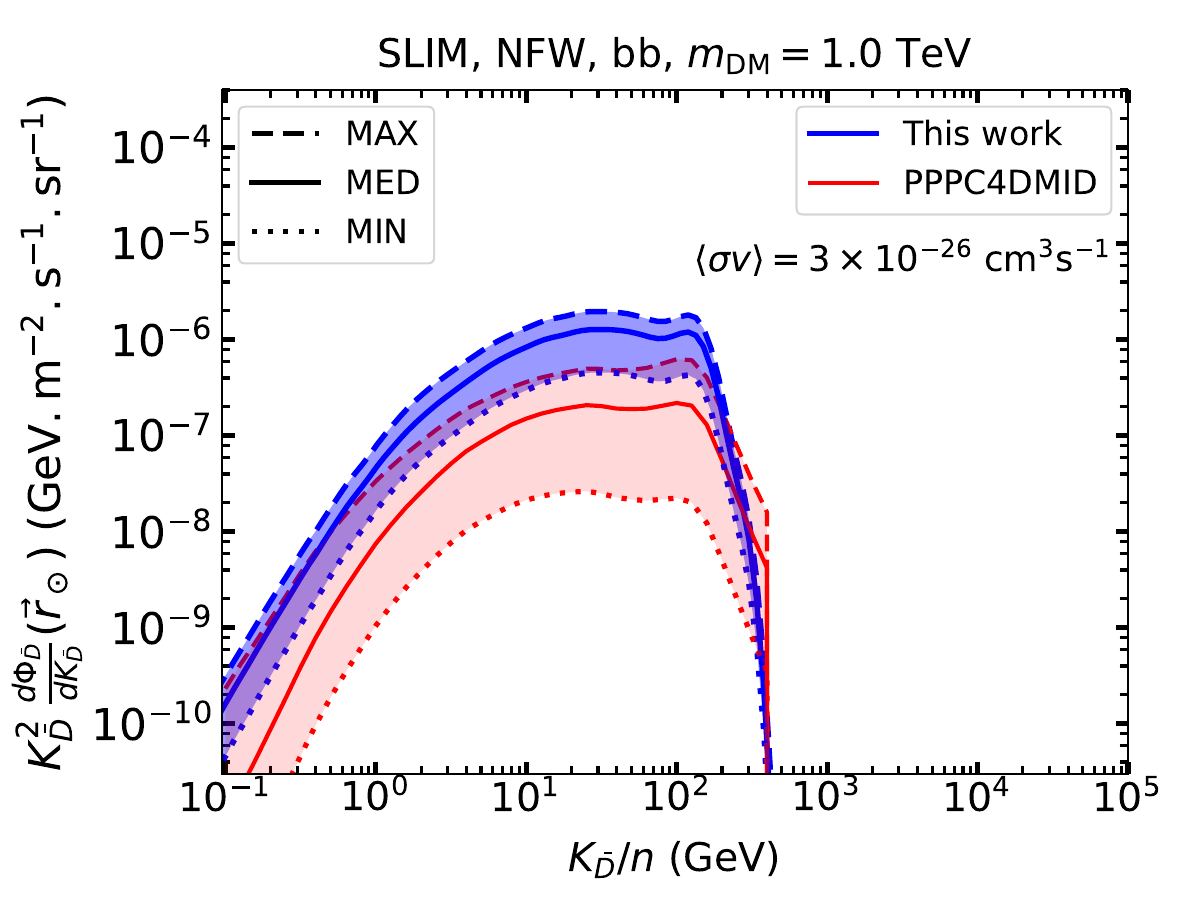} & \hspace{-6mm}
\includegraphics[width=0.36\textwidth]{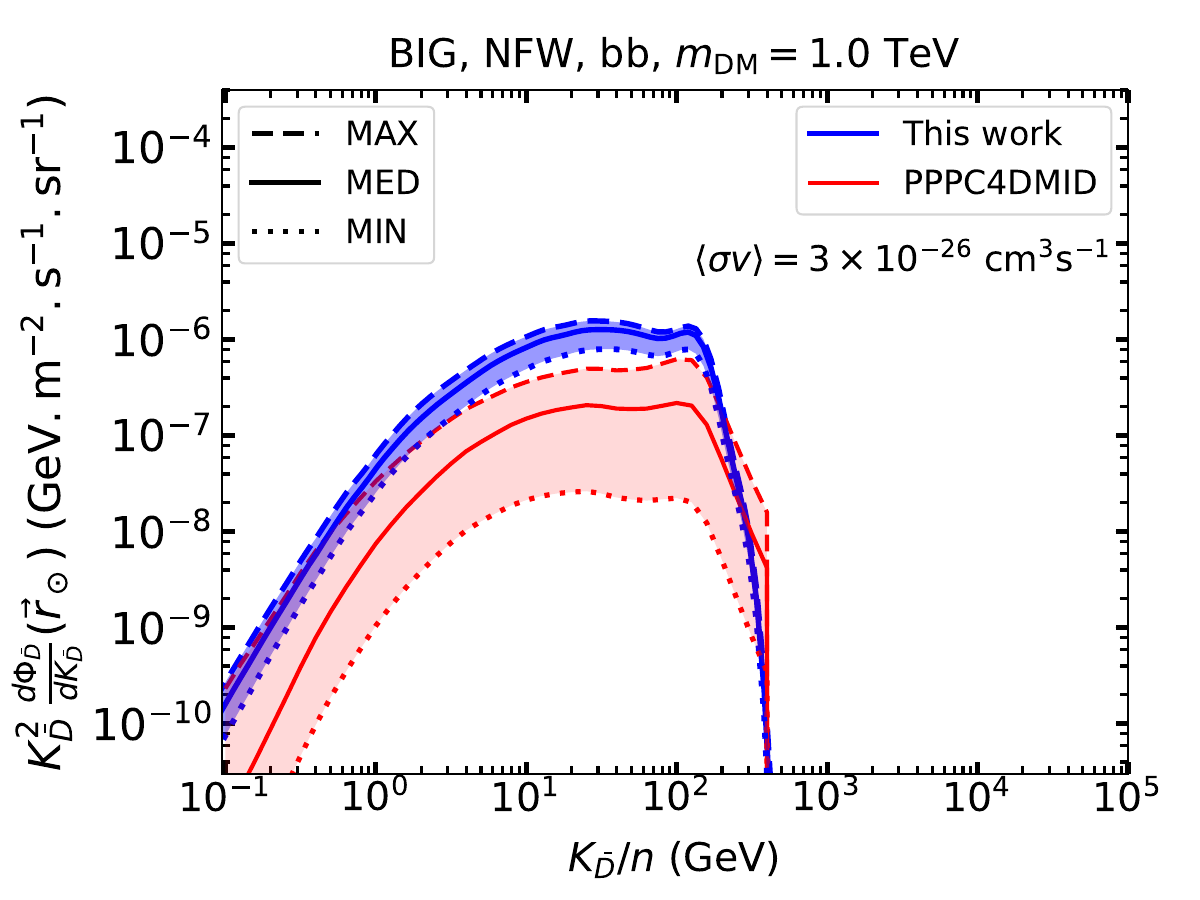} & \hspace{-6mm}
\includegraphics[width=0.36\textwidth]{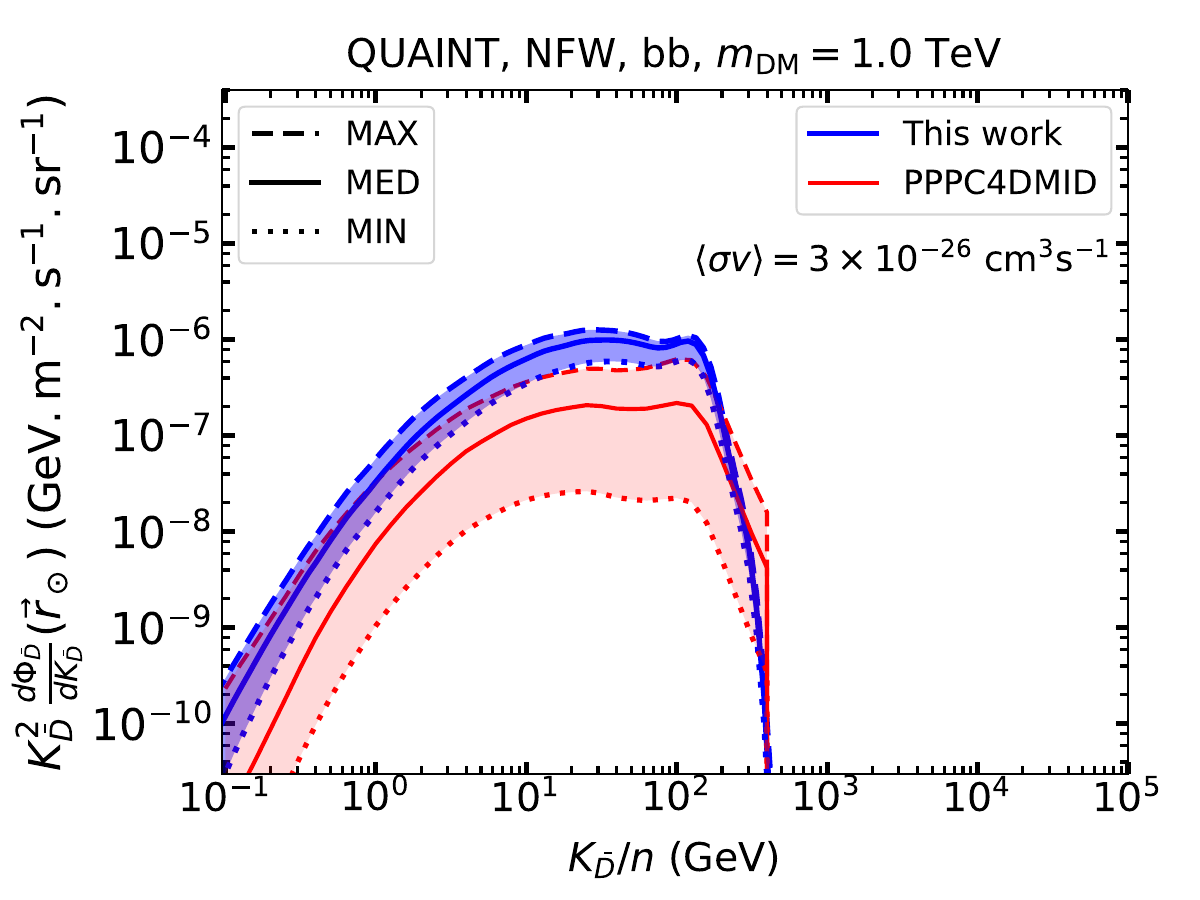}
\end{tabular}

\hspace{-10mm}
\begin{tabular}{ccc}
\includegraphics[width=0.36\textwidth]{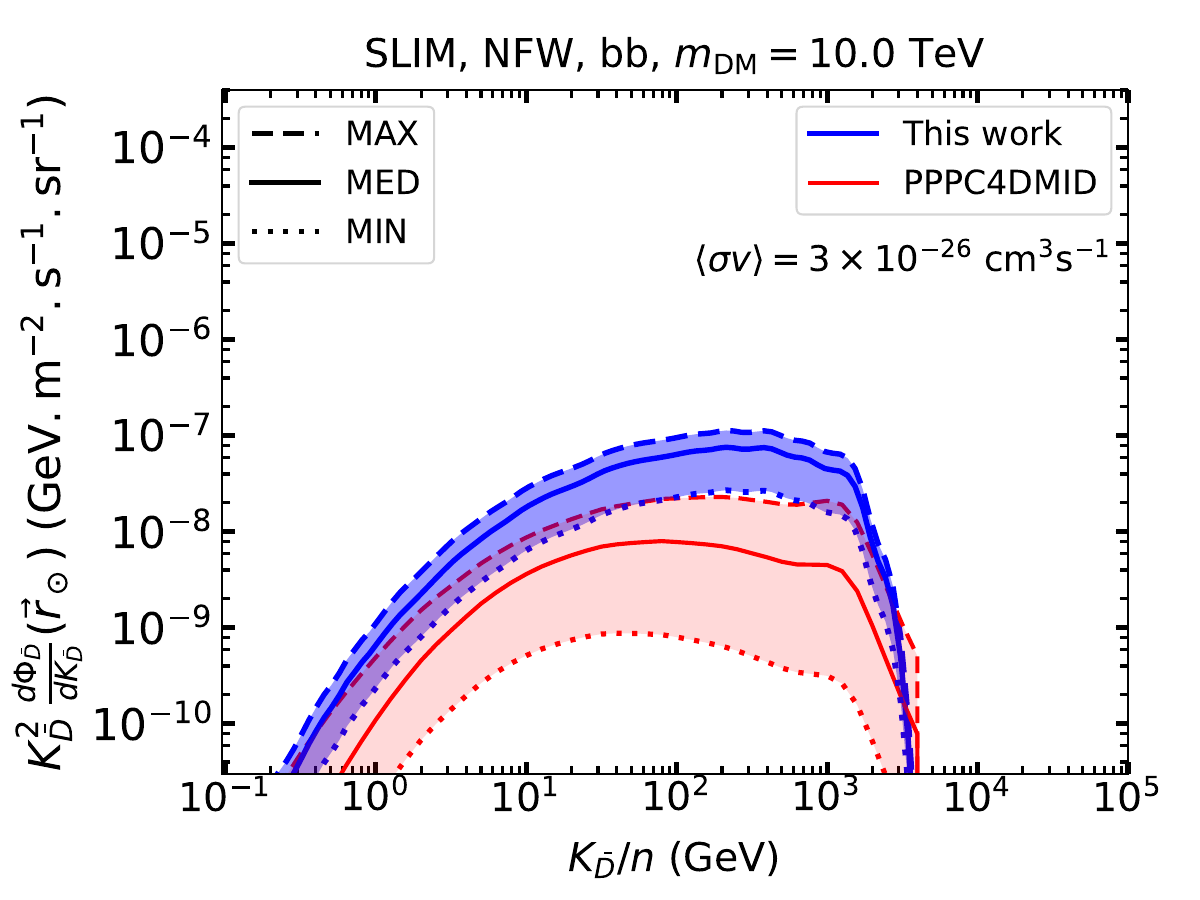} & \hspace{-6mm}
\includegraphics[width=0.36\textwidth]{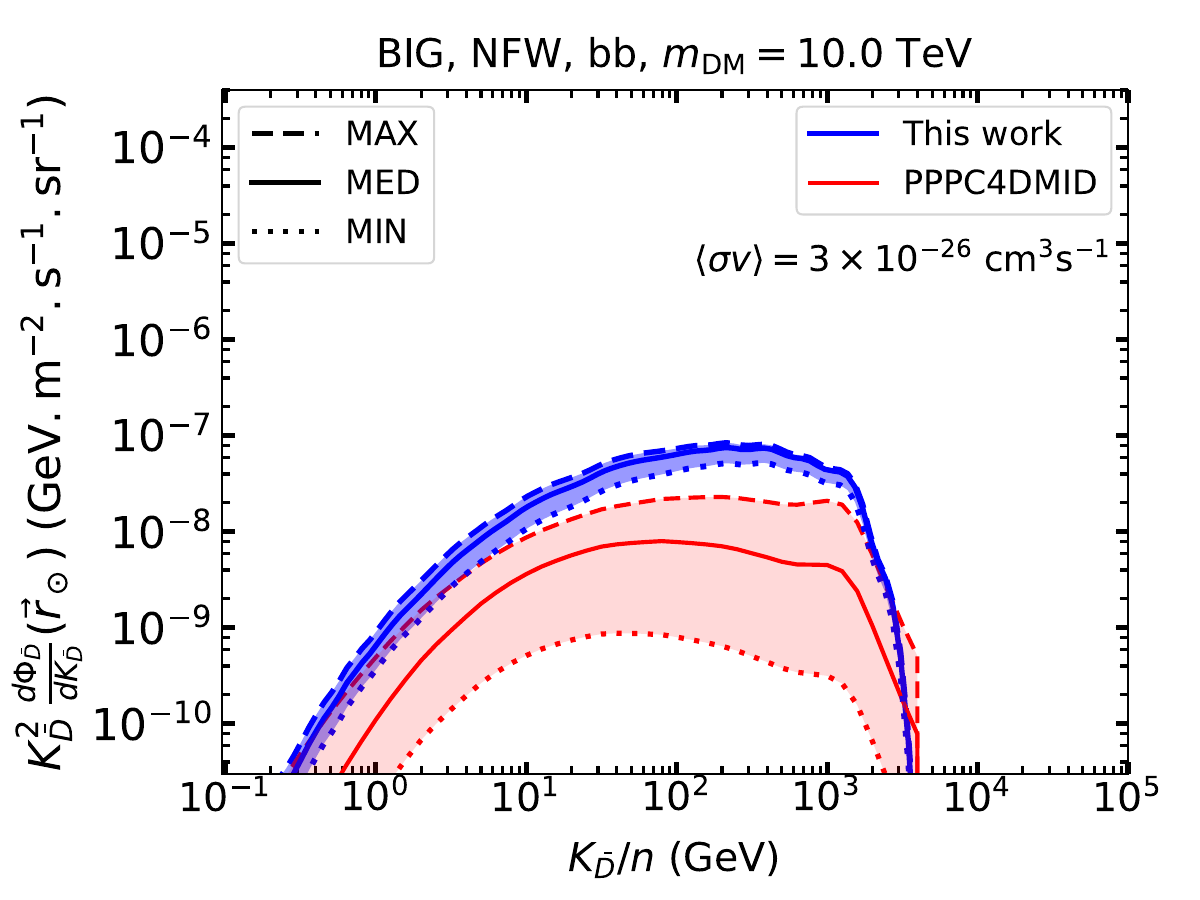} & \hspace{-6mm}
\includegraphics[width=0.36\textwidth]{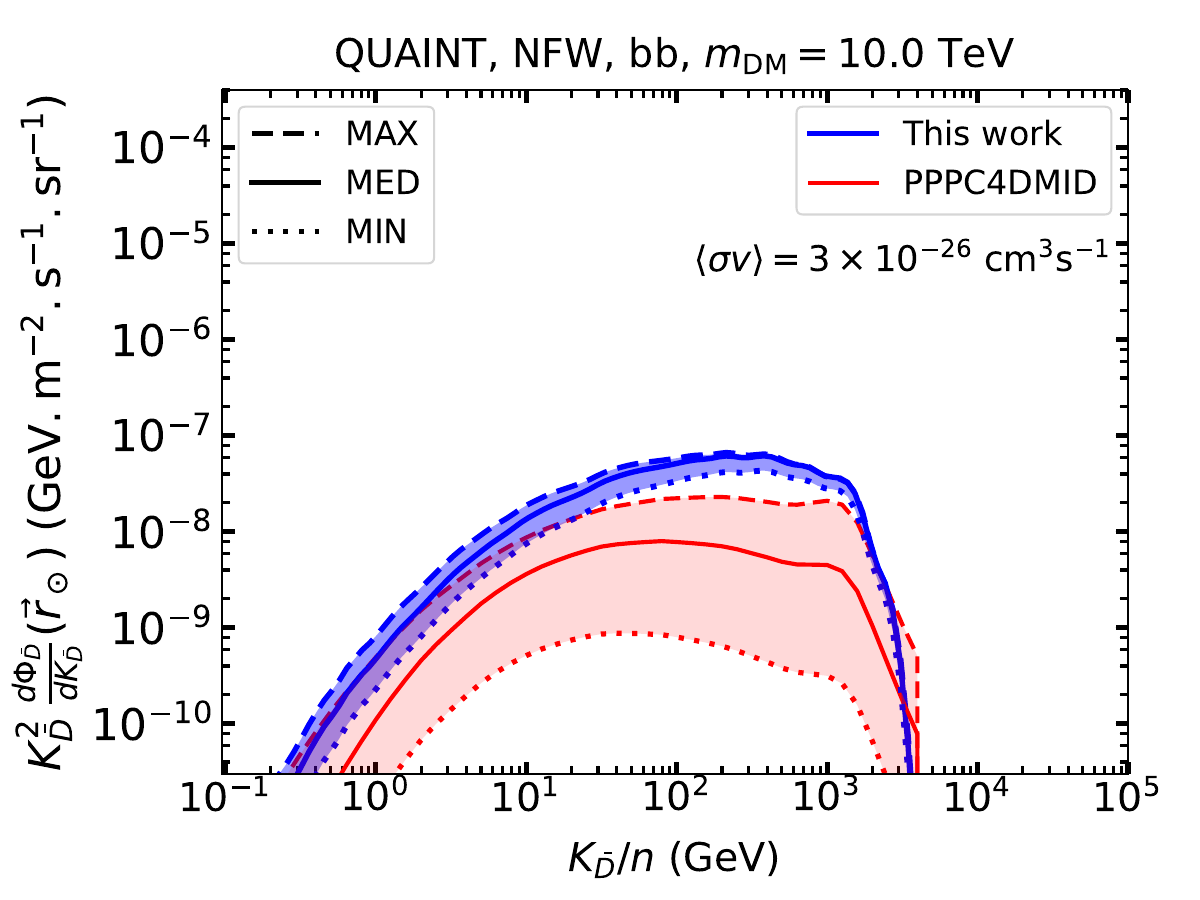}
\end{tabular}

\hspace{-10mm}
\begin{tabular}{ccc}
\includegraphics[width=0.36\textwidth]{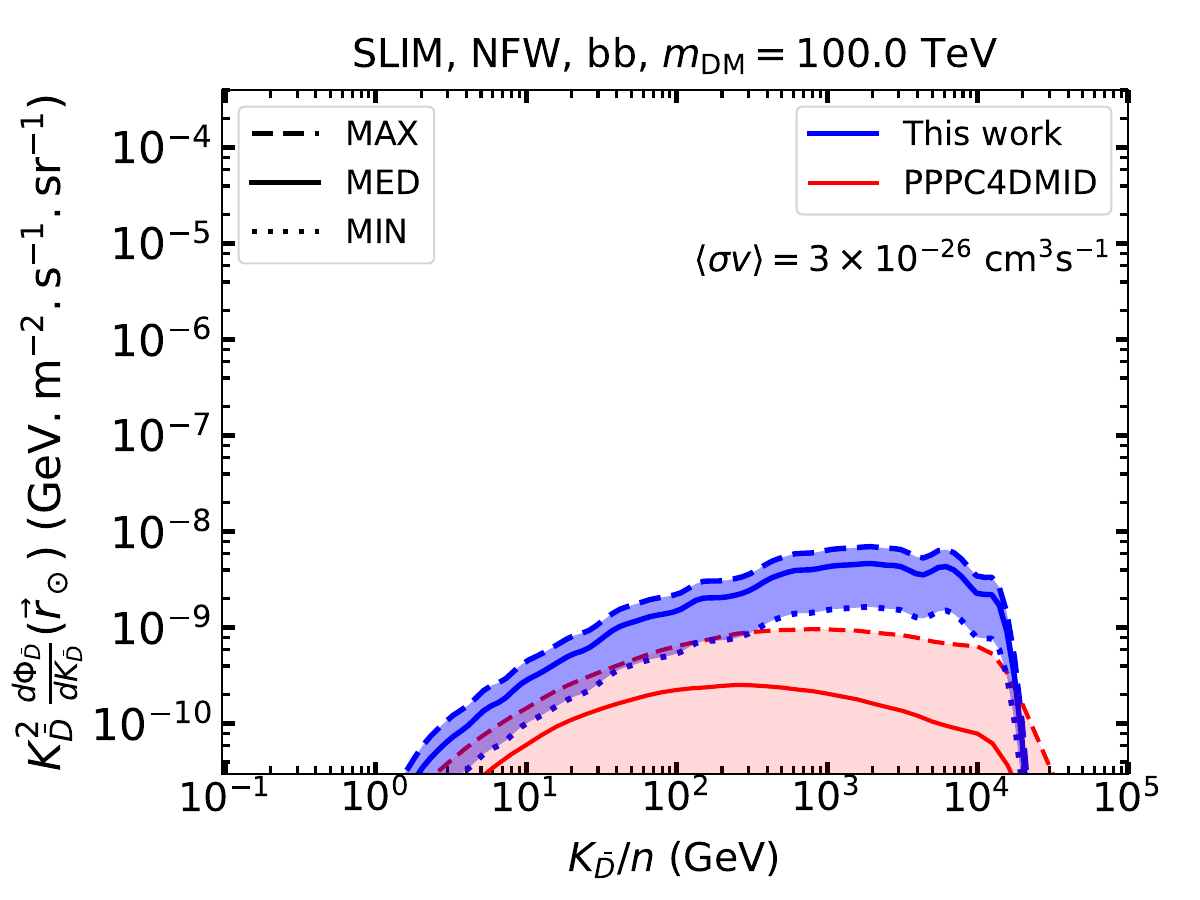} & \hspace{-6mm}
\includegraphics[width=0.36\textwidth]{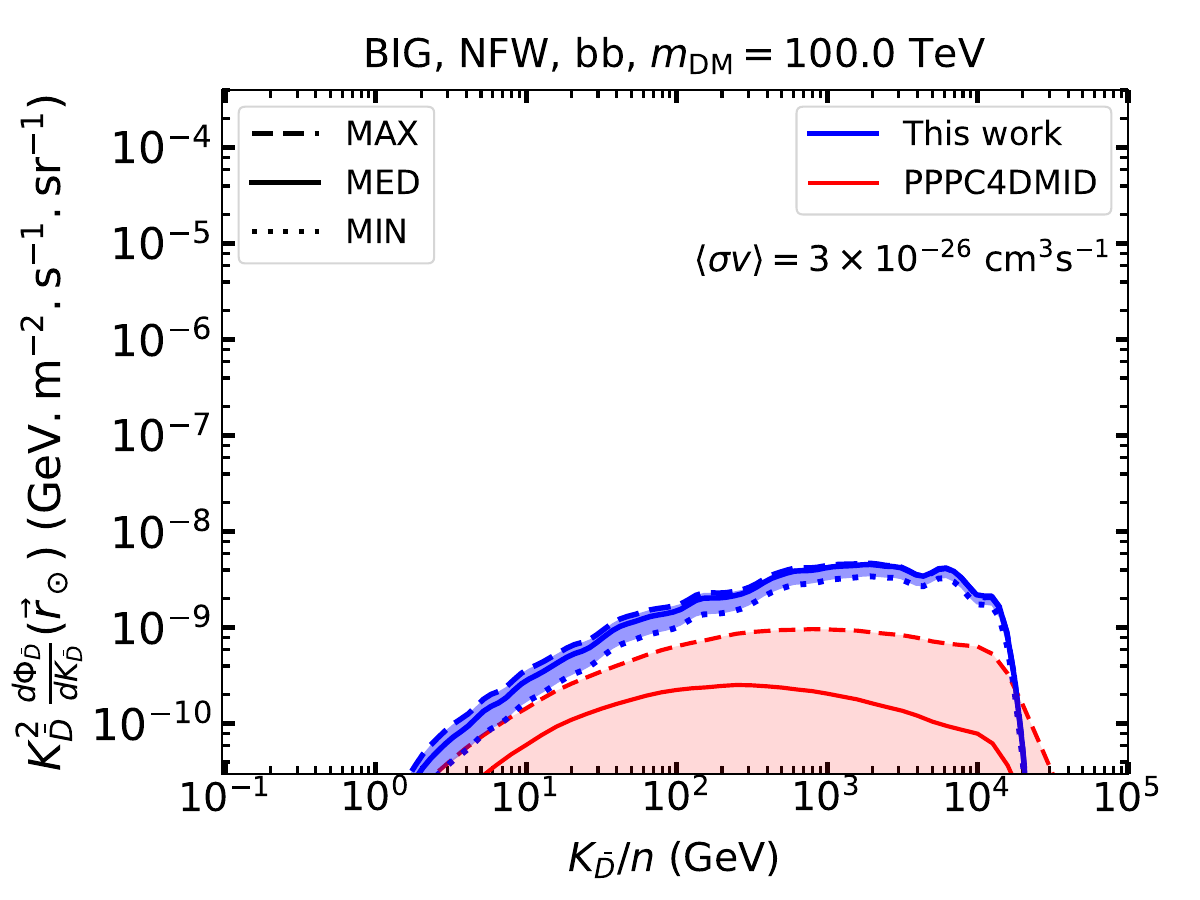} & \hspace{-6mm}
\includegraphics[width=0.36\textwidth]{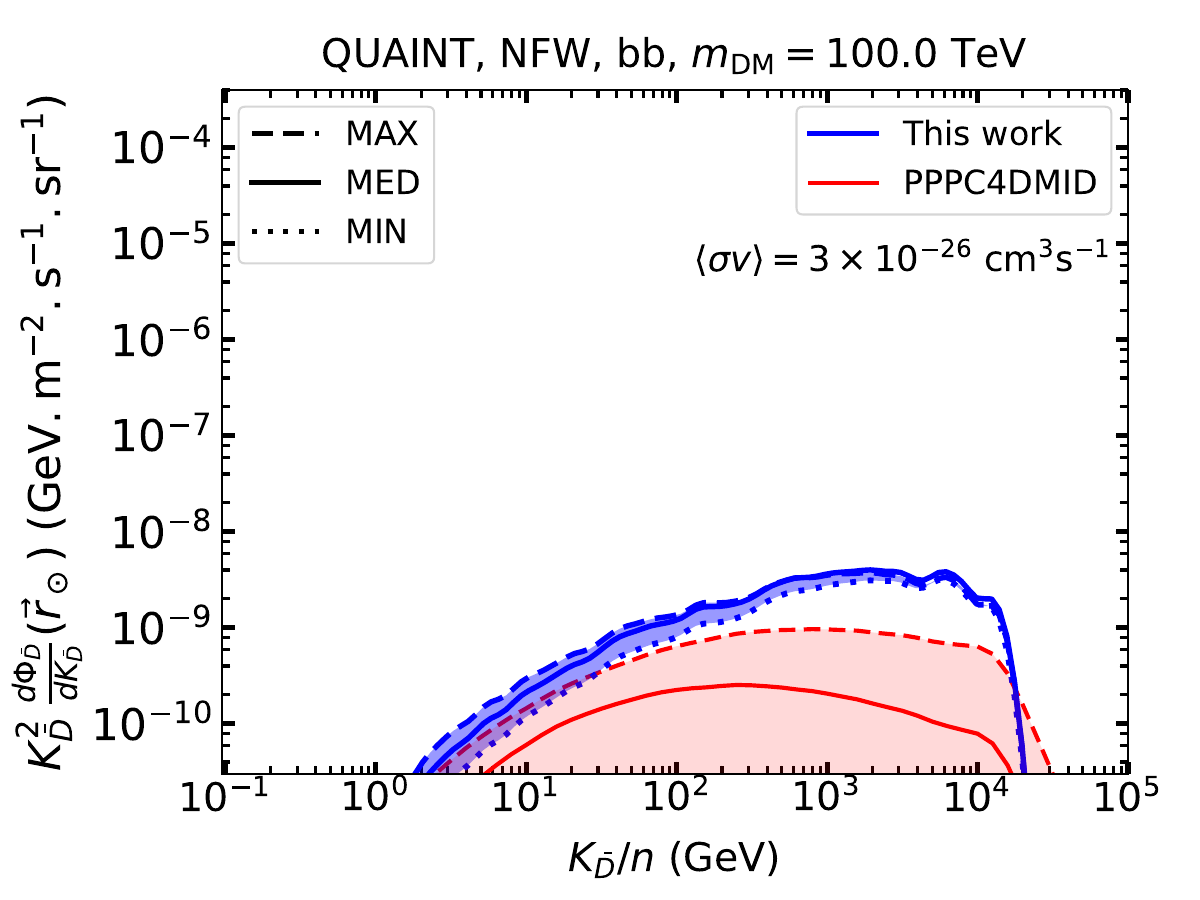}
\end{tabular}

\caption{{\it Anti-deuteron fluxes} (IS) from ${\rm DM \, DM} \rightarrow \bm{b\bar{b}}$, under the {\sc Slim} (left column), {\sc Big} (middle column)
and {\sc Quaint} (right column) propagation  schemes, compared to the previous \texttt{PPPC4DMID} results.
We fix for definiteness an {\bf NFW} DM profile. Like before, the four rows correspond to
DM masses of $10^2$, $10^3$, $10^4$ and $10^5$ GeV, respectively.}
\label{fig:Dbar_bb_Ann}
\vspace{1cm}
\end{figure*}

Figs.~\ref{fig:pbar_SLIM_bb_Ann}, \ref{fig:pbar_BIG_bb_Ann} and \ref{fig:pbar_QUAINT_bb_Ann}
show the comparisons of the $\bar{p}$ flux derived in this work (blue curves) with those
from \texttt{PPPC4DMID} (red curves) for \texttt{SLIM}, \texttt{BIG} and \texttt{QUAINT} models respectively,
considering ${\rm DM \, DM} \rightarrow b\bar{b}$ annihilations.
In each figure, the columns correspond to the NFW, Einasto and Burkert DM profiles respectively,
while the rows show the results for DM masses $10$ GeV, $1$ TeV, $10$ TeV and $100$ TeV, respectively.
In the similar way, Fig.~\ref{fig:pbar_BIG_uu_Ann} and Fig.~\ref{fig:pbar_BIG_WW_Ann} present
the results and comparisons for ${\rm DM \, DM} \rightarrow u\bar{u}$
and ${\rm DM \, DM} \rightarrow W^+W^-$ annihilations, respectively.
Note that for the $W^+W^-$ (and $ZZ$) annihilation/decay channel(s),
we get a significant $\bar{p}$ flux even below the mass threshold of the gauge boson,
i.e., for $m_{\rm DM} < m_W (m_Z)$ (annihilation) and $2\,m_{\rm DM} < m_W (m_Z)$ (decay).
This is due to the off-shell contributions of $W$ and $Z$ produced in the intermediate state of
DM annihilation/decay \cite{Arina:2023eic}. Here the branching fraction of
each of these off-shell states is considered to be 1,
although, in a generalized DM model, one expect this fraction to be suppressed.
Fig.~\ref{fig:pbar_bb_Dec} focuses on the {\em decaying} DM case, showing the $\bar{p}$ flux from ${\rm DM} \rightarrow b\bar{b}$, under a few propagation scenarios and for DM of different masses.
Finally, Fig.~\ref{fig:Dbar_bb_Ann} shows the $\bar{d}$ flux.

\bigskip

There are few important differences between the fluxes presented in this work and the previous ones. We discuss them below.

\begin{figure*}[!t]
\begin{minipage}{0.64\textwidth}
\hspace{-1cm}
\includegraphics[width=\textwidth]{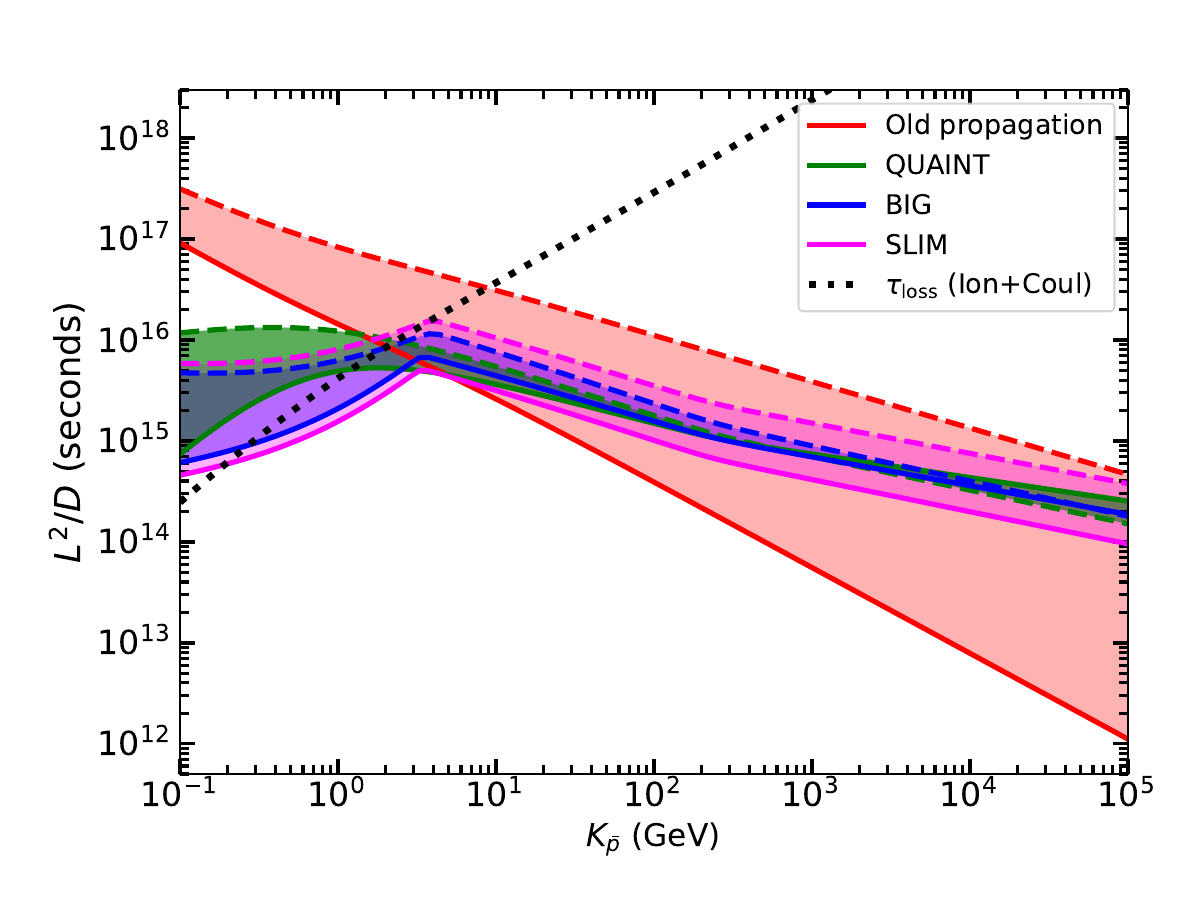}
\end{minipage}
\begin{minipage}{0.32\textwidth}
\caption{The ratio $L^2 / D(K)$ (the typical time-scale for the diffusion)
as a function of the $\bar p$ kinetic energy, for different propagation models.
In each case the band corresponds to a MIN (solid line) to MAX (dashed line) variation.
For comparison, the black dotted line shows the time-scale
for the energy loss (via ionization and Coulomb).}
\label{fig:Lsq_D}
\end{minipage}
\end{figure*}

The main result is that the usage of the new {\sc Max/Med/Min} propagation sets
reduces significantly the uncertainty in the flux at Earth.
Considering the NFW DM profile and the (IS) $\bar{p}$ flux produced from the annihilation
of DM particles of mass 1 TeV into $b\bar{b}$ channel,
the reduction factor $\mathcal{R}_{\bar{p}}$,
defined as:
\begin{equation}
\mathcal{R}_{\bar{p}} =
\frac{\left( {\phi^{\textsc{Max}}_{\bar{p}}}/{\phi^{\textsc{Min}}_{\bar{p}}} \right)_{\rm New}}
{\left( {\phi^{\textsc{Max}}_{\bar{p}}}/{\phi^{\textsc{Min}}_{\bar{p}}} \right)_{\rm Old}} \, ,
\end{equation}
where `New' and `Old' refers to the new and old propagation models,
varies between a factor of 0.24 and 0.08 (for {\sc Slim}), 0.12 and 0.02 (for {\sc Big}) and
0.15 and 0.013 (for {\sc Quaint}) in the kinetic energy range 0.1 GeV -- 700 GeV
(that covers the range of CR experiments like {\sc Gaps} and {\sc Ams-02}).
See, e.g, Figs. \ref{fig:pbar_SLIM_bb_Ann}, \ref{fig:pbar_BIG_bb_Ann} and \ref{fig:pbar_QUAINT_bb_Ann}.
Such a reduction can be understood as follows.
The CR $\bar{p}$ flux roughly scales as $L^2/D$~\cite{Genolini2021MinMedMax},
which for the new propagation models is constrained within a band
much narrower (especially at higher energies)
than the one found in the old propagation model;
see Fig.~\ref{fig:Lsq_D}.
This reduction increases the discovery potential of a possible DM signal, as we will explain in the next section.

Moreover, we find that among the tested propagation models
the {\sc Big} and {\sc Quaint} ones are those for which the flux uncertainty is the smallest.

Another visible difference between the CR flux provided here and those supplied
by \texttt{PPPC4DMID} is an overall shift of the new fluxes in the upward
direction, more prominent for the DM annihilation scenario compared to DM decay.
It is very visible when focusing on the MED line in any of the figures \ref{fig:pbar_SLIM_bb_Ann} to \ref{fig:Dbar_bb_Ann}.
One of the main reasons for this is the use of updated galactic DM density profiles with $\rho_\odot = 0.4$ ${\rm GeV}/{\rm cm}^{3}$
(as mentioned in Sec.~\ref{sec:Models}), while the Old fluxes were produced
using the DM profiles from \cite{Cirelli:2010xx} with $\rho_\odot = 0.3$ ${\rm GeV}/{\rm cm}^{3}$.
For DM annihilation, since the flux is governed by the square of the DM distribution,
this leads to a noticeable enhancement in the flux.
For the decaying DM, since the flux is governed by the DM distribution
(instead of its square), the overall shift in the flux is comparatively less prominent
(see Fig.~\ref{fig:pbar_bb_Dec}).

Another notable point is the suppression of the CR flux in the New estimates (compared to Old ones) for channels like $W^+W^-$ and $ZZ$,
at low energies and for large DM masses,
see Fig.~\ref{fig:pbar_BIG_WW_Ann}.
This is inherited from the spectra $dN/dK$ for $\bar{p}$ (and $\bar{d}$) by \texttt{CosmiXs}, which contain this type of feature (see \cite{Arina:2023eic, DiMauro:2024kml}).

Note that, for schemes with $V_a \neq 0$ ({\sc Big} and {\sc Quaint}), the fluxes can be non-vanishing for $K_{\bar p} > m_{\rm DM}$. This is the imprint of diffusive reacceleration, and it is consistent with previous results. In order to avoid an unphysical run-away behavior at very large kinetic energies induced by numerics, however, we cut the fluxes at a minimum floor chosen to be several orders of magnitude smaller than the peak value.

\section{Applications}
\label{sec:applications}

In this section we illustrate some implications based on the
CR fluxes computed in this work.
The main implication that one can draw is the reduction in the variation
of the DM-induced new CR fluxes (discussed in the previous section),
which in principle increases the discovery potential of a possible DM signal.

\begin{figure*}[!t]
\hspace{-8mm}
\begin{tabular}{ccc}
\includegraphics[width=0.52\textwidth]{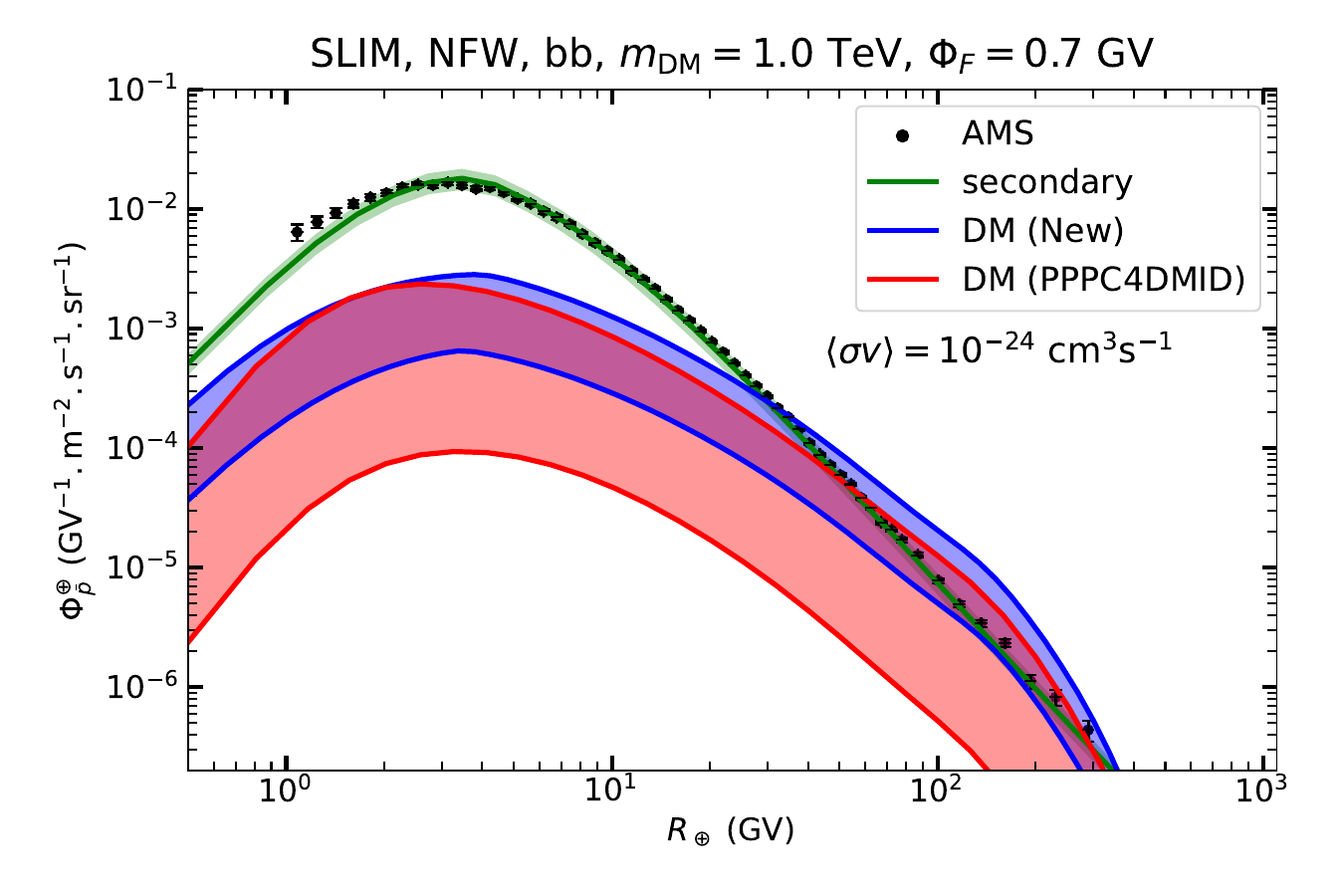}
\includegraphics[width=0.52\textwidth]{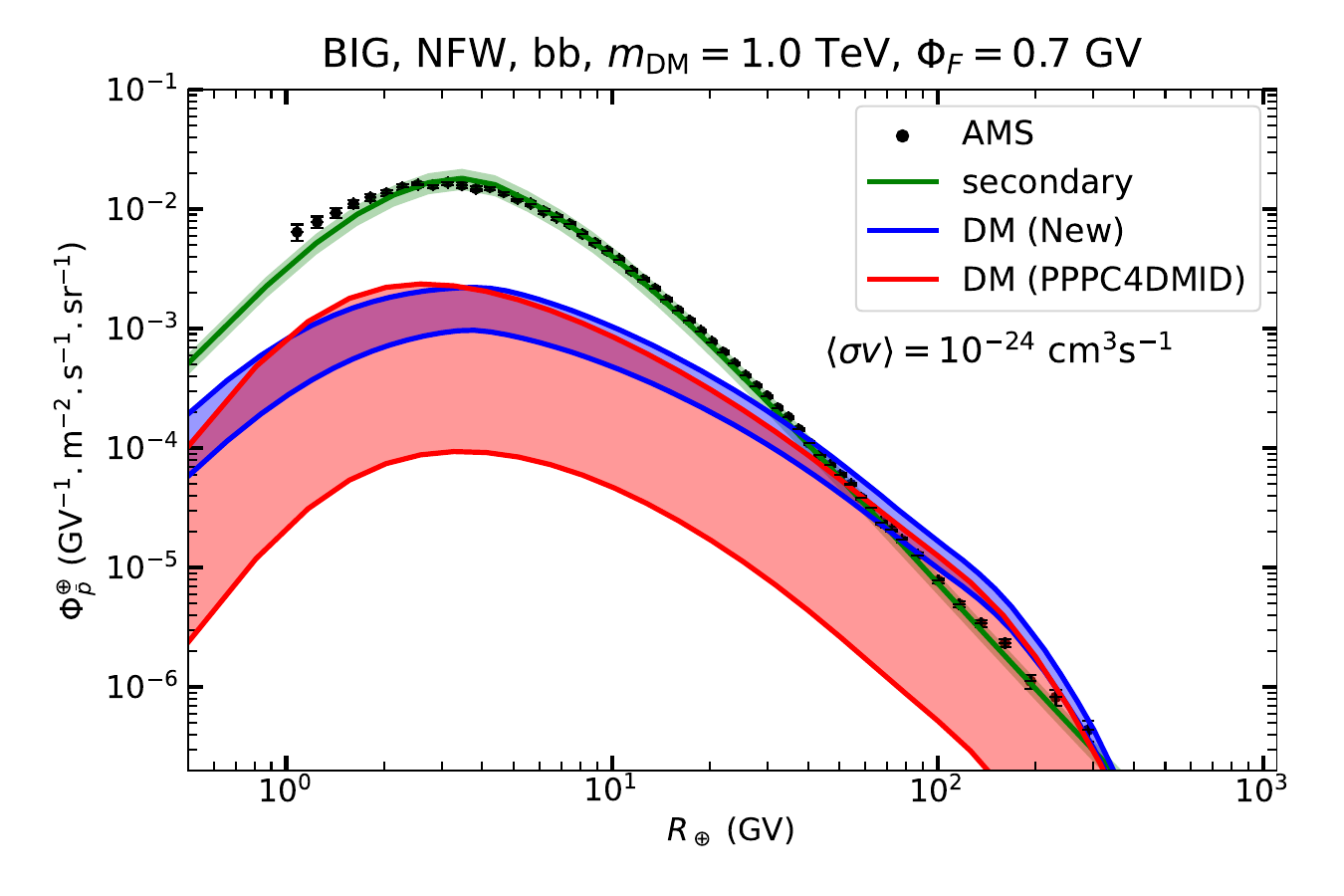}
\end{tabular}

\caption{Comparison of DM-induced and secondary background
$\bar{p}$ fluxes with the {\sc Ams-02} data~\cite{AMS02_data}. For the DM induced
flux we consider the annihilation of 1 TeV DM particles into $b\bar{b}$ and the NFW DM profile.
Here the ToA fluxes are shown considering a Fisk potential of
$\Phi_F = 0.7$ GV. The two panels correspond to the {\sc Slim} and {\sc Big}
propagation schemes. The blue and red lines for the DM-induced flux correspond to
the new propagation models and the old ones from \texttt{PPPC4DMID}, respectively.
In each case, the band corresponds to the
span between the MIN and the MAX scenarios.
The band around the secondary flux (green)
corresponds to a 20\% uncertainty in the flux normalization.
}
\label{fig:residual_pbarflux}
\end{figure*}

In Fig.~\ref{fig:residual_pbarflux}
we illustrate this basic point.
Here we show the comparison of the secondary background and
the DM induced $\bar{p}$ fluxes with the {\sc Ams-02} data~\cite{AMS02_data}.
We consider the ToA fluxes applying a Fisk potential of $\Phi_F = 0.7$ GV.
For the DM-induced fluxes, the bands correspond to the MIN -- MAX propagation sets
for the Old (red band) and New (blue band) propagation models.
The secondary flux (the baseline background) is obtained using
the parametric expression (for the IS flux) provided in \cite{Genolini2021MinMedMax}.
The band associated to the secondary flux is obtained assuming a
20\% uncertainty in the flux normalization, which is roughly the
maximum expected variation in the secondary flux over the transport parameter space
\cite{Genolini2021MinMedMax, Boudaud2020SecondaryOrigin}.

Fig.~\ref{fig:residual_pbarflux}
illustrates clearly that the uncertainties in the DM induced flux due to the
propagation scenarios
(MIN -- MAX) are reduced significantly (compared to those provided earlier by
\texttt{PPPC4DMID}), which increases the discovery potential for DM signature
in terms of the real CR observations provided by
experiments such as, for instance, {\sc Ams-02}~\cite{AMS02_data} or {\sc Gaps}~\cite{GAPS:2022ncd, Tiberio_2025}. In the right panel, for instance, the new fluxes would allow to identify an excess with respect to the secondaries and the data, irrespectively of the propagation uncertainties. The old fluxes, instead, would not, since their uncertainty band spans from a possible excess (for MAX, which corresponds to the higher edge of the stripe) to a negligible contribution (for MIN, which corresponds to its lower edge).

\section{Conclusions}
\label{sec:conclusions}

In this work we compute, {\it based on the updated models} described in Sec.~\ref{sec:Models}, the CR fluxes for antiprotons and
antideuterons induced by the annihilation/decay of weak-scale DM particles
in the Galaxy, and provide them for the users.
We consider the DM mass range 5 GeV -- 100 TeV (10 GeV -- 200 TeV)
and a broad set of annihilation (decay) channels.
The Galactic propagation of DM-induced $\bar{p}$ and $\bar{d}$ is treated in a
semi-analytic method (described in Sec.~\ref{sec:solution})
including energy-losses and diffusive reacceleration.
We provide all our results for the CR fluxes (the IS fluxes
in the kinetic energy range 0.1 GeV -- 100 TeV in a tabulated format)
for the users in the  \texttt{GitHub} repository of the  \href{https://github.com/CosmiXsPPPC}{\texttt{CosmiXsPPPC}} project.

\medskip

Compared to the similar works performed previously by
\texttt{PPPC4DMID}, the present work improves those previous estimates
mainly in the following ways.

\begin{itemize}
\item[$\blacktriangleright$] The first one is the use of the more refined $\bar{p}$ and $\bar{d}$ spectra provided by \texttt{CosmiXs} \cite{Arina:2023eic, DiMauro:2024kml},
which has improved such spectra by utilizing the Vincia shower algorithm in \texttt{Pythia},
as well as electroweak corrections at all orders and off-shell contributions involving the massive bosons.
Also, it tunes the relevant parameters to updated accelerator measurements.
We use the broad sets of annihilation/decay channels and the DM mass bins
provided in \texttt{CosmiXs}.

\item[$\blacktriangleright$] The second and {\it the most important} improvement is the treatment of the propagation of the DM-induced CR fluxes using the new MIN/MED/MAX propagation sets
obtained under the new Galactic propagation schemes: {\sc Slim}, {\sc Big} and {\sc Quaint} from
\cite{Genolini2021MinMedMax, Genolini2019Transport} (see Sec.~\ref{sec:Models}).
Such models were obtained by fitting recent CR data and hence are much more constrained.
Also, for the propagation of the DM-induced CR fluxes we
use updated scattering cross-sections for $\bar{p}$ and $\bar{d}$.

\item[$\blacktriangleright$] The other modification is the use of the new Galactic DM profiles with updated parameters (see Sec.~\ref{sec:Models}).
\end{itemize}

The main point that comes out from our results is the
reduction of the uncertainty band (due to the variation MIN--MAX propagation scenarios)
in the CR fluxes compared to that found in the previous estimates using the Old
propagation models. Such a phenomenon in principle is present for the three updated propagation
schemes considered here, but more prominent for the {\sc Big} and {\sc Quaint} schemes.

We have illustrated in Sec.~\ref{sec:applications}
how the above-mentioned phenomenon can in principle
increases the discovery potential for the DM signals in terms of the measurements of
the present and upcoming CR experiments.

We conclude by saying that the state-of-the-art calculations for
the CR fluxes presented here allow ones to compute robust constraints on a
broad class of weak-scale DM interaction for a wide mass range based on
the CR measurements in the kinetic energy range $\sim$100 MeV -- tens of TeV.
This will be hopefully instrumental in the current era of precision DM indirect searches.

\small
{\subsubsection*{Acknowledgments}
\footnotesize{
M.C and A.K. acknowledge the hospitality of the Institut d'Astrophysique de Paris ({\sc Iap}) where part of this work was done. M.C. also acknowledges the hospitality of the Theory Department at {\sc Cern}. The authors thank Yoann Genolini, David Maurin, Chiara Arina, Roberto De Austri, Nicolao Fornengo, Jan Heisig and Adil Jueid for useful discussions and potential future collaborations.

\noindent Funding and research infrastructure acknowledgments:
Research grant {\sl DaCoSMiG} from the {\sc 4eu+} Alliance (including Sorbonne Université). M.D.M. acknowledges support from the research grant {\sl TAsP (Theoretical Astroparticle Physics)} funded by Istituto Nazionale di Fisica Nucleare ({\sc Infn}). This project has also received financial support from the {\sc Cnrs} through the {\sc Miti} interdisciplinary programs ({\sc GammaNu} project).
}}

\appendix

\section{Validations of the calculations}
\label{sec:consistency_checks}

In this section we provide some validations of our calculations of the CR ($\bar{p}$) flux
by comparing our fluxes with those available in the literature or those
obtained from other publicly available numerical packages.

\begin{figure*}
\hspace{-23mm}
\begin{tabular}{cccc}
\includegraphics[width=0.42\textwidth]{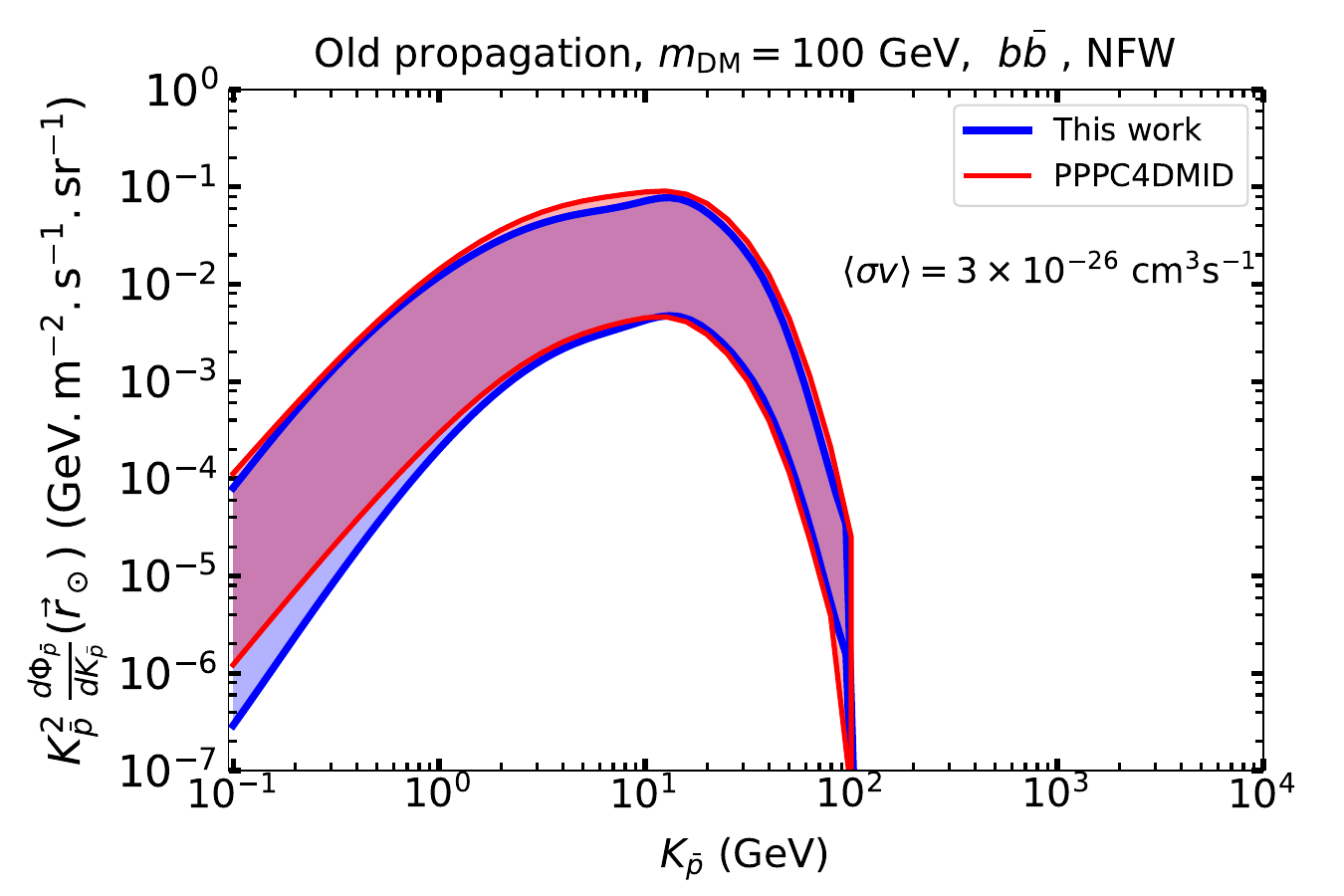} & \hspace{-6mm}
\includegraphics[width=0.42\textwidth]{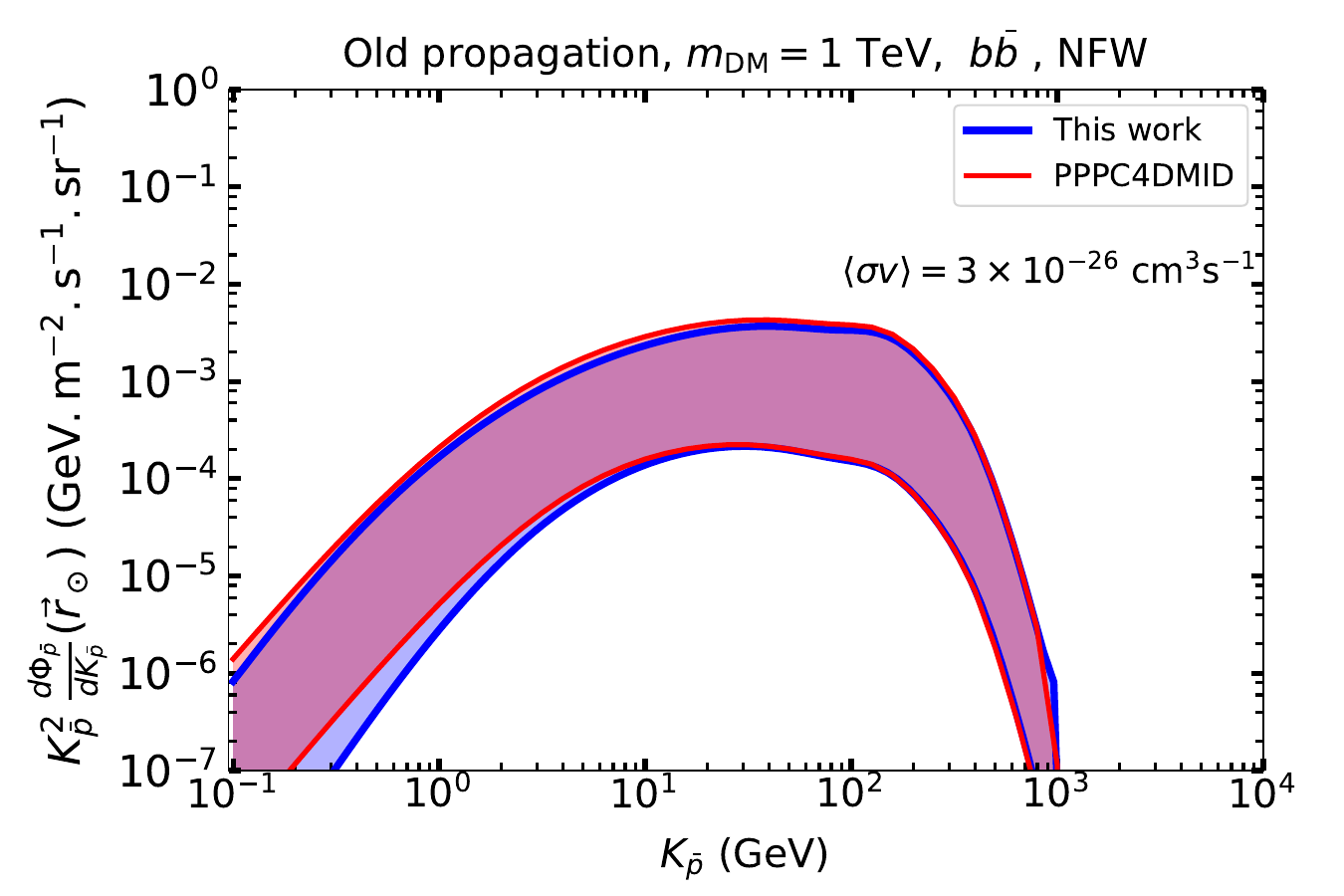} & \hspace{-6mm}
\includegraphics[width=0.42\textwidth]{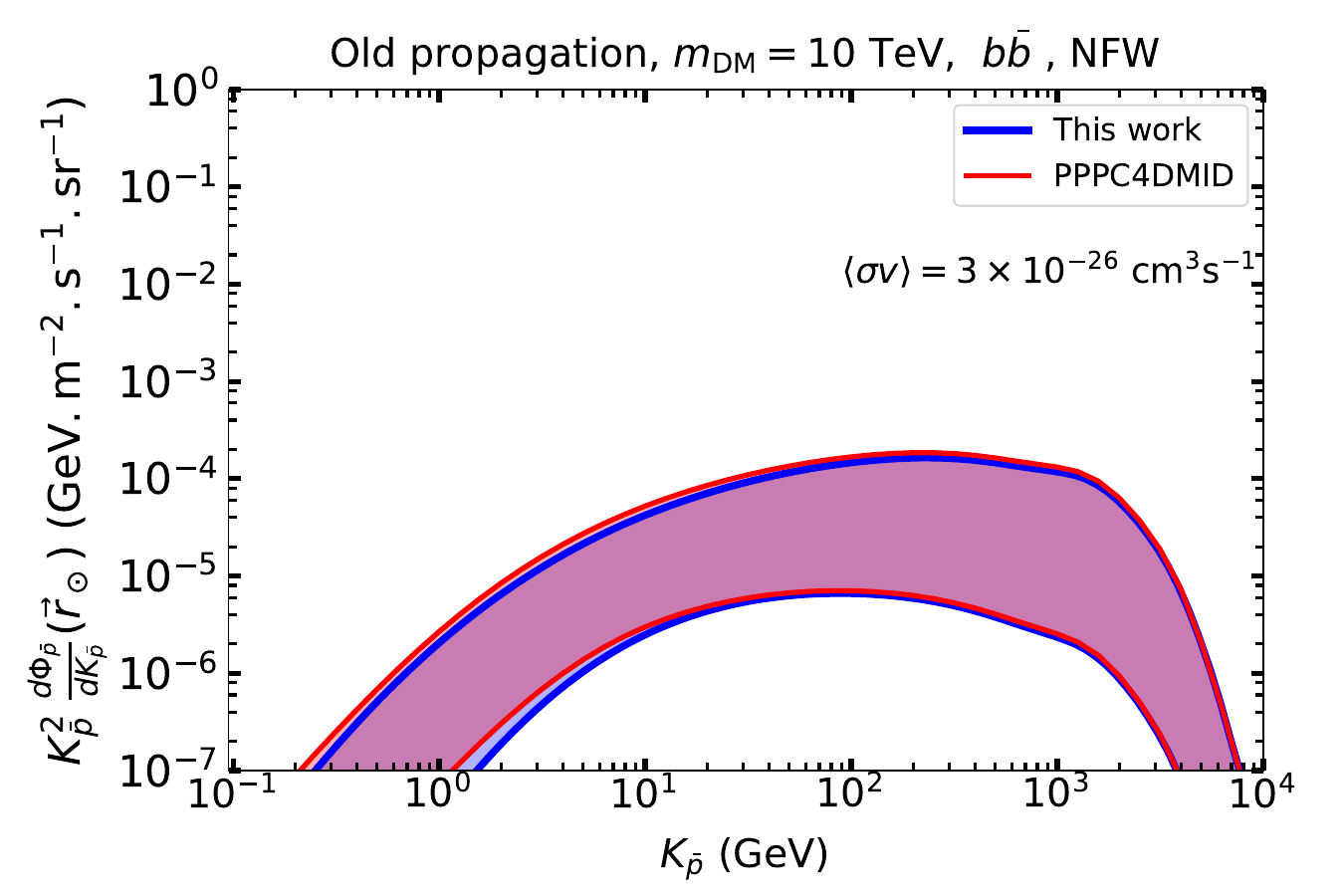}
\end{tabular}
\vspace{-4mm}
\caption{Comparison  between the $\bar{p}$ (IS) fluxes provided by the latest release of \texttt{PPPC4DMID} \cite{Boudaud2015Fussy} (red band) spanning the propagation scenarios MIN -- MAX and those obtained by our current computation (blue band) {\em when employing the same old propagation and DM density parameters as in} \texttt{PPPC4DMID}.
}
\label{fig:PPPC_oldProp_test}
\end{figure*}

\medskip

The first check that we make is the following.
We employ our present computational tools but reverting to the old setup used in the latest release of \texttt{PPPC4DMID} (release 6 based on \cite{Boudaud2015Fussy}, see \href{https://www.marcocirelli.net/PPPC4DMID.html}{\texttt{PPPC4DMID}}). Namely, we use the MIN-MED-MAX sets in Table 1 of \cite{Boudaud2015Fussy}, as well as the simplified expression for the spatial diffusion coefficient and the DM density profiles described there.
We then compare, in Fig.~\ref{fig:PPPC_oldProp_test}, with the fluxes presented in \cite{Boudaud2015Fussy}, finding an excellent agreement.
At low kinetic energies the fluxes estimated here are a bit smaller,
which is attributed to the fact that here we do not include the
`tertiary' antiparticles (as mentioned above in the main text).
This check just allows us to validate our present numerical pipeline and be confident that any difference with respect to the past computations is due to our updated physics.

\medskip

\begin{figure*}[!t]
\centering
\includegraphics[width=0.48\textwidth]{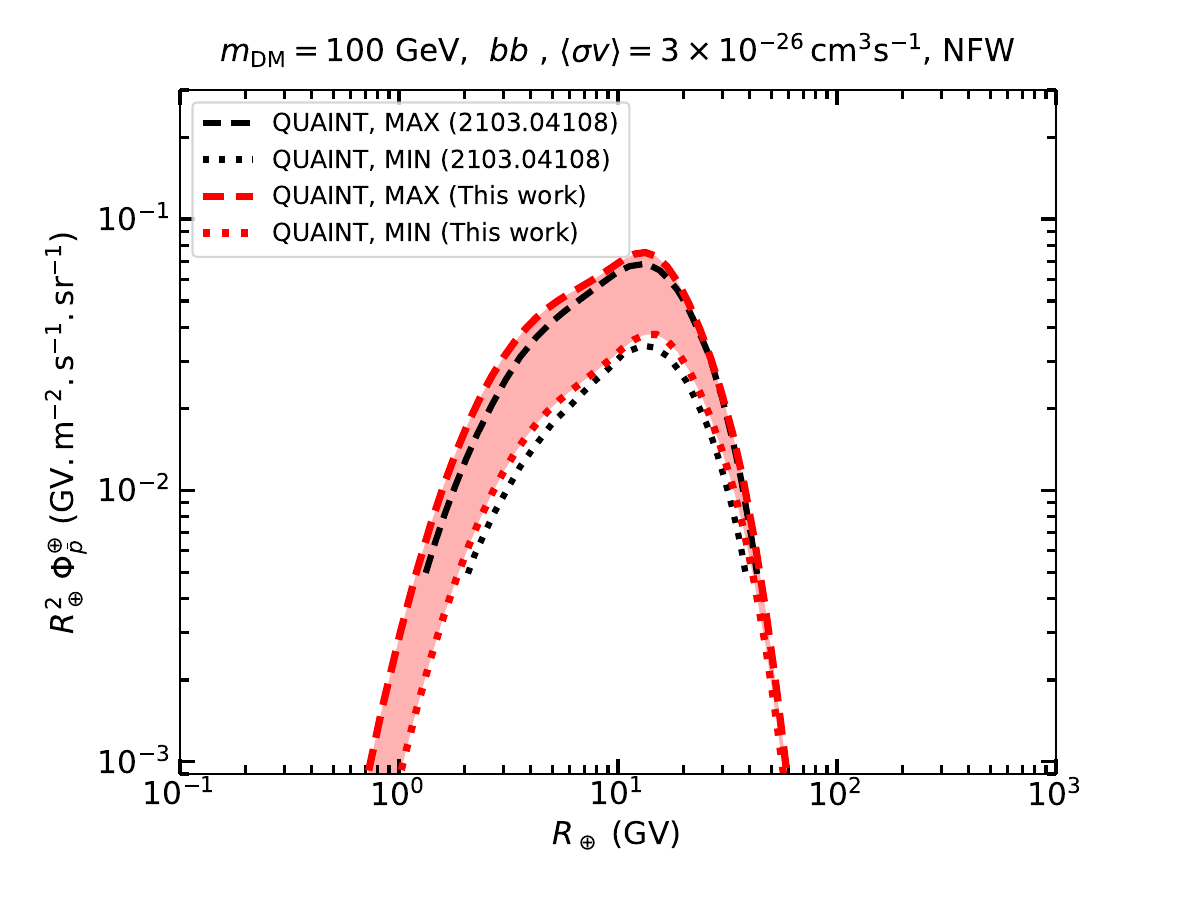} \hspace{-5mm}
\includegraphics[width=0.48\textwidth]{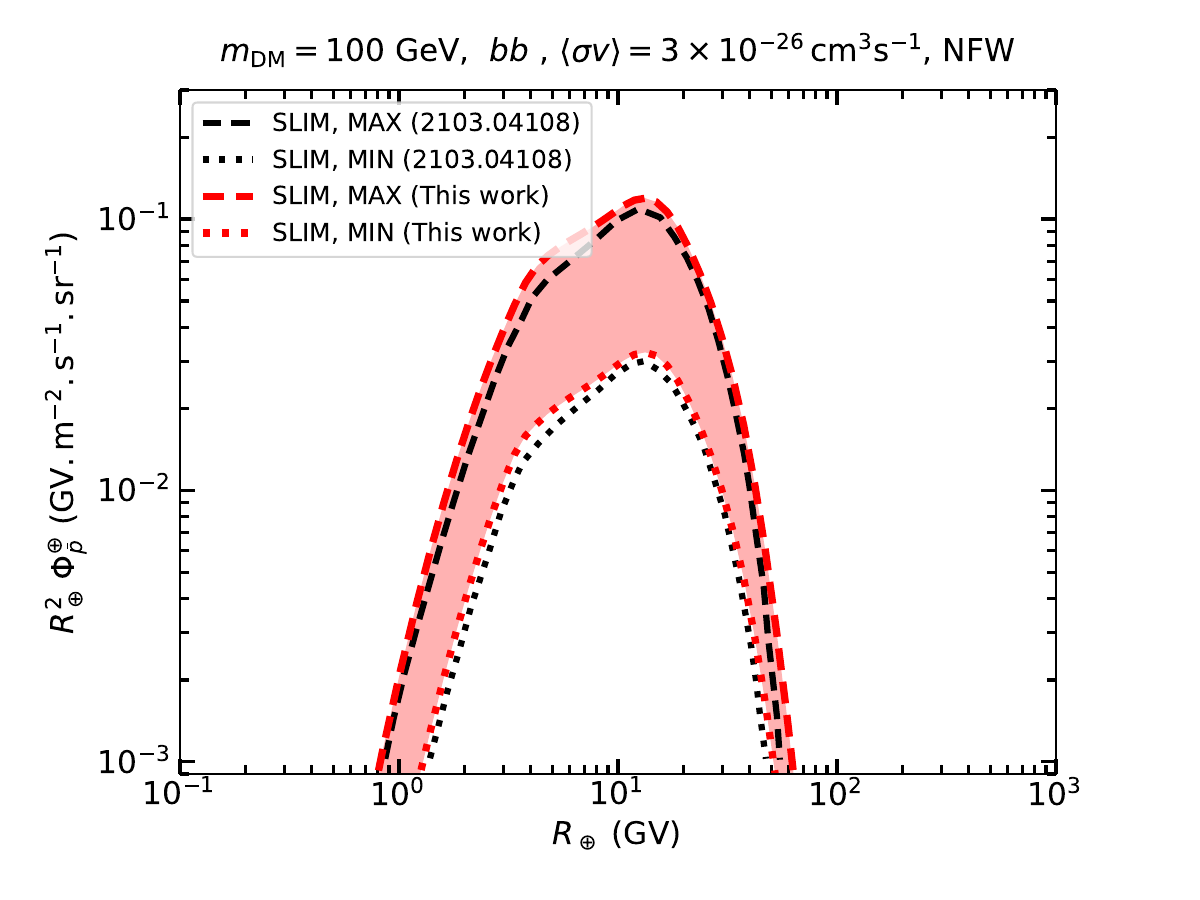}
\caption{The $\bar{p}$ (ToA) fluxes obtained in the present analysis are compared
with those from \cite{Genolini2021MinMedMax}, where such fluxes were obtained
using \texttt{USINE}~\cite{Maurin:2018rmm}. The Fisk potential $\Phi_F = 700$ MV.
Note that, in this particular comparative example,
the same parameter values for different quantities,
as considered in \cite{Genolini2021MinMedMax}, are used.}
\label{fig:Cirelli21_test}
\end{figure*}

The next test consists in comparing, in Fig.~\ref{fig:Cirelli21_test}, the $\bar{p}$ (ToA) fluxes
obtained in the present analysis with those from ref.~\cite{Genolini2021MinMedMax}, where such fluxes were obtained using \texttt{USINE}~\cite{Maurin:2018rmm}.
The Fisk potential $\Phi_F = 700$ MV. Note that, in this particular comparative example,
the same DM models and the same propagation parameter values considered in \cite{Genolini2021MinMedMax} are used.
The excellent agreement found with the refined and dedicated code \texttt{USINE} is not surprising, since the main ingredients of the computations are equivalent, and is reassuring for our systematic analysis.

\medskip

\begin{figure*}
\hspace{-23mm}
\begin{tabular}{cccc}
\includegraphics[width=0.42\textwidth]{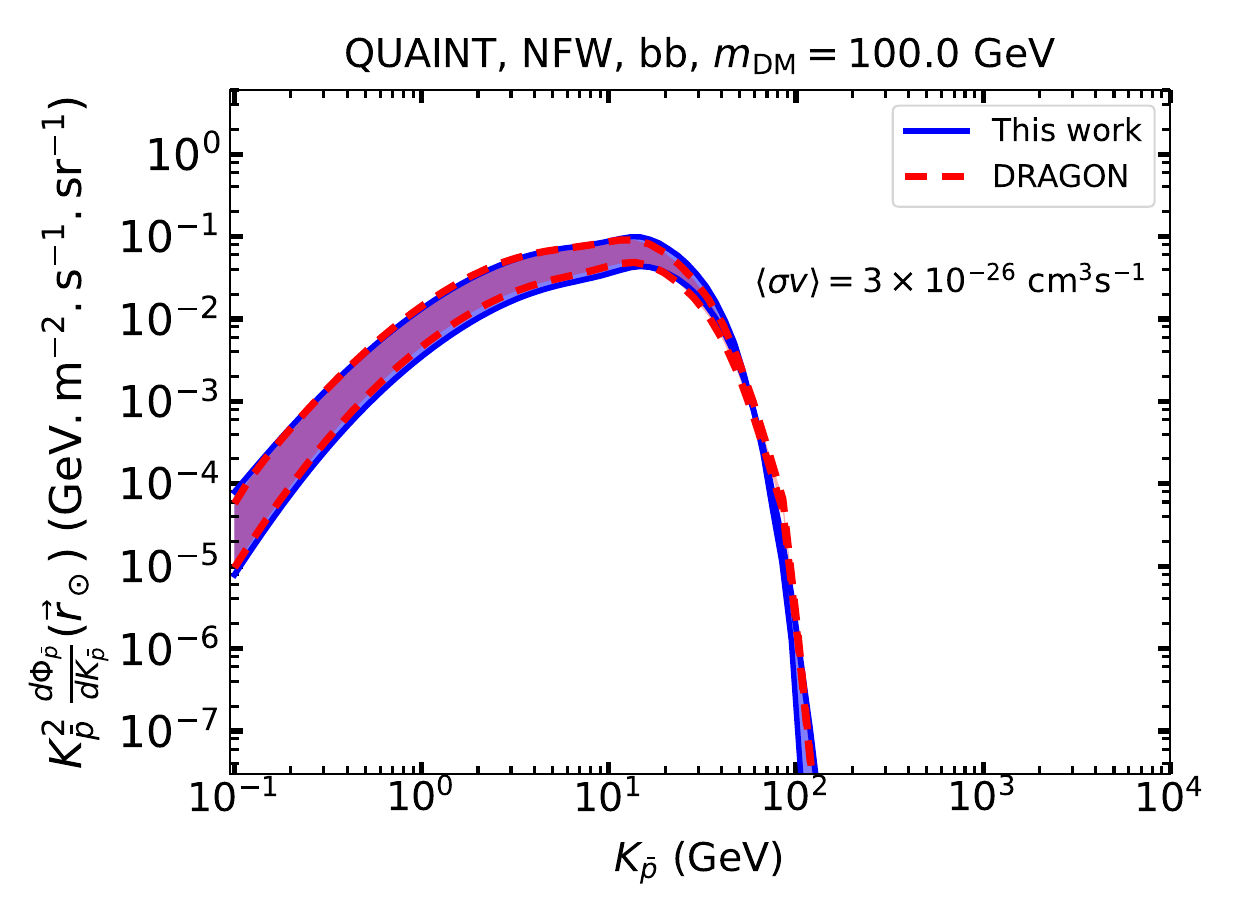} & \hspace{-6mm}
\includegraphics[width=0.42\textwidth]{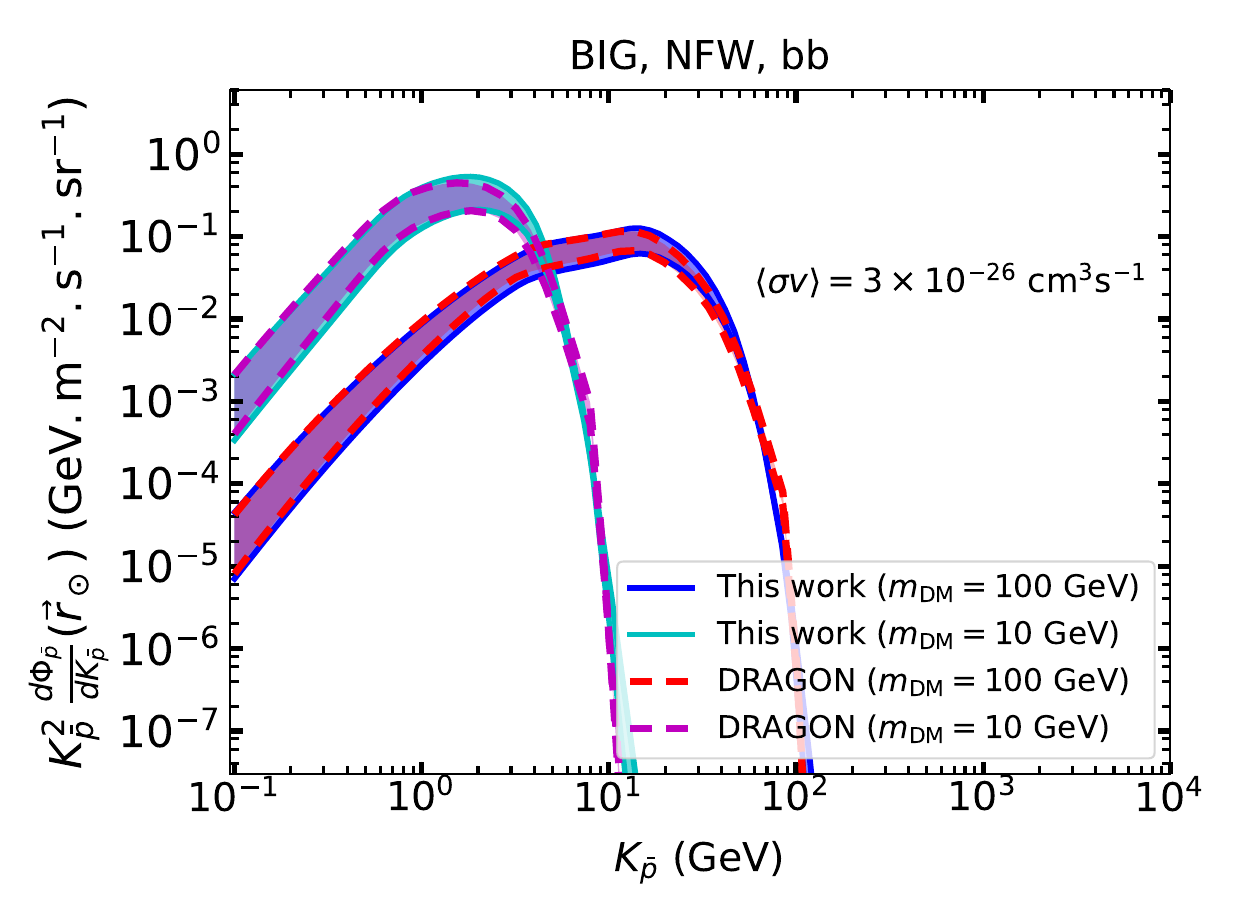} & \hspace{-6mm}
\includegraphics[width=0.42\textwidth]{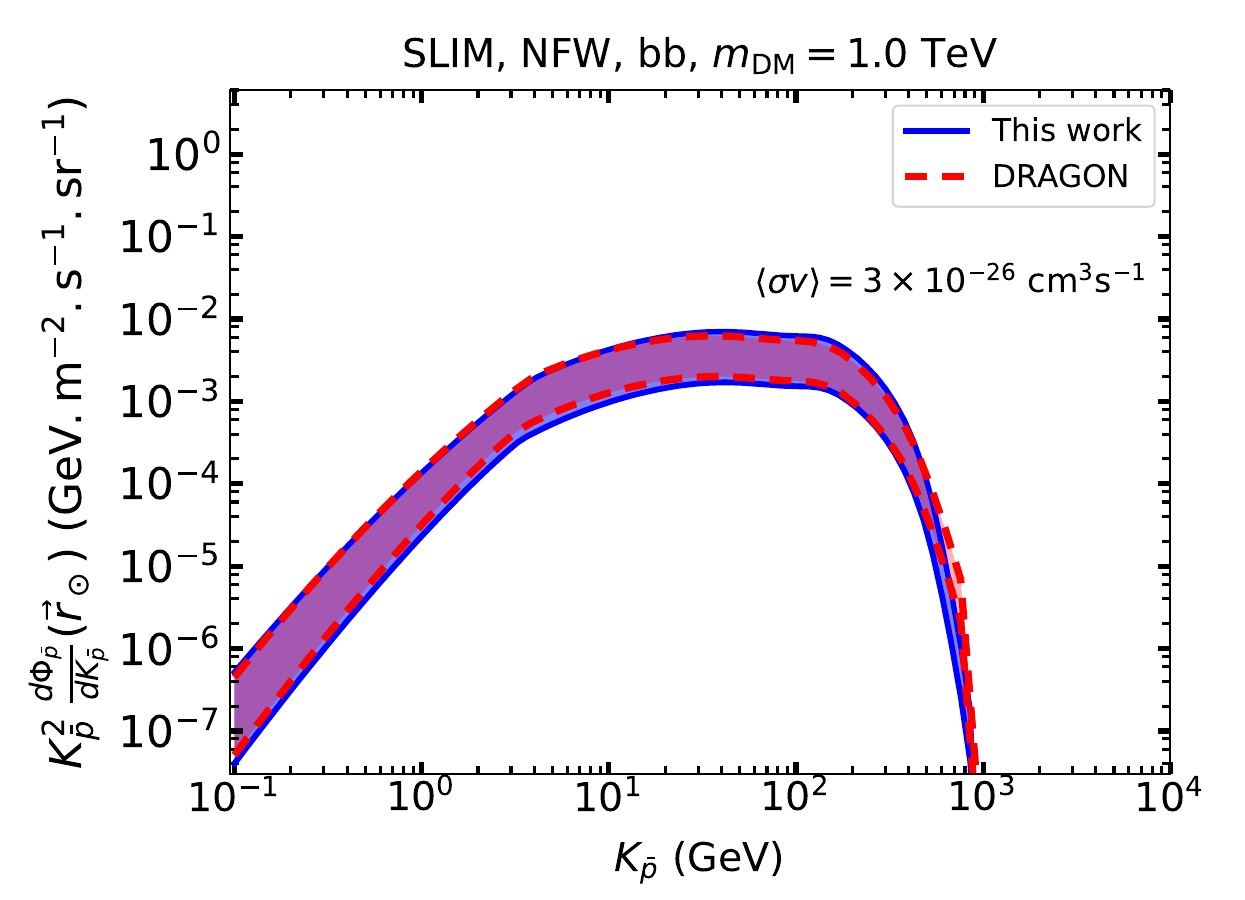}
\end{tabular}
\vspace{-4mm}
\caption{Comparison with \texttt{DRAGON}.
The $\bar{p}$ (IS) flux estimated under different scenarios are compared
with those obtained from \texttt{DRAGON}~\cite{DRAGON},
considering ${\rm DM \, DM} \rightarrow b\bar{b}$ annihilation.
In each case, the band corresponds to spanning the propagation scenarios  MIN -- MAX.
}
\label{fig:DRAGON_test}
\end{figure*}

Finally, in Fig.~\ref{fig:DRAGON_test}, we show the comparisons of the $\bar{p}$ flux estimated here
(under the models discussed in Sec.~\ref{sec:Models}) with those obtained by running the
numerical simulation \texttt{DRAGON}~\cite{DRAGON} (under the same models). The agreement is again excellent.
Note that, as mentioned in the main text, we consider a half-thickness of the Galactic disc $h = 200$ pc
(instead of the usual value $h=100$ pc) in order to get a better
match with \texttt{DRAGON}. This is mainly due to the fact that, in our semi-analytic method,
we consider that the ISM gases (i.e., hydrogen and helium) are
confined within the thin disc of half-thickness $h$ with a homogeneous density.
On the other hand, the numerical solver of \texttt{DRAGON}
considers a spatial distribution of the ISM gases whose tail can
extend in $|z|$ beyond 100 pc. By adopting a larger value of the thickness $h$ we can effectively mimic the detailed \texttt{DRAGON} results.

\section{Inelastic cross section model for $\bar{p}$, $\bar{d}$, and $\overline{{^3}\mathrm{He}}$}
\label{sec:inelastic}

To model the inelastic interaction of antinuclei with the interstellar medium, 
we adopt a
phenomenological Gaussian-profile eikonal ansatz inspired by the optical-limit Glauber
framework~\cite{Glauber:1955qq,Glauber:1970jm,Miller:2007ri,Shukla:2003mb}.
In the optical limit, the reaction cross section for a projectile $P$ on a target $T$ is
\begin{equation}
  \sigma_{R}^{PT}(E)
  = 2\pi \int_0^\infty b\,db\,
  \left[1-\exp\big(-\chi(b;E)\big)\right],
\end{equation}
where $\chi(b;E)$ is the eikonal opacity and depends on the overlap of the projectile and
target thickness functions.
We approximate the opacity by a Gaussian profile in impact
parameter,
\begin{equation}
  \chi(b;T/A) = \tau(T/A)\,\exp\!\left[-\frac{b^2}{2a^2(T/A)}\right],
\end{equation}
which leads to the analytic expression
\begin{equation}
  \sigma_{\rm inel}^{A_P A_T}(T/A)
  = 2\pi a^2(T/A)
  \left[
    \gamma_E
    + \ln\!\left(\frac{A_P A_T\,\sigma_{NN}^{\rm inel}(T/A)}{2\pi a^2(T/A)}\right)
    + E_1\!\left(\frac{A_P A_T\,\sigma_{NN}^{\rm inel}(T/A)}{2\pi a^2(T/A)}\right)
  \right],
  \label{eq:sig_inel_gauss}
\end{equation}
where $\gamma_E$ is Euler's constant and $E_1$ is the exponential integral.
Here $A_P$ and $A_T$ denote the projectile and target mass numbers, while
$\sigma_{NN}^{\rm inel}(T/A)$ is the elementary antiproton--nucleon inelastic
cross section evaluated at the same kinetic energy per nucleon.

\begin{figure*}[!t]
\centering
\includegraphics[width=0.49\textwidth]{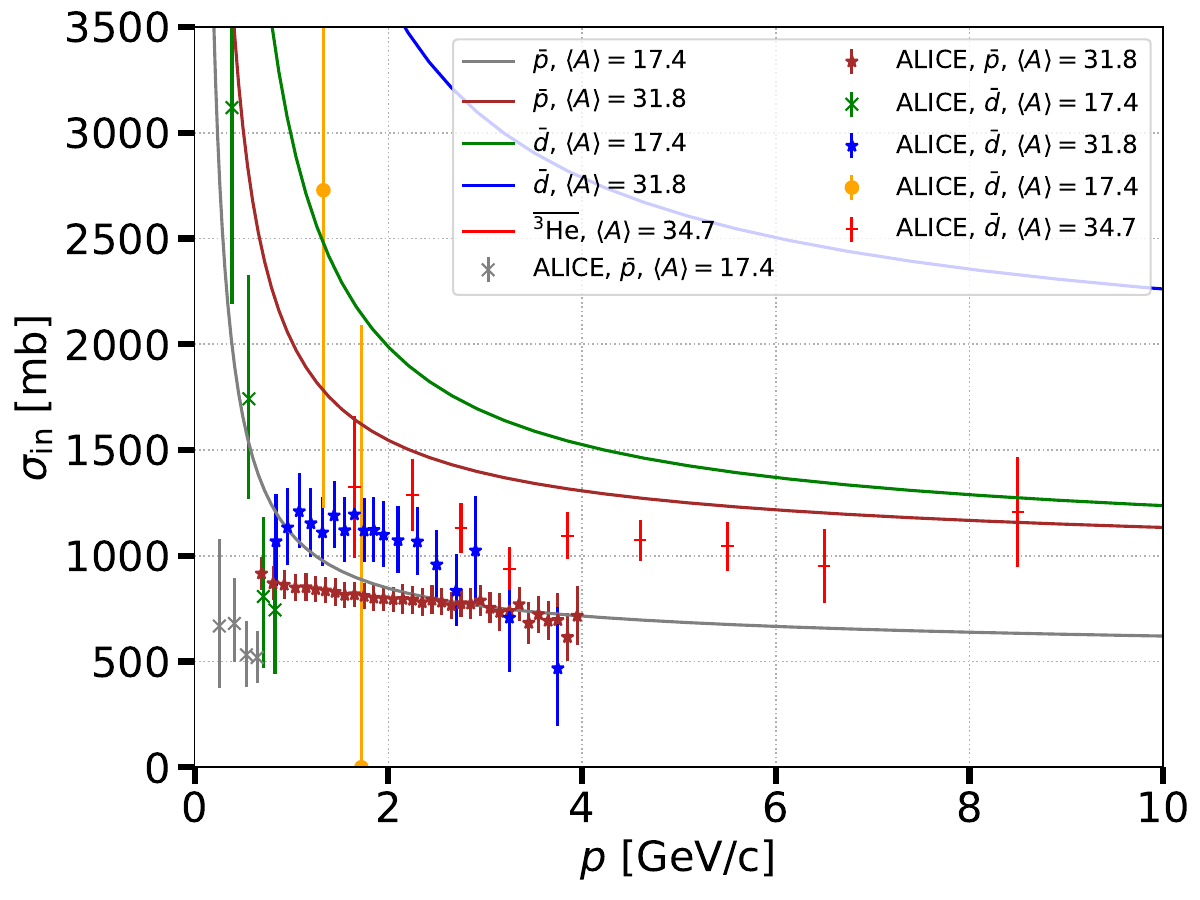}
\includegraphics[width=0.49\textwidth]{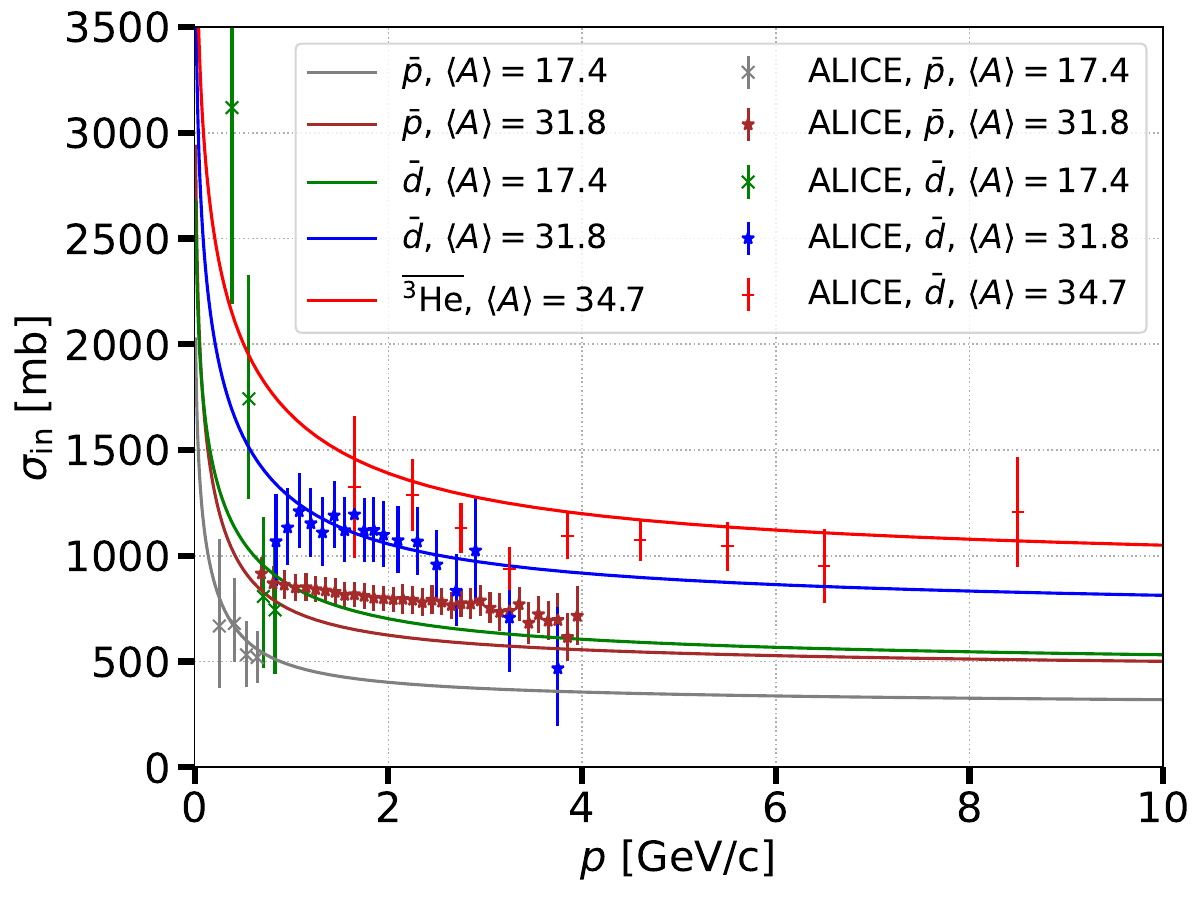}
\caption{The inelastic cross section $\sigma_{\rm inel}$ evaluated with the model in Eq.~(\ref{eq:sig_old}) (left panel) and with our new model explained in Appendix \ref{sec:inelastic} (right panel), compared with the data measured by {\sc Alice} on the detector target with average atomic mass $\langle A \rangle$ of about 17, 32 \cite{ALICE:2020mso,ALICE:2022zuz}.}
\label{fig:Gamma_inel2}
\end{figure*}

For the elementary input we use the traditional parametrization provided in \cite{Tan:1983}:
\begin{equation}
  \sigma_{NN}^{\rm inel}(K)\,[\mathrm{mb}]
  = 24.7\,
  \left[
    1 + 0.584\,K^{-0.115} + 0.256\,K^{-0.566}
  \right],
  \label{eq:sig_ppbar_param}
\end{equation}
where $K$ is the antiproton kinetic energy in GeV. In the present model,
this quantity enters the opacity kernel in Eq.~\eqref{eq:sig_inel_gauss}.

The effective width $a(T/A)$ is taken to be
\begin{equation}
  a^2(T/A)
  \simeq \frac{1}{2}\left(R_P^2+R_T^2\right)
  + k_B\,B_{NN}(T/A)\,(\hbar c)^2,
  \qquad
  R_i = r_0\,A_i^{1/3},
\end{equation}
with $r_0 \simeq 1.2~\mathrm{fm}$ and $k_B=1$ in the numerical implementation.
The first term accounts for the finite transverse sizes of the projectile and target,
whereas the second term incorporates the energy dependence of the elementary interaction
through the forward elastic slope parameter $B_{NN}$.

For $B_{NN}$ we adopt the logarithmic form
\begin{equation}
  B_{NN}(s)
  = B_0 + 2\alpha' \ln\!\left(\frac{s}{s_0}\right),
  \qquad
  s = 2m_p^2 + 2m_p\,(m_p + T/A),
  \qquad
  s_0 = (2m_p)^2,
  \label{eq:BNN}
\end{equation}
with $B_0 = 5.0~\mathrm{GeV}^{-2}$ and $\alpha' = 0.25~\mathrm{GeV}^{-2}$.
This choice provides a mild energy dependence of the effective interaction radius and
prevents an excessively rapid saturation of the inelastic cross section at low energy.

In the current implementation, Coulomb corrections are not included explicitly.
The model is therefore intended as a simple phenomenological interpolation between the
elementary $\bar{p}p$ input and the inelastic cross sections for composite projectiles
and effective nuclear targets.

The model used in Ref.~\cite{Cirelli:2010xx} and in several other works is instead much simple and given by:
\begin{equation}
  \sigma_{\rm{inel}}^{A_T,A_P}(K)\,[\mathrm{mb}]
  = A_T^{\chi_T} A_P^{\chi_P} \sigma_{\rm{inel}}^{N,N}(K),
  \label{eq:sig_old}
\end{equation}
where $\chi_P$ and $\chi_T$ takes into account the scaling of the inelastic $p\bar{p}$ cross section due to the projectile and target atomic numbers. These parameters are of the order of $0.6-1.0$ in analogy with the antiproton production cross sections \cite{Korsmeier:2018gcy,diMauro:2014zea}.

We show in Fig.~\ref{fig:Gamma_inel2} the comparison between our new model and the one in Eq.~(\ref{eq:sig_old}) assuming $\chi_T=2/6$ and $\chi_P=0.85$ with the {\sc Alice} measurement of the $\bar{p}$, $\overline{d}$ and $\overline{^3{\rm He}}$ inelastic cross sections measured for the $p-\rm{Pb}$ and $\rm{Pb}-\rm{Pb}$ collisions at 5 and 13 TeV for the center of mass energies and for particle momenta between 0.3-5 GeV/c$^2$.
These cross sections have been measured for the average {\sc Alice} detector material with atomic mass numbers between $\langle A\rangle = 17.4$ and $\langle A\rangle = 31.8$.

We observe that the new model reproduces well the weak dependence on the effective target mass observed in the {\sc Alice} data, where the cross sections for
$\langle A\rangle = 17.4$ and $\langle A\rangle = 31.8$ differ only moderately, rather
than following a pure $A_T^{2/3}$ scaling~\cite{ALICE:2020mso,ALICE:2022zuz}.

\bibliographystyle{JHEP}
\bibliography{bibliography.bib}

\end{document}